\documentclass{article}
\usepackage{pgf,pgfplots}
\usepackage{algpseudocode}
\usepackage{algorithm} 
\usepackage{graphicx}
\usepackage{amsmath}
\usepackage{amsfonts}
\usepackage{amssymb}
\usepackage{amsthm}
\usepackage{mathtools}
\usepackage{enumerate}
\usepackage{lscape}
\usepackage{longtable}
\usepackage{rotating}
\usepackage{multirow}
\usepackage{color}
\usepackage{url}
\usepackage{subfigure}
\usepackage{rotating}
\usepackage{comment}
\usepackage[utf8]{inputenc}
\usepackage{amssymb}
\usepackage{amsfonts}
\usepackage{tikz}
\usepackage[numbers]{natbib}
\usepackage{enumitem}
\usepackage[hidelinks]{hyperref}
\usetikzlibrary{arrows.meta, shapes, positioning, fit, backgrounds}

\newtheorem{theorem}{Theorem}

\title{Noncooperative Human AI Agent Dynamics}
\author{Dylan Waldner, Vyacheslav Kungurtsev, and Mitchelle Ashimosi}
\date{\today}

\begin{document}

\maketitle

\begin{abstract}
This paper investigates the dynamics of noncooperative interactions between artificial intelligence agents and human decision-makers in strategic environments. In particular, motivated by extensive literature in behavioral Economics, human agents are more faithfully modeled with respect to the state of the art using Prospect Theoretic preferences, while AI agents are modeled with standard expected utility maximization. Prospect Theory incorporates known cognitive heuristics employed by humans, including reference dependence and greater loss aversion relative to utility to relative gains. This paper runs different combinations of expected utility and prospect theoretic agents in a number of classic matrix games as well as examples specialized to tease out distinctions in strategic behavior with respect to preference functions, to explore the emergent behaviors from mixed population (human vs. AI) competition. Extensive numerical simulations are performed across AI, aware humans (those with full knowledge of the game structure and payoffs), and learning Prospect Agents (i.e., for AIs representing humans). A number of interesting observations and patterns show up, spanning barely distinguishable behavior, behavior corroborating Prospect preference anomalies in the theoretical literature, and unexpected surprises. Code can be found at this \href{https://github.com/dylanwaldner/noncooperative-human-AI}{link}.
\end{abstract}

\section{Introduction}
The proliferation of research and development in AI Agents has yielded an explosion of literature studying the nature and possible outcomes associated with their mutual influence. The potential for noncooperative interactions includes pursuit-evasion and tactics in war, bots automating economic activity and trading, and any circumstance wherein an automated AI system is representing some human actor or institution, with unaligned desiderata relative to other stakeholders. This yields the challenging problem of Multi Agent Noncooperative Reinforcement Learning (RL) and RL Nash Games~\cite{wang2022cooperative}. 

However, a natural question that has not been considered with reality-faithful modeling in this domain is what if humans and AI agents are both involved in these interactions? As far as epistemically valid models of human agents, we consider the field of behavioral economics, which has summarily rebutted the \emph{Homo Economicus} model of human decision making as being accurately modeled by maximizing a closed form utility function. While a number of alternatives have been proposed, in this work, we consider Prospect Theory (PT)~\cite{kahneman2013prospect} and its extensions, on account of the fact that it can be considered the leading model, and one whose properties are the most thoroughly axiomatized mathematically, and equilibrium properties understood. 

This paper focuses on empirically exploring the outcomes of noncooperative game play across permutations of a conventional reward-maximizing AI Learning Agent, a Prospect-maximizing Human with a priori complete knowledge of the game's payoffs, and a Prospect-maximizing Learning Agent representing a Human in the game. With comprehensive experiments across a set of classical noncooperative games as well as some games in the PT literature notable for their peculiarities in the presence of Prospect maximization.

The paper is the first paper to study the emergent asymptotic behaviors of mixed populations (PT vs. EU) in noncooperative $2\times2$ games, making the following contributions:

\begin{enumerate}
    \item Models humans and AIs as PT and EU preference agents. 
    \item Discerns between learning and aware agents. 
    \item Formulation and study of Q learning for multiple reference point regimes. 
\end{enumerate}

Section~\ref{sec:back} introduces background material and preliminaries for the mathematical notation used throughout the paper. Section~\ref{sec:Game} introduces the games suite used for the paper, including the payoff matrices and existing equilibria sets that exist in the literature. Section~\ref{sec:problem} presents the experiment structure, including the three types of agents and the algorithm used to define them. Numerical results are presented in Section~\ref{sec:num}, and we conclude with a discussion and future work in Section~\ref{sec:conc}.

\section{Background and Preliminaries}\label{sec:back}
\subsection{Nash Reinforcement Learning} \label{sec: RLNash}
Nash reinforcement learning (RL) is the field concerned with learning policies that converge towards equilibria. The general framework Nash RL builds on is a Markov game, also known as a stochastic game \cite{Shapley, Owen1982}, which is a multi-agent generalization of a Markov Decision Process (MDP), a formal framework for defining agent interactions with an environment. In this section, we will first introduce notation for an MDP, and then build the notation and intuition of a Markov game from the MDP foundation. Then, we will provide some background on Nash RL and what approaches have been conducted to this point. 

An MDP models an agent's interactions with an environment. Time is tracked as discrete time steps, $t = 0, 1, 2, 3, ...$. At each step the agent is presented an environment's state, $S_t \in \mathcal{S}$, and in response selects an action, $A_t \in \mathcal{A}$. As a consequence of that action, a reward $R_{t+1} \in \mathcal{R} \subset \mathbb{R}$ and the next state $S_{t+1} \in \mathcal{S}$ are generated with the transition function \cite{Sutton1998}:

\[(S_{t+1}, R_{t+1}) \sim P(\cdot \mid S_t, A_t)\]

A trajectory, or sequence, through an MDP can then be written as \cite{Sutton1998}:

\[(S_0, A_0, R_1, S_1, A_1, R_2,\dots)\]. 

Agents learn policies, $A_t \sim \pi(A_t \mid S_t)$, that maximize their value functions, which quantify the expected return that comes from being in a state and following a policy. Formally, this is written with the Bellman equation \cite{Bellman:DynamicProgramming}: 

\[V_\pi(S_t) = \mathbb{E}_\pi[R_{t+1} + \gamma V_\pi(S_{t+1})]\] 

where $\gamma \in [0, 1)$ is a discount factor. 

Markov Games are defined by a set of states $\mathcal{S}$ and a \textit{collection} of action sets $(\mathcal{A}_1, \mathcal{A}_2 ,\dots, \mathcal{A}_k)$. Actions are denoted by $A_{t, i} \in \mathcal{A}_i$, and the joint action set can be defined as $\mathbf{A}_t = ({A_{t, 1}},{A_{t, 2}} ,\dots, {A_{t, k}})$, where $k$ is the number of agents in the environment \cite{Littman1994MarkovGA}. Each agent has their own reward function, $R_{i}: \mathcal{S} \times \mathcal{A}_1 \times \mathcal{A}_2 \times \dots \times \mathcal{A}_k\rightarrow \mathbb{R}$, with the joint rewards at time $t$ defined as $\mathbf{R}_t = (R_{t, 1}, R_{t, 2}, \dots, R_{t, k})$. State-reward transitions can then defined as:

\[
(S_{t+1}, \mathbf{R}_{t+1}) \sim P(\cdot \mid S_t, \mathbf{A}_t).
\]

Agents then learn a policy, $\pi_i$ that they follow to maximize their value function $V_{\pi_i}$. Formally, where $\boldsymbol{\pi} = (\pi_1, \pi_2, \dots, \pi_k)$

\[
V_{\pi_i}(S_t) = \mathbb{E}_{\boldsymbol{\pi}}\left[R_{t, i} + \gamma V_{\pi_i}(S_{t+1})\right]
\]

where $\gamma \in [0, 1)$ is the discount factor. We use $\mathbf{\pi}$ as the joint policy because each agent's return depends on every other agent's actions \cite{Shapley}.

The Nash RL framework provides the theoretical foundation for analyzing strategic interactions among rational expected-utility maximizing agents, which we extend to incorporate human agents with Prospect Theoretic preferences in the next section. We use the RL Nash framework to study whether interactions between humans (modeled by Prospect Theory) and AIs (modeled by Expected Utility) converge to an equilibrium with the intention of forecasting potential competitive advantages each party may have over the other. The full formulation is discussed in Section~\ref{sec:ptne}.

\subsection{Prospect Theory}\label{sec:prospecttheory}

Behavioral Economics, using experiments of human decision making under scarcity, has revealed that the maximization of closed form utility functions is an incorrect model of human decision making. While the field of cognitive heuristics and biases is rich in a variety of sub-rational phenomena and alternative models, the most well respected theoretical framework comes from Prospect Theory as developed by Kahneman and Tversky, see, e.g.~\cite{kahneman2013prospect}. 

Prospect Theory (PT) emerged from extensive experimental evidence demonstrating systematic violations of expected utility theory. The theory introduces several key psychological principles: decisions are made relative to reference points rather than final states; losses loom larger than equivalent gains (loss aversion); people exhibit diminishing sensitivity to changes further from reference points; and probabilities are transformed through weighting functions that overweight small probabilities and underweight large ones.

Mathematically, the properties of a preference function $V$ as given in~\cite{page2022optimally} exhibits, given a human $h$ at a given time $t$:
\begin{enumerate}
    \item With any agreeable economic good or service $x$, more of $x$ would increase $V$, and losing $x$ from the possession of $h$ would decrease $V$ (existence of local preferences)
    \item However, the decrease is steeper than the increase (risk aversion)
    \item Decisions are made with respect to increasing local preference, rather than maximization or minimization
\end{enumerate}

Cumulative Prospect Theory (CPT) \cite{Tversky1992} extends Prospect Theory \cite{kahneman2013prospect} by introducing cumulative weighting functions that measure relative probability instead of independent probabilities. The general structure with a nonlinear value function, nonlinear probability weighting function, and a kink at the reference point remain constant, but the cumulative weighting function sorts outcomes and takes cumulative probabilities to preserve first order stochastic dominance. Intuitively, CPT asks "How likely is \textit{at least this good} of an outcome". 

Define a CPT \emph{Prospect} as a set of outcomes $o$ with corresponding probabilities $p$, $P=(o_1,p_1,o_2,p_2,\cdots, o_M,p_M)$, with outcomes ranked in ascending order, $o_1<o_2<\cdots <o_M$. A reference point $r$ defines an agent's baseline for comparison, splitting outcomes into gains and losses relative to $r$. This dynamic is expressed in the value function \(v_i(o, r)\), which formalizes the loss aversion and diminishing sensitivity to change in values:
\begin{equation} \label{eq:valuefunction}
v_i(o, r) = 
\begin{cases} 
(o - r)^{\alpha}, & o \geq r \\
-\lambda (r - o)^{\beta}, & o < r
\end{cases}
\end{equation}

where \(\alpha, \beta \in (0,1]\) control the curvature (diminishing sensitivity) and \(\lambda > 1\) captures loss aversion. 

Furthermore, Cumulative Prospect Theory shows that humans have non linear perceptions of probabilities. To model the 'inverse s-shape', we implement the weighted probability formulation from \cite{Tversky1992}:

\begin{equation}
w^+(p) = \frac{p^{\gamma}}{(p^{\gamma} + (1-p)^{\gamma})^{1/\gamma}}, \quad w^-(p) = \frac{p^{\delta}}{(p^{\delta} + (1-p)^{\delta})^{1/\delta}}
\end{equation}

with parameters \(\gamma, \delta \in (0,1]\). Based on empirical studies, typical parameter values are \(\alpha = \beta = 0.88\), \(\lambda = 2.25\), \(\gamma = 0.61\), and \(\delta = 0.69\). 

To bring the value function and weighted probability function into one Cumulative Prospect Theoretic transformation, for an agent $i$ with reference value $r_i$, the value of a prospect is given by~\cite{leclerc2014prospect},
\begin{equation}\label{eq:val}
v_i(P) = \sum\limits_{o_k<r_i}M^{-}(o_k) v_i(o_k,r_i)+\sum\limits_{o_k>r_i} M^{+}(o_k)v_i(o_k,r_i)
\end{equation}
with
\[
M^{-}(o_k) = w^{-}_i(\Phi(o_k))-w^{-}_i(\Phi(o_{k-1})),\,\,\, M^{+}(o_k) = 
w^{+}_i(\bar{\Phi}(o_k))-w^{+}_i(\bar{\Phi}(o_{k+1}))
\]
with $\Phi$ the cumulative distribution function and $\bar{\Phi}$ the decumulative distribution function, and $w^+$ and $w^-$ are both strictly increasing, differentiable and have ``inverse S'' shapes ($\frac{d^2 w^+(p)}{dp^2}$ is negative for $p<c$ and positive for $p>c$), and $v_i(o,r)$ is differentiable for all $o\neq r$, decreasing in $r$ and increasing in $o$. 

In repeated game settings with learning, Q-learning is used to estimate the payoffs associated with different multi-period strategies, while the greedy decision maximizes the expected Prospect, which is a function of this quantity, instead of the expected reward directly.

\subsection{Prospect Theory and Nash Equilibrium}\label{sec:ptne}

A natural question to ask is in the presence of noncooperative interactions between different agents, if one or more of the agents has preferences modeled by Prospect Theory, what best-response fixed points exist, and how do they relate to the classical Nash Equilibrium concept. 

The application of Prospect Theory in Nash Games appears in several recent works. A comprehensive PhD thesis on the subject is given in~\cite{leclerc2014prospect}. The recent papers \cite{phade2019geometry,cadre2025irrationality,keskin2016equilibrium} investigate some of the formal properties of the structure of Equilibria under Prospect Theory. Leclerc~\cite{leclerc2014prospect} explains that traditional Nash Equilibrium may be pathological, i.e. no equilibrium exists, when PT preferences are introduced because its non-linearities break Nash Equilibrium existence guarantees. Specifically, the non-linearities break the betweenness property from EU, and creates nonconvexities downstream in the strategy space that can lead to no solutions existing. This problem has led to multiple equilibrium concepts including Pure Action Nash Equilibrium, Mixed Action Nash Equilibrium, and various Black-Box Strategy equilibria as identified in recent literature~\cite{phade2020blackboxstrategiesequilibriumgames}.

However, in this paper, we borrow the Prospect Theoretic Equilibrium in Beliefs (PT-EB) concept as presented in~\cite{leclerc2014prospect},
\[
\sigma_i\in \mathop{co}\left\{\arg\max_{\sigma'_i\in \Delta_i} V_i(\sigma'_i,\sigma_{-i};r_i)\right\}
\]
and variations are discussed based upon the value of $r_i$. Notably, in PT-EB agents mix over opponent actions, not their own, and the convex hull is taken to convexify the strategy space. Theorem~\ref{thm:2.3} from~\cite{leclerc2014prospect} shows that PT-EB resolves the Prospect Theory Nash Equilibrium (PT-NE) issue and restores equilibria to games:

\begin{theorem}[Adapted from \cite{leclerc2014prospect}]\label{thm:2.3}
For any finite, simultaneous $k$-player game with PT preferences, complete preferences, 
and fixed vectors $(r_i^{\min})_{i=1}^k$, $(r_i^{\max})_{i=1}^k \in \mathbb{R}^k$ 
(with $r_i^{\min} \le r_i^{\max}$ for each $i$), and given continuous functions 
$f_i : D \times [r_i^{\min}, r_i^{\max}] \rightarrow [r_i^{\min}, r_i^{\max}]$, 
there exists $s \in D$ such that for each player $i$, $
s_i \in \operatorname{co}\!\left(\arg\max_{s_i' \in D_i} v_i(s_i', s_{-i}; r_i)\right),
$ and $
r_i = f_i(s, r_i).
$
\end{theorem}

Theorem~\ref{thm:2.3} certifies existence under certain properties of the functional form. It can be observed that anomalies as far as equilibrium nonexistence or absurdity arise in the general case. Importantly, \citeauthor{leclerc2014prospect} defines the PT transformation after \cite{Tversky1992}, with cumulative probability weighting functions and sorted outcomes, so PT-EB can be effectively understood to be CPT-EB. 

\paragraph{Reference Points}

CPT rewards are relative to some reference point. This reference point is associated with the psychological observation of adaptation in affect to exposure to rewards. This definition is flexible and can correspond to real historical rewards or expectations based on external calculations. The nature and existence of equilibria then can depend, depending on the game structure, on the manner in which the reference point is computed. 

When $r_i=\mathbb{E}[X^{\sigma}]$, that is, the expected state outcome at the equilibrium, we can observe from~\cite{leclerc2014prospect} that:
\begin{itemize}
    \item Pure strategies are equivalently classical Nash Equilibria and Prospect Theoretic Equilibrium in Beliefs
    \item Mixed strategies, however, become notably distinct, even for one-off games
\end{itemize}

\subsection{Applications}

The integration of Prospect Theory into multi-agent systems reveals important implications across several contemporary application domains where humans and AI systems interact strategically. In algorithmic trading environments, AI systems typically optimize for expected returns while human traders exhibit well-documented behavioral patterns including loss aversion and non-linear probability weighting \cite{disposition, investorreluctantodean}. This fundamental mismatch in decision-making processes can lead to market phenomena that deviate from traditional rational expectations models, including the disposition effect where humans hold losing investments too long and sell winning investments too quickly \cite{disposition}, and excess volatility that cannot be explained by standard financial models.

In autonomous vehicle navigation and interaction, the mixed autonomy scenario presents unique challenges. Self-driving cars employing expected utility maximization must coexist with human drivers who make decisions based on reference-dependent preferences and framing effects \cite{fridmanmitcars}, a problem partially addressed by \cite{SadighAutonomous} and potentially made more robust with PT preferences. Human drivers may accept different risk profiles when situations are framed as "gaining time" versus "avoiding being late," \cite{Tversky1992,kahneman2013prospect} and their responses to near-miss events may be influenced more by the emotional impact of almost experiencing a loss rather than the actual probability calculations (\cite{fridmanmitcars} shows human drivers are suboptimal and difficult to model). This creates complex interaction dynamics that pure reinforcement learning approaches may fail to capture adequately.

Cybersecurity represents another critical domain where behavioral considerations matter significantly. AI defense systems often face human attackers whose risk assessment follows Prospect Theory patterns rather than expected utility maximization \cite{aggRiskAverse, AGGARWAL2022102671}. 
This leads to attack patterns and security vulnerabilities that may not emerge in simulations where all agents are modeled as rational expected utility maximizers \cite{Shogren1990, secureGrosslags}.

Resource allocation in shared computing environments and network management systems provides a fourth important application area. AI schedulers and resource managers typically optimize for system-wide efficiency metrics, while human users evaluate allocation decisions relative to reference points \cite{Tversky1992, kahneman2013prospect} and exhibit strong loss aversion when facing reductions from their current allocation levels \cite{Tversky1992, kahneman2013prospect}. This can lead to resistance to otherwise efficiency-improving changes \cite{KimKankanhalli2009} and requires careful consideration of the behavioral aspects of human users in system design \cite{FehrFair}.

Finally, modeling the interactions between AIs and Prospect Theory agents provides general insight into the competitive dynamics between AIs and humans. AIs are increasingly being designed to be autonomous and share the world with humans, and goal oriented AIs may need to compete with humans to succeed. Applyng PT preferences to human agents better models the cognitive biases that will have downstream effects on equilibria and convergence points between human and AI dynamics. Importantly, better models yield better information for which humans can plan, govern, and integrate AI into society.

\subsection{Previous Work and Contributions}

The foundation for integrating Prospect Theory into game-theoretic analysis was established through several key works. Leclerc's comprehensive dissertation~\cite{leclerc2014prospect} provides the fundamental framework for Prospect Theory preferences in noncooperative games, proving existence results for various equilibrium concepts and analyzing classical games under different reference point formulations. This work demonstrates that pure strategy equilibria under Prospect Theory often align with classical Nash equilibria, while mixed strategy equilibria can differ substantially due to the non-linear probability weighting characteristic of human decision-making.

Other research has expanded this foundation in several directions. Recent work by \cite{phade2019geometry} has examined the geometric properties of Nash and correlated equilibria under CPT, demonstrating that the set of correlated equilibria can be disconnected—a structural property that differs fundamentally from classical game theory under expected utility. This geometric insight has important implications for learning dynamics, as it suggests that convergence to equilibrium may follow more complex pathways in CPT environments. Keskin~\cite{keskin2016equilibrium} develops specialized equilibrium concepts for agents with CPT preferences and establishes conditions for their existence and computational properties. More recently, Cadre and colleagues~\cite{cadre2025irrationality} explore how various forms of irrationality, including Prospect Theory preferences, shape Nash equilibria in strategic settings. 

The intersection of multi-agent reinforcement learning with CPT preferences presents particular complexities~\cite{ghaemi2024namg}. Standard RL algorithms based on expected return maximization require fundamental modifications to accommodate reference-dependent valuation, loss aversion, and probability weighting~\cite{a2016cumulativeprospecttheorymeets, lepel2025prospecttheoreticpolicygradientframework}. Furthermore, in mixed populations where AI agents employ expected utility maximization while human agents follow CPT principles, the learning dynamics become coupled in ways that existing MARL theories cannot adequately address.

Despite these theoretical advances, significant gaps remain in the literature. Most existing work focuses on one-shot games~\cite{keskin2016equilibrium, HOTA2016135, METZGER2019396} or homogeneous populations~\cite{cadre2025irrationality, METZGER2019396, ghaemi2024namg} where all agents share the same decision-making framework. The dynamics of repeated interactions between heterogeneous agents—specifically between AI systems employing expected utility maximization and human decision-makers following Prospect Theory—remain largely unexplored.

Extending these approaches to larger, multi-agent settings is particularly challenging given that the violation of the betweenness axiom by CPT necessitates more nuanced equilibrium concepts \cite{phade2020blackboxstrategiesequilibriumgames}. Similarly, while evolutionary game theory has explored the (semi-)stability of prospect theory preferences in population dynamics~\cite{RIEGER20141}, the learning-theoretic aspects of how individual agents with CPT preferences adapt their strategies over time have been limited to small, tabular, and homogeneous settings~\cite{ghaemi2024namg} and single agent settings~\cite{lepel2025prospecttheoreticpolicygradientframework, ramasubramanian2021reinforcementlearningexpectation}.

Furthermore, the learning and adaptation processes in mixed populations (CPT/EU) have received limited attention, with most analysis focusing on equilibrium characterization~\cite{phade2020blackboxstrategiesequilibriumgames, phade2020learninggamescumulativeprospect}, algorithmic (non-game theoretic) convergence~\cite{Borkar_2021, ramasubramanian2021reinforcementlearningexpectation, lalmohammed2025modeling}, or non PT-equilibria in homogeneous populations~\cite{danisMARL2023, ghaemi2024namg}.

Our work aims to address this gap by empirically studying the asymptotic outcomes of repeated interactions between expected-utility maximizing AI agents and CPT-based human decision-makers (heterogeneous). We contribute to the literature by formalizing the problem of heterogeneous human-AI strategic interactions, studying agents with fundamentally different preferences, and demonstrating through empirical analysis the emergent behaviors that arise specifically from the interaction of different decision-making frameworks. 

Furthermore, we study the impact of dynamic reference point adaptations. While \cite{leclerc2014prospect} establishes equilibrium existence with fixed reference points, most real-world strategic interactions involve evolving reference points based on historical outcomes, social comparisons, or aspiration levels. The interaction between reference point dynamics and equilibrium convergence specifically in the context of CPT/EU (approximating Human/AI) provides a useful model for the systemic investigation of mixed human-AI interactions. The model yields insight as to how PT preferences, and thus realistic Human decisions and behavior, interact with standard expected reward maximizing AI Agents. 

We consider this model to be a complementary work to the broader field of AI safety as a model of how humans and AIs may interact in non-cooperative settings. In particular, the model allows us to examine how AI agents that optimize expected utility interact strategically with human decision-makers whose preferences follow Prospect Theory. Such interactions may create systematic advantages or vulnerabilities arising from human behavioral biases. The goal of this work is not to provide a predictive model of human–AI behavior, but rather to offer an early analytical framework for comparing human and AI preferences and studying how psychological features such as reference points may influence human–AI strategic relationships.

\section{Game Suite}\label{sec:Game} 
In this section we present the game suite for our experiments. We split the games into three types: \nameref{sec:classical} contains five classical game theory games and  \nameref{sec:anomalous} contains two PT anomaly inducing games. The point of the game selection is to ground our agents in classical theory before exploring pathology inducing games to search for equilibria.


\subsection{Classical Games} \label{sec:classical}

We analyze five classical games that demonstrate the full range of Prospect Theory effects and provide the foundation for our experimental analysis. These games allow us to systematically study how Prospect Theory preferences interact with different game structures and how these interactions affect learning dynamics in mixed human-AI populations. All equilibria are presented in the for $(p, q)$, where $p$ is the probability that player one plays the first action, and $q$ is the probability player two plays the first action. For example, in prisoner's dilemma, $(p, q)$ refer to the probability that players one and two cooperate. 

\subsubsection{Prisoner's Dilemma}

The Prisoner's Dilemma (PD) is a dominance-solvable game where strategic incentives create social dilemmas, with a canonical equilibrium at $(0, 0)$. The payoff matrix we use is: 

\[
\begin{pmatrix}(-1,-1)&(-3,0)\\ (0,-3)&(-2,-2)\end{pmatrix}
\]

With PT players, \cite{leclerc2014prospect} demonstrates that dominance-solvable games like PD are robust to reference point selection. He demonstrates analytically that PD has a $(0, 0)$ equilibrium in Nash Equilibrium, Prospect Theoretic Nash Equilibrium (PT-NE), and in Prospect Theoretic Equilibrium of Beliefs (PT-EB) \cite{leclerc2014prospect}.

\subsubsection{Matching Pennies (MP)}

Matching Pennies is a zero-sum game with a canonically unique mixed strategy equilibrium at $(0.5, 0.5)$. The payoff matrix we use is:

\[
\begin{bmatrix}(1,0)&(0,1)\\ (0,1)&(1,0)\end{bmatrix}
\]

Leclerc finds that, with PT players, Matching Pennies has an equilibrium at $(1/2, 1/2)$ in PT-NE and PT-EB \cite{leclerc2014prospect}. 

\subsubsection{Battle of the Sexes}

Battle of the Sexes is a coordination game with three canonical equilibria: $(0, 0)$, $(1, 1)$, and $(3/5, 2/5)$. The payoff matrix we use is:

\[
\begin{bmatrix}(3,2)&(0,0)\\ (0,0)&(2,3)\end{bmatrix}
\]

Leclerc \cite{leclerc2014prospect} does not analyze Battle of the Sexes, so we leave PT-NE and PT-EB equilibria as open questions.  

\subsubsection{Stag Hunt}

Stag Hunt captures the tension between risk-dominant and payoff-dominant equilibria, which makes it particularly relevant to PT research as PT players are loss averse. The canonical equilibria are $(0, 0)$, $(1, 1)$, and, for this payoff matrix, $(1/2, 1/2)$:

\[
\begin{bmatrix}(3,3)&(0,2)\\ (2,0)&(1,1)\end{bmatrix}
\]

Leclerc \cite{leclerc2014prospect} does not analyze Stag Hunt, so we leave the PT-NE and PT-EB as open questions.

\subsubsection{Chicken}

Chicken (also known as Hawk-Dove) is an anti-coordination brinksmanship game that has canonical equilibria at $(0, 1)$ and $(1, 0)$, and the mixed equilibrium specific to our payoff matrix is $(9/10, 9/10)$. The payoff matrix we use is:

\[
\begin{bmatrix}
    (0,0)&(-1,1)\\ (1,-1)&(-10,-10)
\end{bmatrix}
\]

Leclerc does not analyze Chicken, so we leave its PT-NE and PT-EB equilibria as open questions.


 \subsection{Anomalous Games} \label{sec:anomalous}

Leclerc's thesis \cite{leclerc2014prospect} provides several critical examples where Prospect Theory preferences lead to anomalies in classical game-theoretic predictions. These examples demonstrate how PT preferences can break Nash Equilibria and contextualize our learning agent analysis as an exploration for equilibrium. 

\subsubsection{Ochs' Game}

The Ochs' game \cite{OCHS1995202} represents a pathological case where PT-NE breaks down. We use Ochs' game with $c=4$:

\[
\begin{bmatrix}(4, 0)&(0, 1)\\(0, 1)&(1, 0)\end{bmatrix}
\]

classical game theory predicts a mixed strategy Nash equilibrium at $(0.2, 0.5)$ \cite{OCHS1995202}. However, Leclerc demonstrates that PT-NE fails to exist, and that PT-EB restores existence, yielding $(p, q) \approx (0.5, 0.05)$.

\subsubsection{Crawford's Counterexample}

Crawford's Counterexample \cite{CRAWFORD1990127} is explicitly designed to break equilibrium if players have preferences that violate the Von Neumann-Morgenstern independence axiom --- which PT preferences do. When both players have EU preferences, the Nash Equilibrium is $(1/2, 1/2)$. The payoff matrix is: 

\[
\begin{pmatrix}(2,-2)&(0,0)\\(0,0)&(-1,1)\end{pmatrix}
\]

Leclerc shows that PT-NE does not exist for Crawford's Counterexample, because quasi-convexity of one player's preferences (as is the case with PT) implies a player will use a pure strategy, but there is clearly no pure equilibrium for this game \cite{leclerc2014prospect}. Leclerc does not analyze PT-EB for Crawford's Counterexample, so we leave PT-EB open.

\section{Problem Formulation and Algorithms}\label{sec:problem}

We consider a two-player Markov game defined by the tuple $(\mathcal{S}, \mathcal{A}_1, \mathcal{A}_2, P, R_1, R_2)$, where one player is an AI agent and the other is a human agent. The state space $\mathcal{S}$, action spaces $\mathcal{A}_1, \mathcal{A}_2$, and joint action $\mathbf{A}_t = (A_{t,1}, A_{t,2})$ follow the Markov game formalism established in Section~\ref{sec: RLNash}. We suppress $i$ in the action notation $A_{i, t}$ when $i$ is clear for brevity. State-reward transitions are generated as:

\[
(S_{t+1}, \mathbf{R}_{t+1}) \sim P(\cdot \mid S_t, \mathbf{A}_t).
\]

Each learning agent $i$ learns a policy $A_{t} \sim \pi(A_{t} \mid S_t)$ and uses average reward \cite{Sutton1998}:

\begin{equation}\label{eq:avgrew}
r(\pi_i) = \lim_{T \to \infty} \frac{1}{T} \sum_{t=0}^{T-1} \mathbb{E}_{\boldsymbol{\pi}}[R_{t,i}]
\end{equation}

Learning agents search for the optimal policy $\pi_i^*$ to maximize the average reward:

\[
\pi_i^* = \arg\max_{\pi_i} r(\pi_i)
\]

Under the average reward formulation, where $\boldsymbol{\pi} = (\pi_1, \pi_2)$ is the joint policy, the differential value function is defined as:

\[
Q_{\pi_i}(S_t, A_t) = \mathbb{E}\left[\sum_{k=0}^{\infty} \big(R_{t+k+1} - r(\pi_i)\big) \mid S_t, A_t \right]
\]

from which we derive the bellman update \cite{Sutton1998}: 

\[
Q_{t+1}(s,a)=(1-\alpha_t) Q_t(s,a)+\alpha_t(R_t+\max_a Q_t(\bar{s},a))
\]

where $\bar{s}$ is the sampled next state from the average state distribution. 


In repeated game settings, we consider two variants based on the notion of a state history, or joint action memory of previous rounds, where each combination of previous $n$ joint actions constitutes a state. We consider a variant where $n = 0$, an iterated one-off game with strategies independent of history, and a variant where $n = 2$, which captures the relevant game theoretic patterns (e.g. tit for tat) while constraining the history for tractability. 

\subsection{State Representation} \label{sec:staterep}
Consider a two player repeated \(m \times m\) matrix game. At each stage \(t\), the two players choose actions $a_0^{(t)}, a_1^{(t)} \in \{0,1,\dots,m-1\}, t = 0,1,\dots,n-1$. Let the history of length \(n\) be $H = \big((a_0^{(0)},a_1^{(0)}), (a_0^{(1)},a_1^{(1)}), \dots, (a_0^{(n-1)},a_1^{(n-1)})\big)$. Each action pair is encoded as a single digit in base \(m^2\): $p_t = m a_0^{(t)} + a_1^{(t)}, p_t \in \{0,1,\dots,m^2-1\}$. The state corresponding to the history \(H\) is then defined as $s(H) = \sum_{t=0}^{n-1} p_t (m^2)^t$. Substituting the definition of \(p_t\) gives $s(H) = \sum_{t=0}^{n-1} \left(m a_0^{(t)} + a_1^{(t)}\right)(m^2)^t$. 

So, as an example consider the case where $n=2$ in a $2 \times 2$ game and the past two actions $((0, 1), (1, 1))$. Each action gets encoded to $(1, 3)$, and summed together as $1 \times (2^2)^0 + 3 \times (2^2)^1 = 13$, so the state id for that joint action pair is 13. 

We extend the state definition for the Learning Human (Section~\ref{sec:learninghuman}) to include the reference points normalized into $[0, 1]$ and binned into $B$ discrete bins. Specifically, we use $B=5$ to get $b(r) \in \{0, 1, 2, 3, 4\}$. To synthesize the joint action integer, $s(H)$, and reference bin, $b(r)$ into a state integer, we use $S_{id} = s(H)B + b(r)$, or $S_{id} \in \{0, \dots, (m^2)^n B - 1\}$. 

Continuing the example above, we get $S_{id} = 13 \times 5 + \{0, 1, 2, 3, 4\}$, yielding $\{65, 66, 67, 68, 69\}$ as the state ID range for the joint action pair $((0, 1), (1, 1))$ and $B = 5$ for the Learning Human. 

\subsection{Tie Breaks} \label{sec:tiebreaks}
All agents include a tie breaking logic in their policies that check whether the Value transformed outcomes (EU for AI, CPT for Human) are near ties, measured as whether the distance between the action values are less than a threshold $\tau = 0.1$. If the actions are labeled a tie, we take the softmax between them. 

Formally, if agent $i$'s actions are given by $A_i = \{a_{i}^1, a_{i}^2, \dots, a_i^m\}$, their values are given by $V(A_i) = \{V(a_i^1), V(a_i^2), \dots, V(a_i^m)\}$. The top two actions (in descending order) are then given by $a_i^{(1)}, a_i^{(2)}$, and we compare their difference to $\tau$: $| V(a_i^{(1)}) - V(a_i^{(2)})| < \tau$, and if less than $\tau$ the agent samples an action from the policy $\pi_i = \operatorname{Softmax}(V(A_i))$. Otherwise, the agent selects $a_i^{(1)}$. 

\subsection{Utility Maximizing AI Agents}\label{sec:AI}

AI agents employ standard expected utility maximization with reinforcement learning, seeking policies that maximize expected discounted returns. States are defined by a history of joint actions, where a state is formally defined as:

Recall $r(\pi_i)$ from Eq.~\ref{eq:avgrew}. Formally, each AI agent $i$ aims to find:

\[
\pi_i^* = \arg\max_{\pi_i} r(\pi_i)
\]

AI agents implement Q-learning based on the Bellman equation with updates:

\[
Q_i^{AI}(S_t,A_t) \leftarrow (1-\alpha)Q_i^{AI}(S_t,A_{i, t}) + \alpha\left[R_i(S_t,\mathbf{A}_t) - r(\pi_i) +\max_{A_{t+1}} Q_i^{AI}(S_{t+1},A_{t+1})\right]
\]

using learning rate $\alpha \in (0,1]$. During training, AI agents employ $\epsilon$-greedy exploration with $\epsilon$ decaying from 0.3 to 0.01 over the training process. This formulation represents the baseline multi-agent reinforcement learning approach without behavioral modifications, providing the maximizing competitor to Prospect Theoretic players. 

\subsection{Human Agents with Prospect Theory Preferences} \label{sec:human}

We implement two distinct human agent types capturing different levels of strategic sophistication. Aware Human (AH) agents possess complete knowledge of the game structure from the outset, computing exact CPT best responses using opponent best replies over the payoff table. Learning Human (LH) agents employ vanilla Q-learning with average reward, not discounting, and then apply the CPT transformation (Section~\ref{sec:prospecttheory}) to Q values in the policy to reflect human behavior. The CPT transformation employs the Kahneman and Tversky parameterization \(\alpha = \beta = 0.88\), \(\lambda = 2.25\), \(\gamma = 0.61\), and \(\delta = 0.69\) \cite{Tversky1992}. 

Human agents form beliefs over opponent actions $b_{i, t}(\cdot \mid S_t)$ that is updated as an Exponential Moving Average (EMA) with $\eta_{b} = 0.95$ controlling the update speed. The Aware Human (AH) is the degenerate case, and so beliefs reflect the opponent's best reply. 

We implement four models of reference points, denoted as $\bar{R}_{i, t}$ (with $i$ suppressed where obvious), with update parameter \(\eta_{ref} = 0.95\) controlling adaptation rate. 

\begin{itemize}
    
    \item \textbf{EMA (Adaptive Expectations)}: Reference points adapt through an EMA of experienced payoffs:
    \[\bar{R}_t^i = \eta_{ref} \bar{R}_{t-1}^i + (1 - \eta_{ref}) R_{i,t-1}\]
    This captures psychological adaptation where recent outcomes gradually update aspiration levels.

    \item \textbf{V-Based}: Reference points are derived from learned Q-values for the Learning Human, and from the payoff table for the Aware Human to form state value reference points:
    \[\bar{R}_t^i =  \sum_{A_{i, t}} A_{i, t} \sum_{A_{-i, t}} b_{i, t}(A_{-i, t} | S_t) Q_i(S_t, A_{i, t}, A_{-i, t})\]
    where \(b(\cdot)\) represents beliefs about opponent actions. This models rational expectations based on strategic understanding of the game.
    
    \item \textbf{EMAOR (Social Comparison)}: Reference points track opponent payoffs through exponential moving average:
    \[\bar{R}_t^i = \eta_{ref} \bar{R}_{t-1}^i + (1 - \eta_{ref}) R_{-i,t-1}\]
    This captures relative performance evaluation, where agents compare their outcomes against their opponent's.
\end{itemize}

The Fixed and EMA models represent individual-centric reference formation, while Q-Based and EMAOR incorporate strategic and social factors. 

\subsubsection{Aware Human} \label{sec:aware}
Aware human agents represent expert human players or scenarios with perfect information, computing best reply dynamics over the raw payoff table at decision time. Importantly, they do not take in the state history, they compute the one shot best response. The policy is presented in Algorithm~\ref{alg:aware}:

\begin{algorithm} [H] 
    \begin{algorithmic}
        \State Initialize Opponent Best Replies list $OBR \gets \emptyset$, Player values list $PV \gets \emptyset$, CPT value function $v(\cdot)$, Opponent Value Function $v_{opp}(\cdot)$
        \State Let AH actions be $A_i = \{a_i^0, a_i^1, \cdots, a_i^{n-1}\}$
        \State Let Opp actions be $A_j = \{a_j^0, a_j^1, \cdots, a_j^{n-1}\}$
        \State Let AH, opp payoff matrices be $R_{AH}, R_{opp} \in \mathbb{R}^{n \times n}$
        \For{Each AH action $a_i^k \in A_i$}
            \State $OBR^k \gets \arg \max_{a_j^\ell \in A_j}(v_{opp}(R_{opp}(a^k_{i}, a_j^\ell))) $ 
            \EndFor
        \For{Each AH action $a_i^k \in A_i$}
            \State $PV^k \gets v(R_{AH}(a^k_i, OBR^k))$
            \EndFor
        \State \Return $\arg \max_k (PV^k)$
    \end{algorithmic}
    \caption{Aware Human Best Reply Policy}
    \label{alg:aware}
\end{algorithm}

where the opp. value function $v_{opp}(\cdot)$ is EU for AI opponents (nothing changes) and the value function (eq~\ref{eq:valuefunction}) for PT opponents. We use the value function and not the full transformation because the AH agent assumes best reply behavior, making the lottery degenerate and so we drop the probability weighting function.

\subsubsection{Learning Human} \label{sec:learninghuman}

In contrast, Learning Human (LH) agents lack prior knowledge and must discover game dynamics through interaction, combining CPT preferences at decision time with average reward ($r_i(\pi_i)$) Q-learning over untransformed rewards. Importantly, to incorporate beliefs as a part of Q -learning, the Q-table for LH agents is expanded from $Q(S_t, A_{i, t})$ to $Q(S_t, A_{i, t}, A_{-i, t})$, condensed notationally to $Q_i(S_t,\mathbf{A}_t)$. During decision time and within the update, we take the expectation of opponent actions with respect to the beliefs $\mathbb{E}_{b_i}[Q_i(S_{t+1},\mathbf{A}_{t+1})]$ to get $Q_i(S_t, A_t)$. 

\[
Q_i^{LH}(S_t,\mathbf{A}_t) \leftarrow (1-\alpha)Q_i^{LH}(S_t,\mathbf{A}_t) + \alpha\left[R_i(S_t,\mathbf{A}_t) - r(\pi_i) +\max_{\mathbf{A}_{t+1}} \mathbb{E}_{b_i}[Q_i^{LH}(S_{t+1},\mathbf{A}_{t+1})]\right]
\]

where the constant learning rate $\alpha = 0.01$. The LH optimal policy uses the prospect $P = R_i(S_t, \mathbf{A}_t) - r(\pi_i), b_i(A_{-i, t})$:

\[
\pi_i^* = \arg\max_{\pi_i} \mathbb{E}\left[\sum_{t=0}^{\infty} V_i\big(P\big)\mid \boldsymbol{\pi} \right]
\]

where $V_i(\cdot)$ is the CPT transformation (Eq~\ref{eq:val}). 

\subsection{Learning Algorithms for Mixed Populations}

We synthesize Section~\ref{sec:problem} in Algorithm \ref{alg:heterogeneous_q}. 

\begin{algorithm}[H]
\caption{Heterogeneous Q-Learning for Human-AI Agents}
\label{alg:heterogeneous_q}
\begin{algorithmic}[1]
\Require Initialize $Q_i^{AI}(S_t,A_t)$ for AI agents and
$Q_i^{LH}(S_t,A_i,A_{-i})$ for Learning Human agents; payoff matrices
$R_i(A_i,A_{-i})$ and $R_{-i}(A_i,A_{-i})$; value lists
$\mathcal{V}^{AH}, \mathcal{V}^{LH} \gets \emptyset$; reference point
$r_i$; CPT value transformation $V_i(\cdot)$ from Eq.~\ref{eq:val}; and
belief function $b_i(A_{-i}\mid S_t)$ for human agents.
\Require Pathology threshold $\tau$, belief update $\lambda_b$, reference update $\lambda_r$

\For{episode $= 1$ to $M$}
    \For{$t = 1$ to $T$}
        \For{each agent $i$}
            \If{agent $i$ is AI}
                \State With probability $\epsilon$: $A^{AI}_{i, t} \gets \text{random action}$
                \State Otherwise: $A^{AI}_{i, t} \gets \arg\max_{A_{i, t}} Q_i^{AI}(S_t,A_t)$
            \ElsIf{Aware Human (AH)}
                \For{each $a_i \in \mathcal{A}_i$ \Comment{Static payoff matrix}}
                    \State $OBR(a_i) \gets \arg\max_{a_{-i} \in \mathcal{A}_{-i}} v_{opp}(R_{-i}(a_i,a_{-i}))$
                    \State $V^{AH}(a_i) \gets v(R_i(a_i,OBR(a_i)))$
                \EndFor
                \State $a_{AH} \gets \arg\max_{a_i \in \mathcal{A}_i} V^{AH}(a_i)$
            \Else \Comment{Learning Human (LH)}
                \If{$\mathrm{rand}(0,1) < \epsilon$}
                    \State $a_{LH} \gets \text{uniform random action}$
                \Else
                    \For{each $a_i \in \mathcal{A}_i$}
                        \State $P_i(a_i) \gets \{(Q_i^{LH}(s,a_i,a_{-i}) - r_i,\; b_i(a_{-i}\mid s))\}_{a_{-i}\in\mathcal{A}_{-i}}$
                        \State $V^{LH}(a_i) \gets V_i(P_i(a_i))$
                    \EndFor
                    \If{$|V^{LH}(a_i^{(1)}) - V^{LH}(a_i^{(2)})| < \tau$} \Comment{Tie break}
                        \State $\forall a_i:\; p(a_i) \propto \exp(V^{LH}(a_i)/\text{temp})$
                        \State $a_{LH} \gets \text{sample from } p(a_i)$
                    \Else
                        \State $a_{LH} \gets \arg\max_{a_i \in \mathcal{A}_i} V^{LH}(a_i)$
                    \EndIf
                \EndIf
            \EndIf
        \EndFor
        \State Execute joint action $\mathbf{a}$, observe rewards $\mathbf{r}_t^{rew}$ and next state $S_{t+1}$
        \For{each agent $i$}
            \If{agent $i$ is AI}
                \State $Q_i^{AI}(S_t,A_t) \gets (1-\alpha)Q_i^{AI}(S_t,A_t) + \alpha[\mathbf{r}_{i, t}^{rew} - r(\pi_i)+ \max_{A_{t+1}} Q_i^{AI}(S_{t+1},A_{t+1})]$
            \Else \Comment{Learning Human}
                \State $b_i(A_{-i, t+1} \mid S_{t+1}) = \lambda_b\,b_i(A_{-i, t} \mid S_{t+1}) + (1 - \lambda_b)\,\mathbf{1}[A_{-i} = A_{-i, t}]$
                \State $r_i = \lambda_{ref}\,r_i + (1 - \lambda_{ref})\,\mathbf{r}^{rew}$ \Comment{EMA; fixed or $Q^{EU}$ also viable}
                \State $y^{LH}_i \gets \mathbf{r}^{rew}_{i, t} - r(\pi_i) + \max_{A_{i, t}} \sum_{A_{-i, t}} b_i(A_{-i, t + 1} \mid S_{t+1}) Q_i^{LH}(S_{t+1},A_{i, t+1},A_{-i, t+1})$ 
                \State $Q_i^{LH}(S_t,\mathbf{A}_t) \gets (1-\alpha)Q_i^{LH}(S_t,\mathbf{A}_{t}) + \alpha y^{LH}_i$
            \EndIf
        \EndFor
    \EndFor
\EndFor
\end{algorithmic}
\end{algorithm}

\subsection{Experimental Configuration}

Our experimental framework systematically tests all pairwise combinations of the three agent types (AI, AH, LH) across the complete game suite established in Section~\ref{sec:Game}. Table~\ref{tab:expconfig} summarizes the six population configurations. Notably, experiments with heterogenous agents and asymmetric payoff tables are run with each agent playing row/column for completeness, so AI-LH will be repeated as LH-AI for asymmetric payoff matrices.

\begin{table}[h]
\centering
\caption{Experimental Population Configurations}
\label{tab:expconfig}
\begin{tabular}{|c|c|c|}
\hline
\textbf{Configuration} & \textbf{Agent 1 Type} & \textbf{Agent 2 Type} \\
\hline
AI-AI & Expected Utility Maximizer & Expected Utility Maximizer \\
AI-AH & Expected Utility Maximizer & Aware Human (PT) \\
AI-LH & Expected Utility Maximizer & Learning Human (PT) \\
AH-AH & Aware Human (PT) & Aware Human (PT) \\
AH-LH & Aware Human (PT) & Learning Human (PT) \\
LH-LH & Learning Human (PT) & Learning Human (PT) \\
\hline
\end{tabular}
\end{table}
Additionally, we repeat each experiment with each of the reference point models from Section~\ref{sec:human} (Fixed, EMA, V-Based, EMAOR). Each experiment is split into $500$ episodes to group metrics, but learning is conducted as a continuous game so terminal states retain their bootstrapped value when states are present. 

\begin{table}[h]
\centering
\caption{Experimental Parameters}
\begin{tabular}{|l|c|}
\hline
\textbf{Parameter} & \textbf{Value} \\
\hline
Learning rate \(\alpha_Q\) & 0.01 \\
Initial exploration \(\epsilon\) & 0.3 \\
Final exploration \(\epsilon_{\min}\) & 0.01 \\
Exploration decay & 0.995 per episode \\
Belief update rate \(\lambda_b\) & 0.95 \\
Reference update rate \(\lambda_r\) & 0.95 \\
Pathology threshold \(\tau\) & 0.1 \\
Softmax temperature \(T\) & 1.3 \\
\hline
\end{tabular}
\end{table}

Ablations are conducted across game, matchup, ref point type, and finally across the length of the state history (see Section~\ref{sec:staterep}). Specifically, our analysis investigates state histories of lengths $0$ and $2$, interpreted as a one off game between random agents and a Markov game with the length truncated for tractability. 


\section{Numerical Results}\label{sec:num}

This section presents the numerical results from our experiments. We split comparisons based on game to set a frame, but focus our inquiry on the matchups within each game. Within each matchup, we split analysis into state history = $0$ and state history = $2$, and then compare the results across reference types (EMA, V-based, and EMAOR). Recall that each experiment runs for $50,000$ steps. Each run starts with a reference point at 0. 

Recall that the Aware Human (AH) is a best response agent that does not learn (Section~\ref{sec:aware}). The AH vs AH matchup is presented as a CPT baseline for the experiment suite, as the agents do not succumb to noise and behave deterministically. We then compare each matchup to the AH vs AH baseline to infer the effect of learning and EU preference types on game outcome. 

Recall that equilibria, if extant, are denoted in the form $(p, q)$, where $p$ is the probability that player 1 plays action 1 and $q$ is the probability that player 2 plays action 1. 


\subsection{Prisoner's Dilemma} 
Prisoner's Dilemma (PD) is a baseline game for the experiment, as extensive analysis in classical EU theory and in \citeauthor{leclerc2014prospect}'s treatment for PT theory have established that in one off games, PD has a pure equilibria at (0, 0), or mutual defection. However, in repeated games a variety of behavior patterns can emerge (e.g. tit for tat) depending on the game structure and player preferences. 

\subsubsection{One Off Game (State History = 0)}
In the One Off Prisoner's Dilemma, we expect policies to converge to mutual defection. The AH vs. AH baseline matchup meets this expectation (see Figure~\ref{fig:PDAHAHJA}), but we did discover an anomaly in the AH vs. LH matchup with EMAOR reference points. First, we start with the AH vs. AH baseline:

\begin{figure}[H]
    \centering
    \begin{minipage}[t]{0.32\linewidth}
        \centering
        \includegraphics[width=\linewidth]{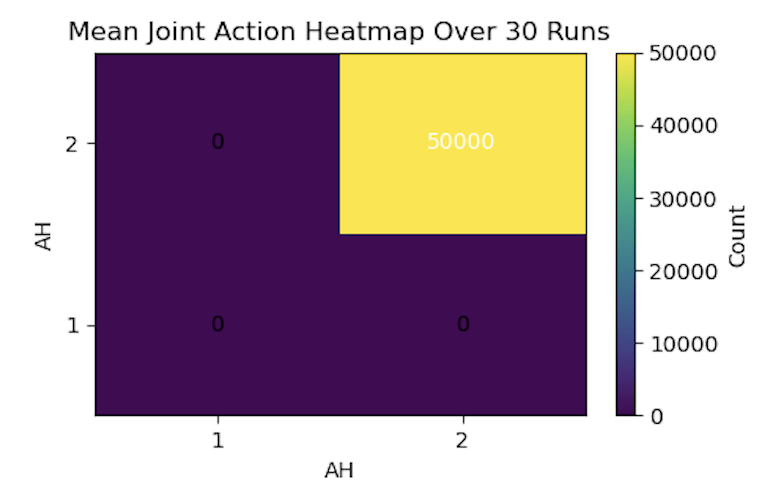}
        \caption{EMA}
        \label{fig:PDAHEMAJATab}
    \end{minipage}
    \begin{minipage}[t]{0.32\linewidth}
        \centering
        \includegraphics[width=\linewidth]{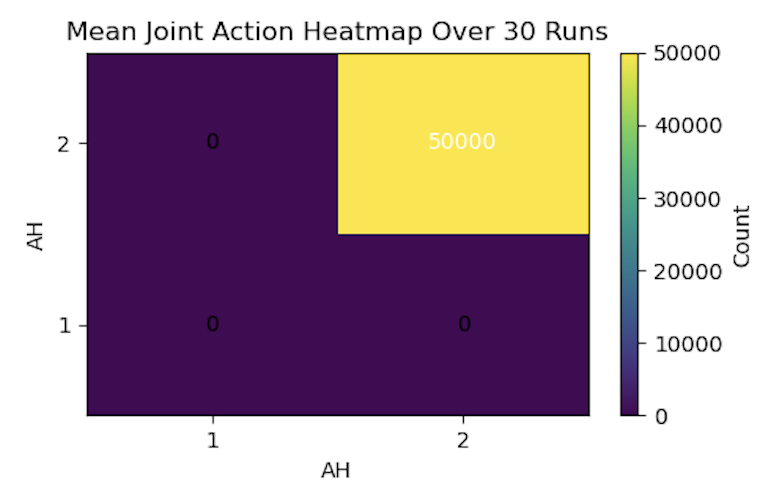}
        \caption{V-based}
        \label{fig:PDAHVJATab}
    \end{minipage}
    \begin{minipage}[t]{0.32\linewidth}
        \centering
        \includegraphics[width=\linewidth]{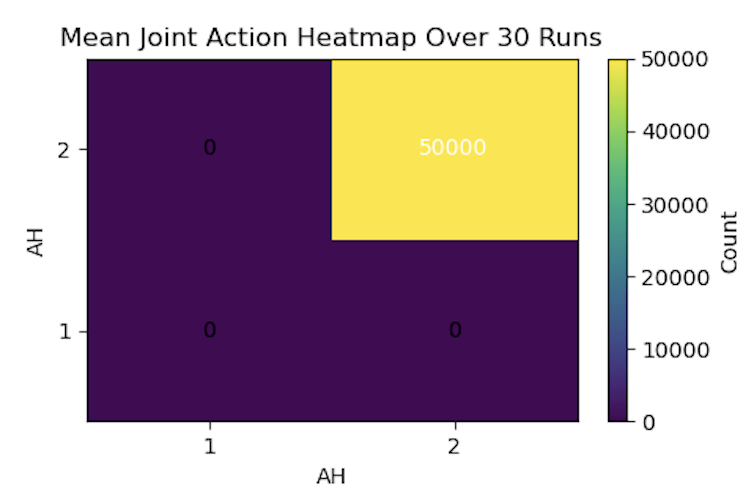}
        \caption{EMAOR}
        \label{fig:PDAHEMAORJATab}
    \end{minipage}
    \caption{AH vs AH PD Joint actions by Ref Type}
    \label{fig:PDAHAHJA}
\end{figure}

\begin{figure}[h]
    \centering
    \begin{minipage}[t]{0.32\linewidth}
        \centering
        \includegraphics[width=\linewidth]{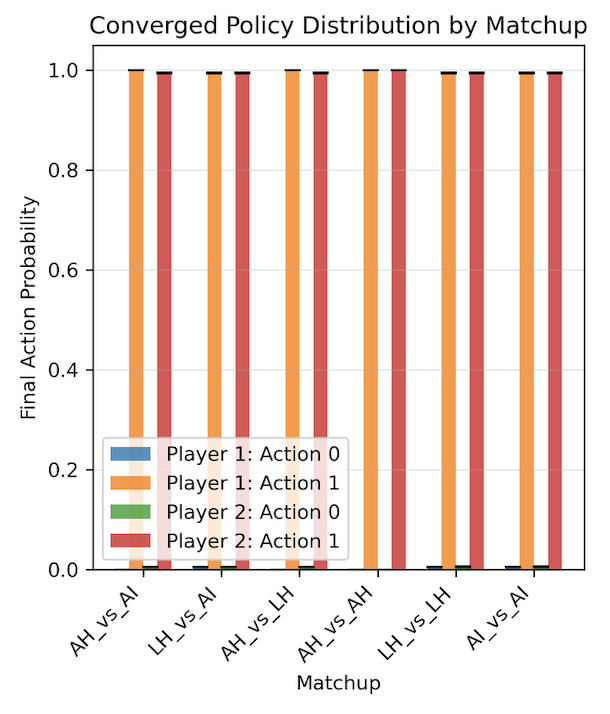}
        \caption{EMA}
        \label{fig:PDAHEMAJATab}
    \end{minipage}
    \begin{minipage}[t]{0.32\linewidth}
        \centering
        \includegraphics[width=\linewidth]{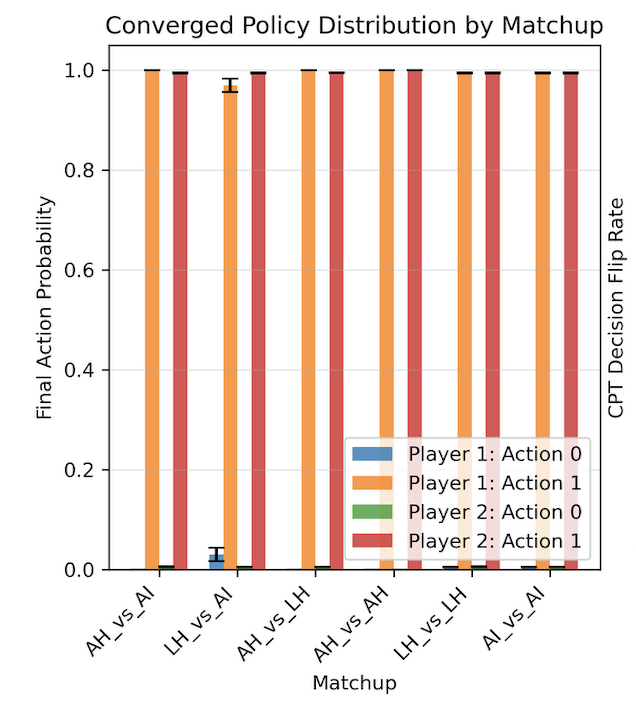}
        \caption{V-based}
        \label{fig:PDAHVJATab}
    \end{minipage}
    \begin{minipage}[t]{0.32\linewidth}
        \centering
        \includegraphics[width=\linewidth]{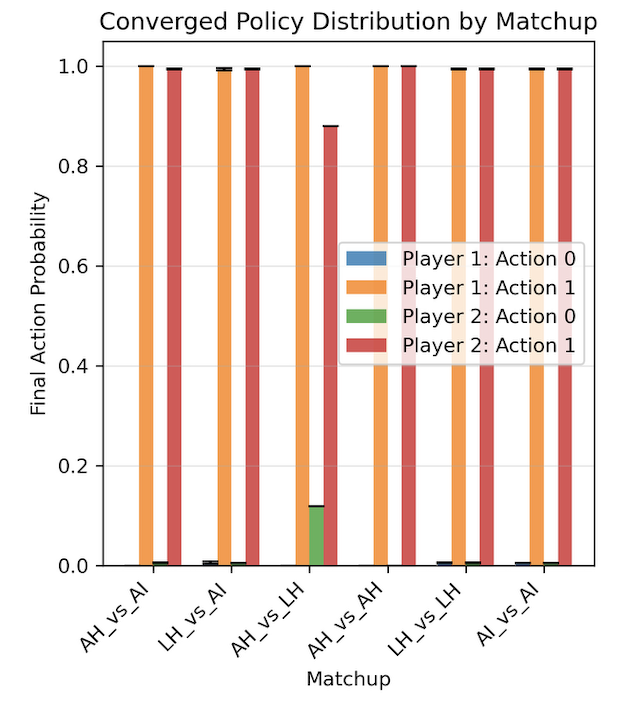}
        \caption{EMAOR}
        \label{fig:PDOSAPEMAOR}
    \end{minipage}
    \caption{All Matchup Policies (Last $5,000$ steps)}
    \label{fig:Policies1PD}
\end{figure}

Each AH vs AH matchup in one off PD was in agreement with unanimous equilibria at (0, 0). The AH agents were able to slightly outperform the learning agents because the learning agents maintained slight noise in their policy, but behaviorally they all converged to (0, 0) (see Figure~\ref{fig:Policies1PD}). However, we discovered an anomaly in the AH vs LH EMA over Opponent Rewards (EMAOR) matchup, where the LH cooperated $\approx 10\%$ of the time. 

\begin{figure}
        \centering
        \includegraphics[width=0.7\linewidth]{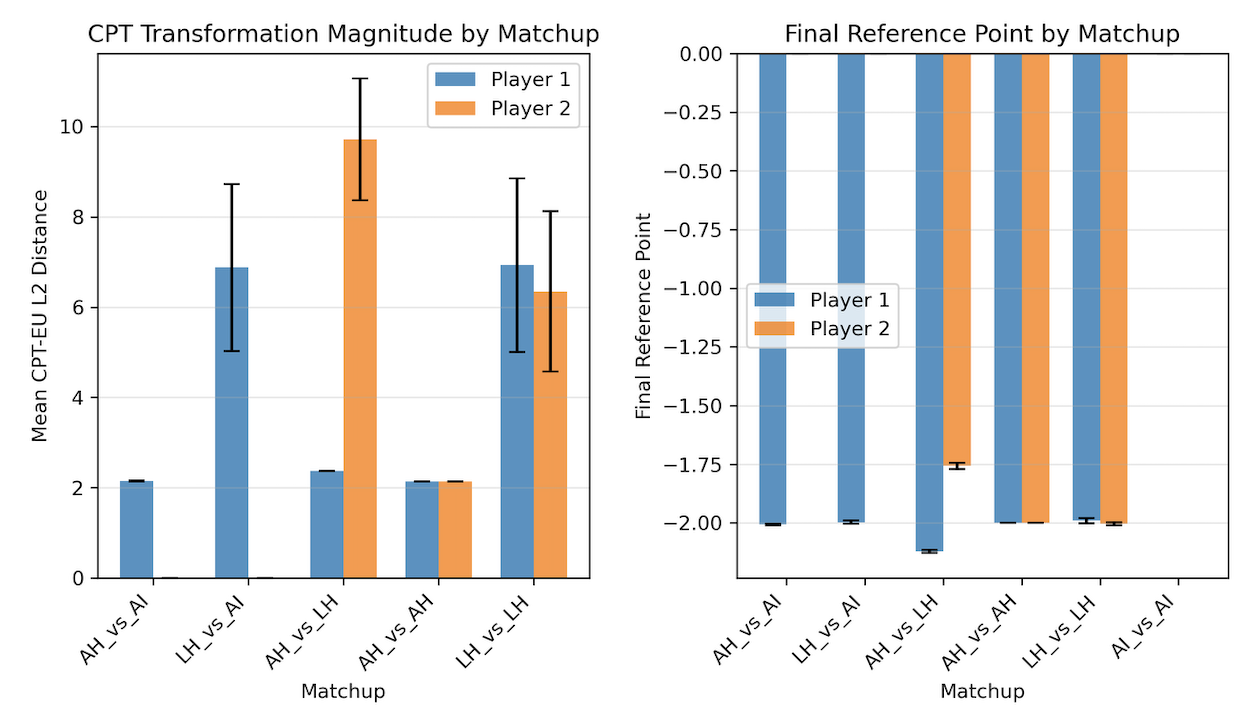}
        \caption{Ref Point and CPT Transform Magnitude (EMAOR,  Last $5,000$ steps)}
        \label{fig:ref1pdAHLH}
\end{figure}

It's unclear why the AH vs LH matchup deviated, but we can search for the cause by taking a look at the reference point chart between agents in each matchup (Figure~\ref{fig:ref1pdAHLH}). The LH reference point is much higher, reflecting the higher rewards that the AH received, that is perhaps contextualized by the magnitude of CPT transformation that the LH agent reported. For context, the CPT transform magnitude graph is obtained by finding the L2 distance between the EU action values and the CPT transformed action values at each time step. Notably, in the AH vs. LH matchup the LH agent set the maximum transformation magnitude for all matchups by $\approx 50\%$. 
\vspace{-1.5em}

\subsubsection{Repeated Game (State History = $2$)}
In a repeated Prisoner's Dilemma, we expect greater variance in the behaviors that agents engage in. Note that the AH agents do not have states, they interact with the payoff matrix and not the state history, so the baseline they set in One Shot at (0, 0) holds for repeated games.

\begin{figure}[h]
    \centering
    \begin{minipage}[t]{0.32\linewidth}
        \centering
        \includegraphics[width=\linewidth]{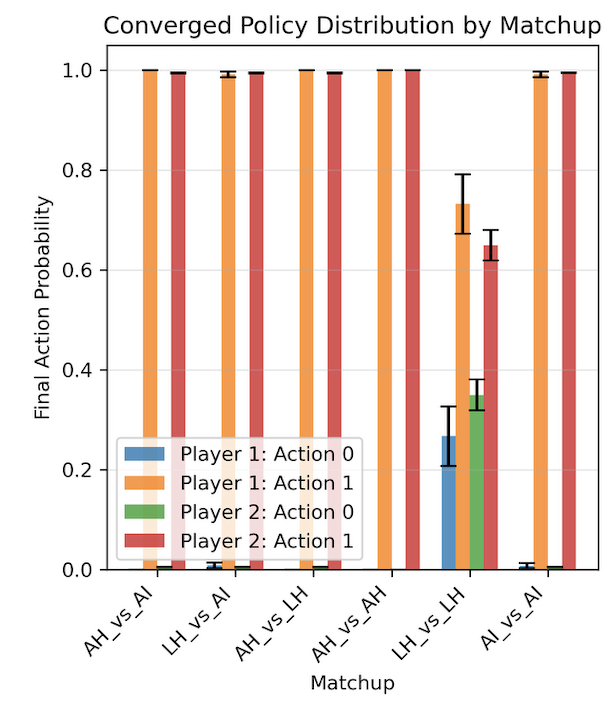}
        \caption{Policies}
        \label{fig:}
    \end{minipage}
    \begin{minipage}[t]{0.32\linewidth}
        \centering
        \includegraphics[width=\linewidth]{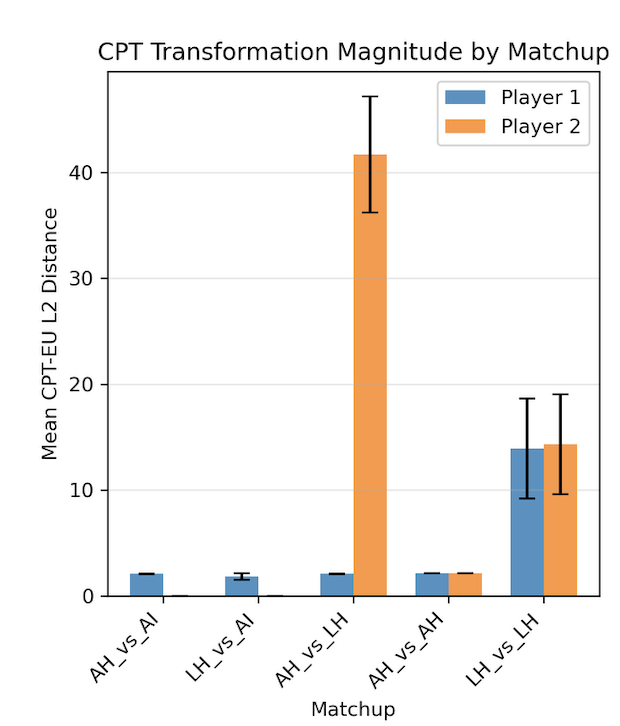}
        \caption{L2: EU CPT}
        \label{fig:}
    \end{minipage}
    \begin{minipage}[t]{0.32\linewidth}
        \centering
        \includegraphics[width=\linewidth]{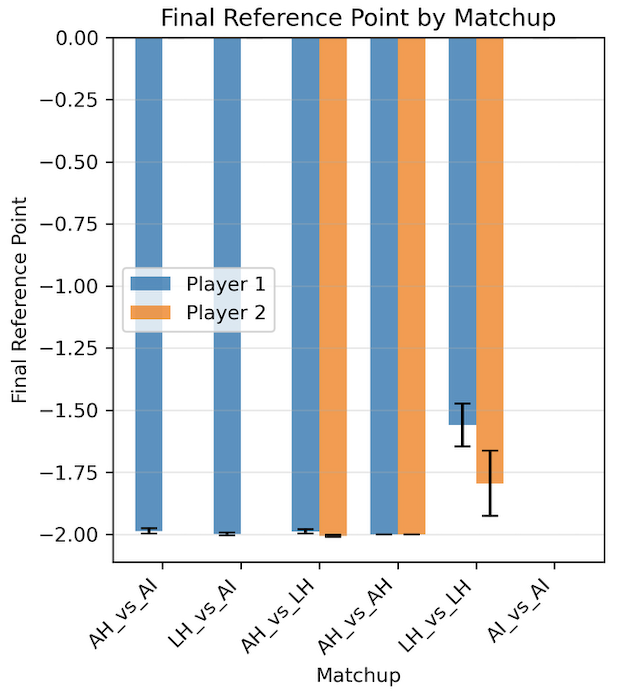}
        \caption{Ref. Points}
        \label{fig:}
    \end{minipage}
    \caption{LH Anomaly: Matchup Policies, CPT/EU L2, and Ref. Points (EMA, Last $5,000$ steps)}
    \label{fig:LHAnom}
\end{figure}

Once again, LH exhibited anomalous behavior, but this time with the EMA reference setting. In EMAOR and V-based reference point settings, all matchups converged to $\approx (0, 0)$, so the anomaly does not seem to be coupled with the reference type. Furthermore, in the one off game the anomaly occurred in the AH vs LH matchup, but in this repeated games anomaly it occurs in the LH vs LH matchup. That being said, it does seem to be connected to the LH agent.

\subsubsection{PD Concluding Remarks}
Most of the matchups converged to defect/defect, as expected. However, LH diverged from the expectation twice, vs AH with EMAOR reference point type in the one shot PD, and vs LH with EMA reference point type in repeated PD. 
\begin{figure}[H]
    \centering
    \includegraphics[width=0.8\linewidth]{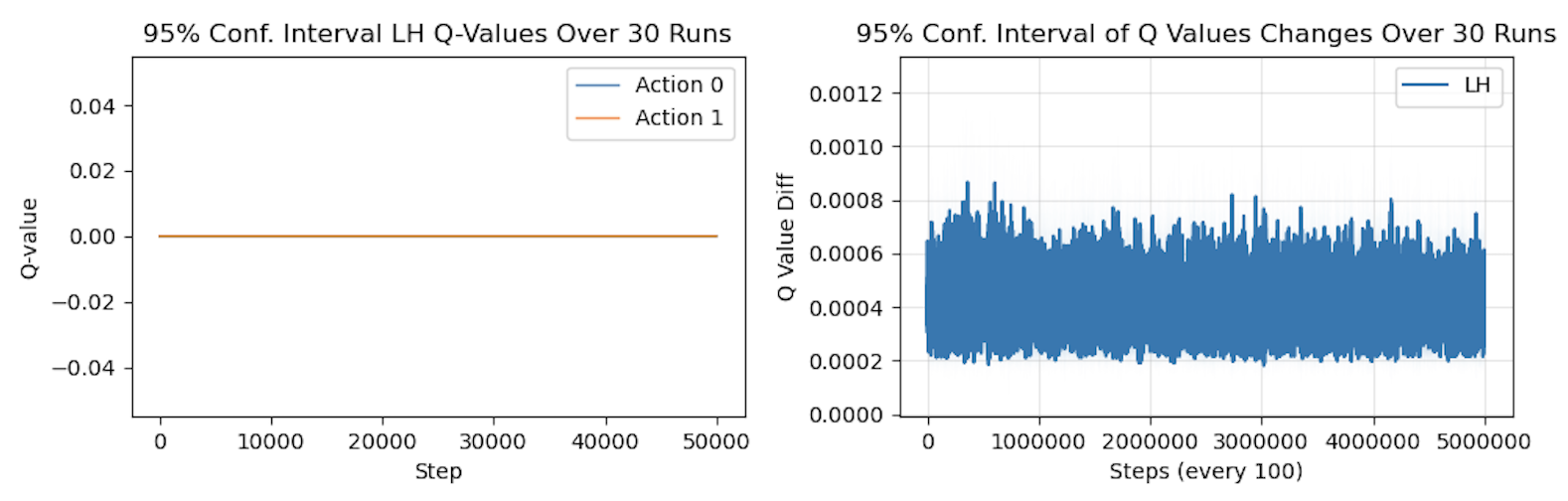}
    \caption{One Shot Anomalous LH Q Values and Q Value Change (EMAOR vs AH)}
    \label{fig:LHQvalpd1}
\end{figure}
Interestingly, the learning dynamics for each case were divergent. In the one shot game, the Q values across the 5 reference bin states oscillated around 0 (Figure~\ref{fig:LHQvalpd1}). 

However, in the repeated setting, the LH Q values skyrocketed with significant noise between runs despite low volatility in Q values during runs: 

\begin{figure}
    \centering
    \includegraphics[width=\linewidth]{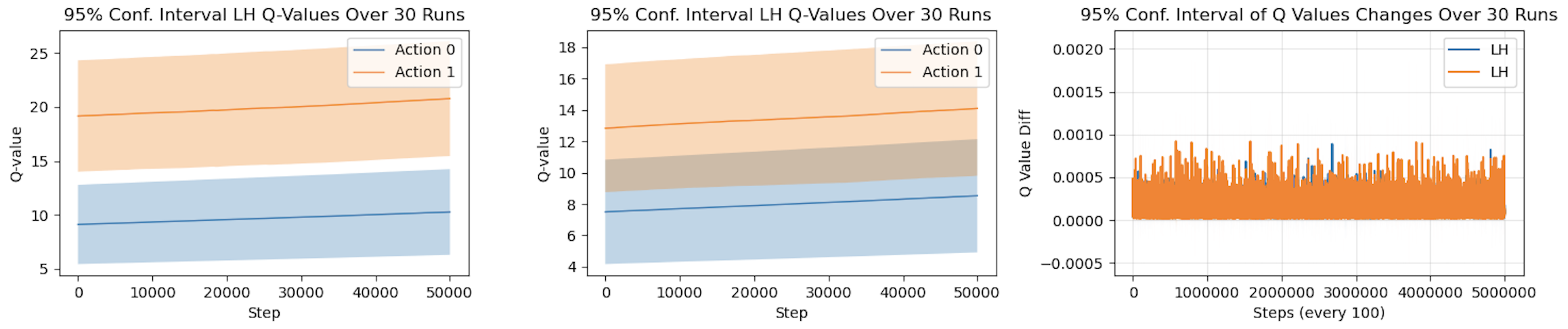}
    \caption{Repeated Anomalous LH Agents (Player 1, left; Player 2, middle) Q vals and Q value change (EMA)}
    \label{fig:pdrepeatLHQ}
\end{figure}

The strong deviation between the learning metrics suggests that the cause of the anomaly is not a clearcut learning pathology, especially given LH's regression to the norm in other matchups. Furthermore, the pathology does not correlate one to one with high CPT transformation magnitudes or reference point magnitude/direction. The V-based reference point setting makes the lack of correlation clear, with each CPT agent eclipsing 30 in the L2 distance and driving reference points to $r > +20$ and $r <-30$. 

\subsection{Matching Pennies}
Matching Pennies (MP) is another classic and widely studied game with an EU/Nash Equilibrium at $(\frac{1}{2},\frac{1}{2})$ and a PT-NE and PT-EB equilibrium at $(\frac{1}{2},\frac{1}{2})$ \cite{leclerc2014prospect}. Once again, we will start with the AH vs AH baseline and analyze anomalies and interesting findings. 

\subsubsection{One Off Game (State History = $0$)}
\begin{figure}
    \centering
    \includegraphics[width=0.75\linewidth]{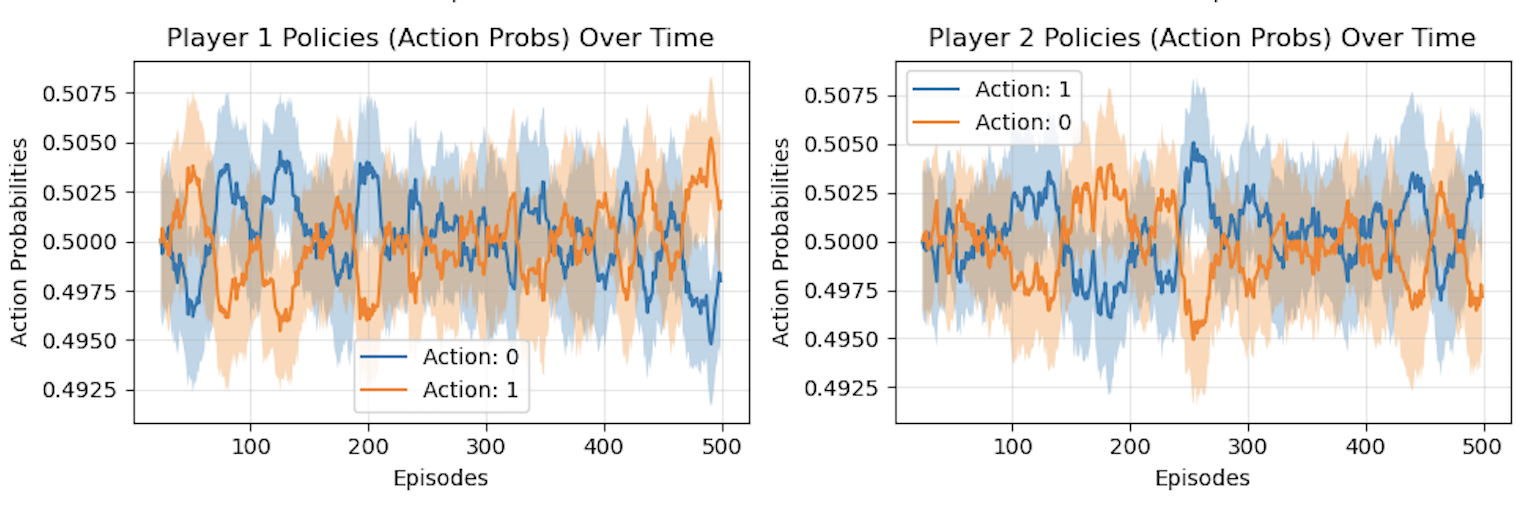}
    \caption{AH vs AH Baseline (EMA)}
    \label{fig:MPAHAH}
\end{figure}

The AH vs AH matchup finds the $(\frac{1}{2},\frac{1}{2})$ equilibrium oscillating around the $50\%$ action mark in Figure~\ref{fig:MPAHAH}, which shows the action frequencies with the EMA reference point type. V-based and EMAOR were nearly identical. 

Next we expect the converged policies (final $5,000$ steps) to see what behavioral trends emerged. 

\begin{figure}[h]
    \centering
    \begin{minipage}[t]{0.32\linewidth}
        \centering
        \includegraphics[width=\linewidth]{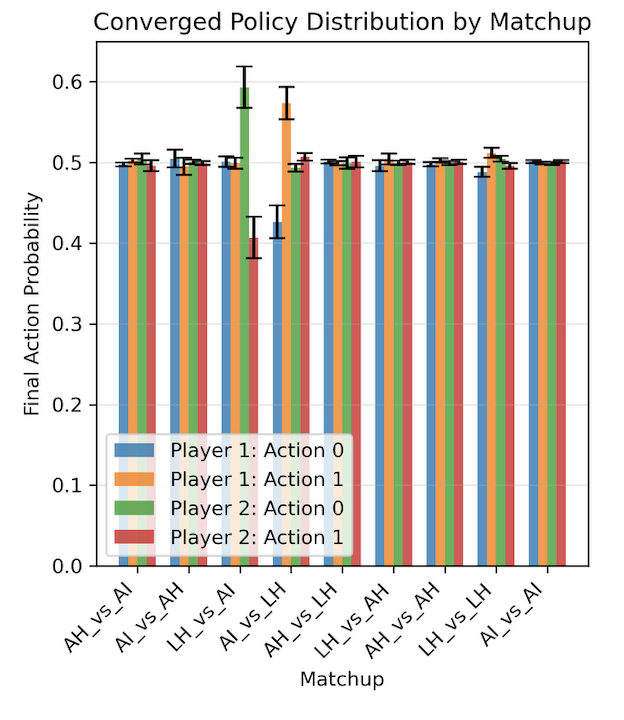}
        \caption{EMA}
        \label{fig:}
    \end{minipage}
    \begin{minipage}[t]{0.32\linewidth}
        \centering
        \includegraphics[width=\linewidth]{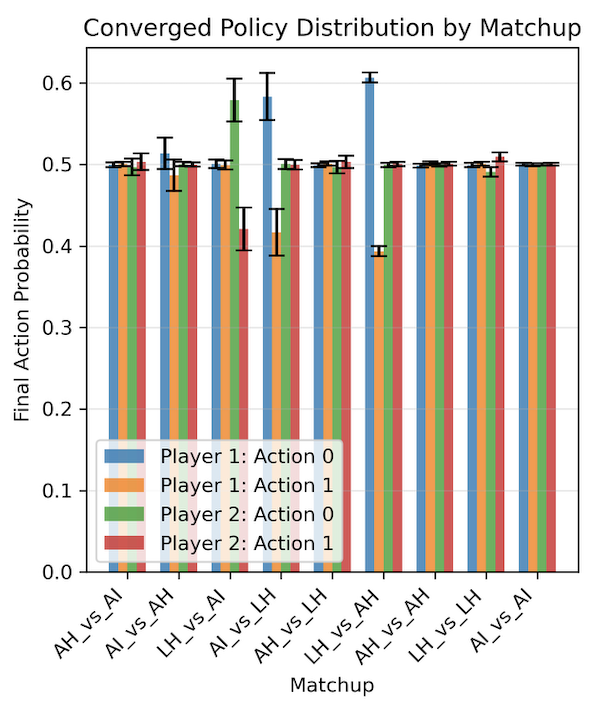}
        \caption{EMAOR}
        \label{fig:}
    \end{minipage}
    \begin{minipage}[t]{0.32\linewidth}
        \centering
        \includegraphics[width=\linewidth]{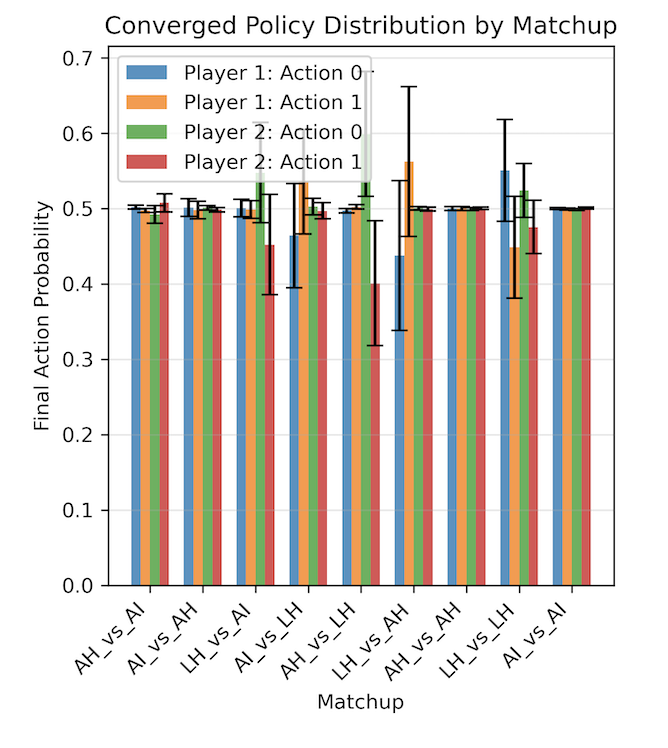}
        \caption{V-based}
        \label{fig:}
    \end{minipage}
    \caption{Converged Policies One Shot MP (Last $5,000$ steps)}
    \label{fig:pi1MP}
\end{figure}

Interestingly, this time the AI agent demonstrated anomalous behavior in all three reference types, specifically in its matchups with LH. The deviation from equilibrium allowed LH to outcompete AI, and the consistency across runs and reference types suggests that it is a real result. That being said, LH showed flashes of its anomalous behavior in the EMAOR and V-based reference point types, jumping up to $\approx p = \frac{3}{5}$ in EMAOR --- a significant departure from the equilibrium. 

\begin{figure}[H]
    \centering
    \begin{minipage}[t]{0.32\linewidth}
        \centering
        \includegraphics[width=\linewidth]{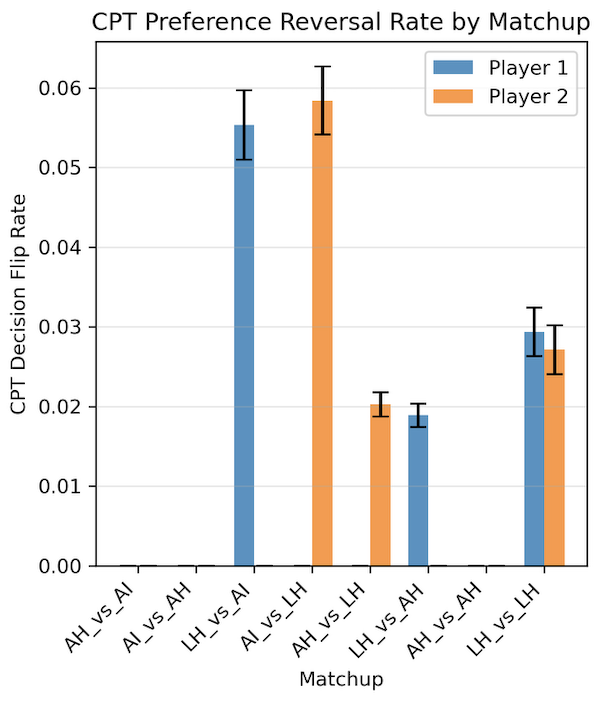}
        \caption{EMA}
        \label{fig:}
    \end{minipage}
    \begin{minipage}[t]{0.32\linewidth}
        \centering
        \includegraphics[width=\linewidth]{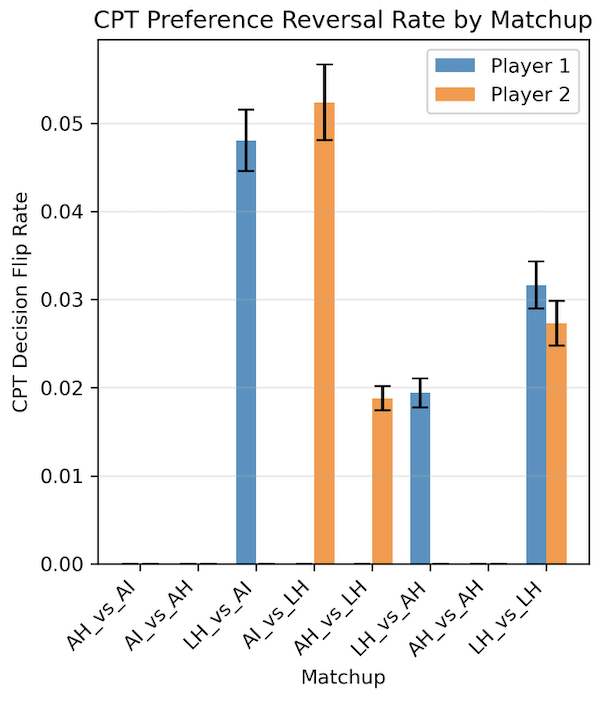}
        \caption{EMAOR}
        \label{fig:}
    \end{minipage}
    \begin{minipage}[t]{0.32\linewidth}
        \centering
        \includegraphics[width=\linewidth]{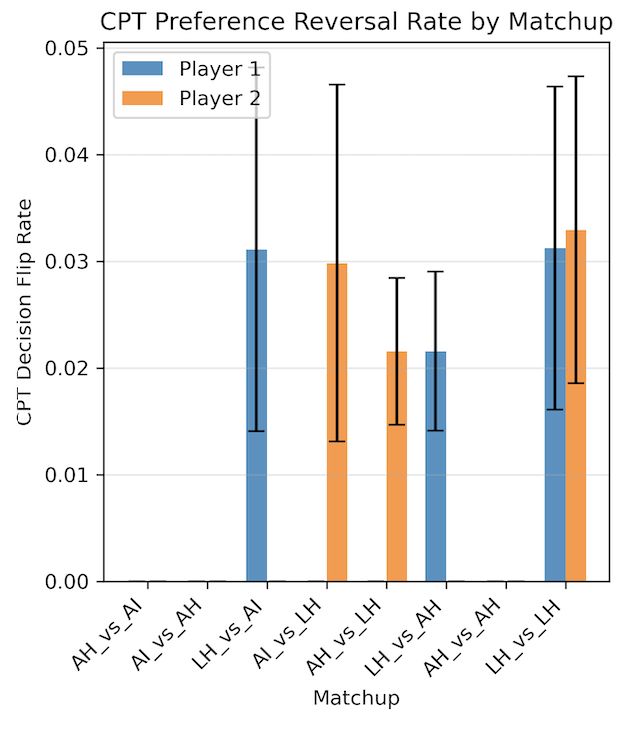}
        \caption{V-based}
        \label{fig:}
    \end{minipage}
    \caption{Action Changes One Shot MP (Last $5,000$ steps)}
    \label{fig:change1MP}
\end{figure}

Tying the LH and AI anomalous behavior together is an analysis of the CPT induced action changes, defined by a deviation between the EU best action and the CPT best action. In Prisoner's Dilemma, action change rates were negligible. However, Figure~\ref{fig:change1MP} shows that across all reference types LH's CPT preferences changed its behavior. 
\vspace{-1em}

The LH showed exhibited significant influence from CPT preferences on its decision making, surely aided by the similarly valued actions. However, it should be noted that actions were tested for tie breaks by finding the difference between actions and skipping if $\textbf{diff} < 1e^{-8}$. This check prevents numerical noise or random tie breaks from counting as preference induced change. 

Relevantly, the LH's action change coincided with AI's erratic behavior. While the correlation does not imply causation, the LH did exhibit high change rates across all reference point types, and in the same matchup the AI exhibited erratic behavior and lost relative reward. 

\subsubsection{Repeated Game (State History = $2$)}
\begin{figure}[H]
    \centering
    \begin{minipage}[t]{0.32\linewidth}
        \centering
        \includegraphics[width=\linewidth]{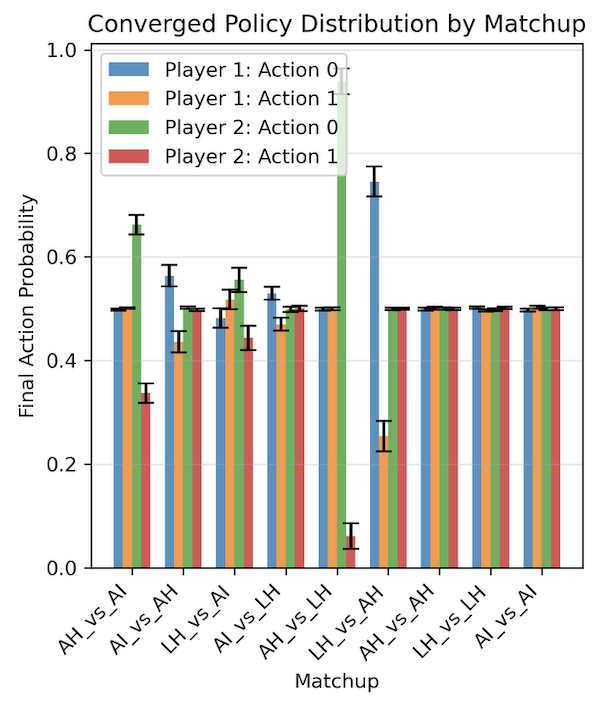}
        \caption{EMA}
        \label{fig:}
    \end{minipage}
    \begin{minipage}[t]{0.32\linewidth}
        \centering
        \includegraphics[width=\linewidth]{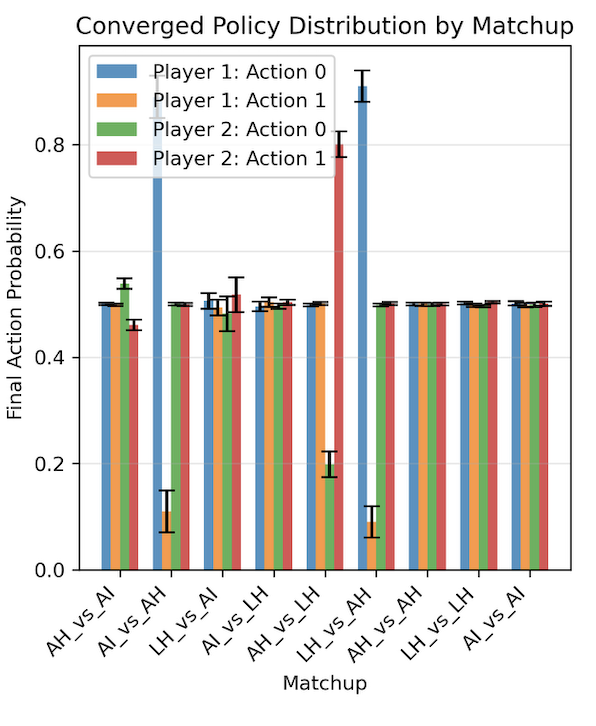}
        \caption{EMAOR}
        \label{fig:}
    \end{minipage}
    \begin{minipage}[t]{0.32\linewidth}
        \centering
        \includegraphics[width=\linewidth]{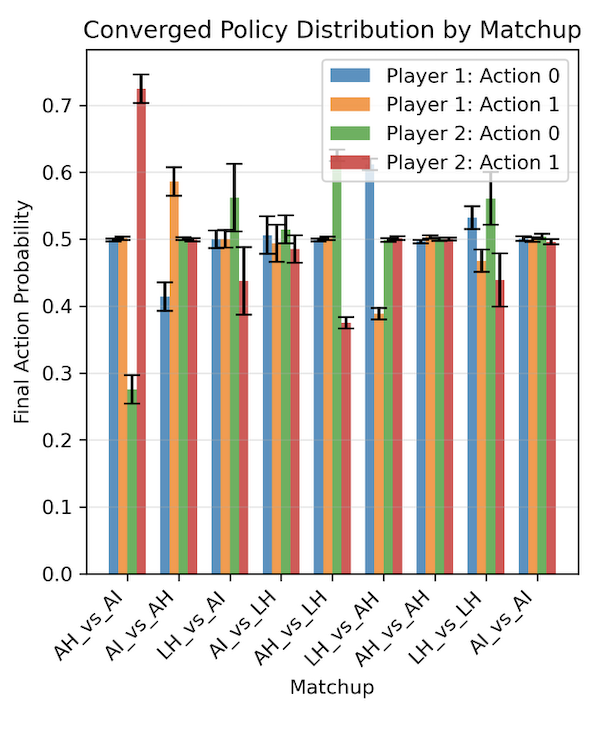}
        \caption{V-based}
        \label{fig:}
    \end{minipage}
    \caption{Policies Repeated MP (Last $5,000$ steps)}
    \label{fig:policyRMP}
\end{figure}

The repeated Matching Pennies game exhibited even more extreme, anomalous behavior. Figure~\ref{fig:policyRMP} visualizes the converged policies across reference types, where the most extreme result was LH exhibiting relative action ratios as high as $9:1$ (EMA, EMAOR). Notably, across 30 runs the variance within each policy is relatively low, suggesting that these results are statistically relevant. AH's equilibrium play \textit{seemed} to induce significant variance from opponent play, shown by both AI and LH variance against AH versus their relative stability when playing another learning agent. 

Interestingly, despite extreme policy variance, the rewards reeived in each matchup stayed even between players (the only outlier was a $10\%$ AI gain over LH in LH vs AI EMA). It is somewhat surprising that the significant deviance from equilibrium was all but invisible in the rewards, which could be a cause and an explanation as to how those policies were converged to in the first place. 

\begin{figure}[H]
    \centering
    \begin{minipage}[t]{0.32\linewidth}
        \centering
        \includegraphics[width=\linewidth]{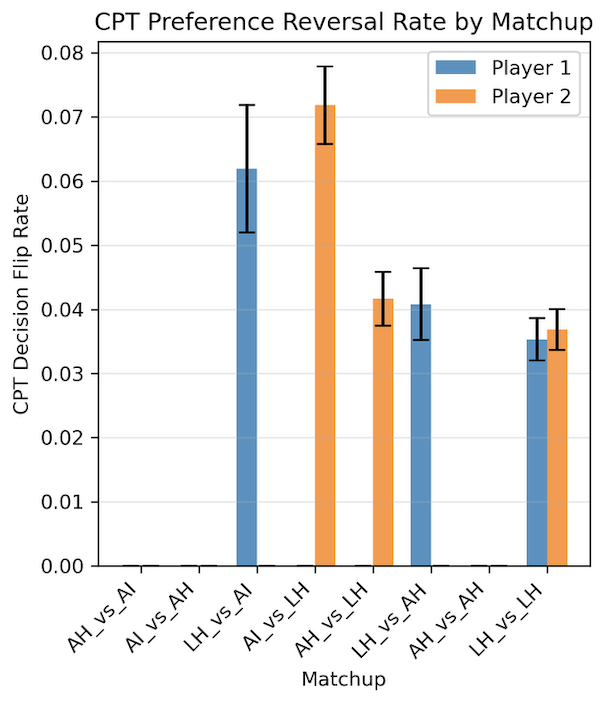}
        \caption{EMA}
        \label{fig:}
    \end{minipage}
    \begin{minipage}[t]{0.32\linewidth}
        \centering
        \includegraphics[width=\linewidth]{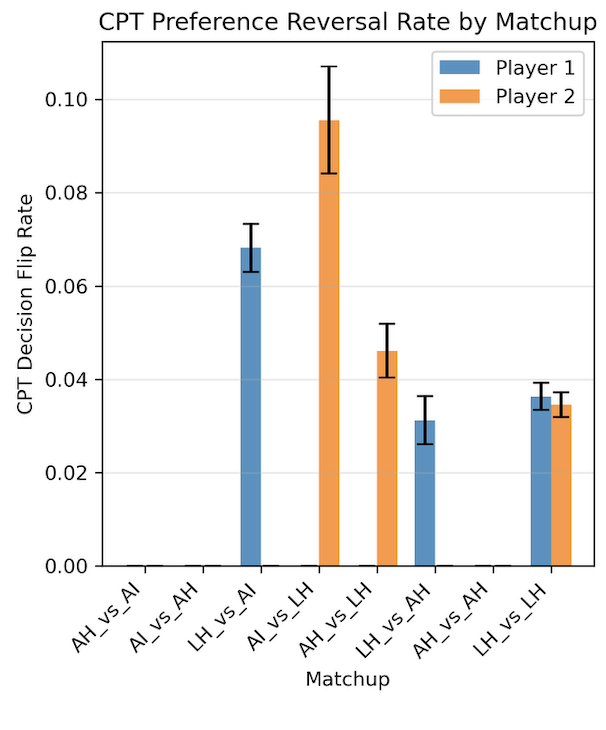}
        \caption{EMAOR}
        \label{fig:}
    \end{minipage}
    \begin{minipage}[t]{0.32\linewidth}
        \centering
        \includegraphics[width=\linewidth]{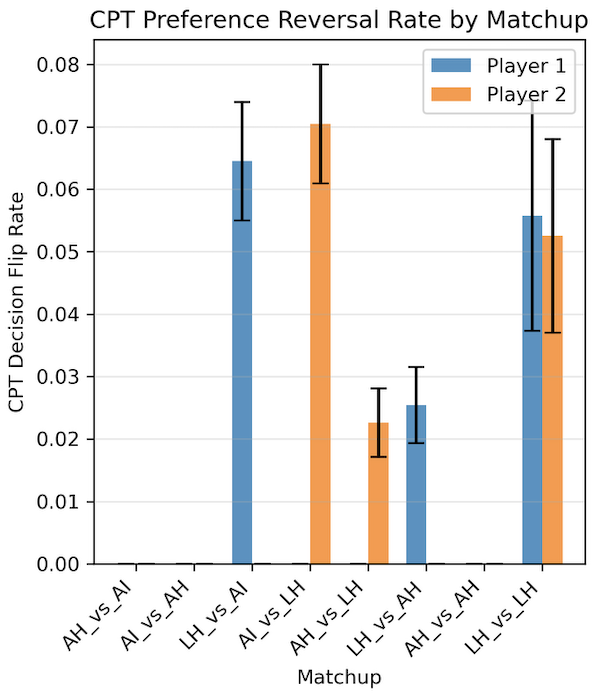}
        \caption{V-based}
        \label{fig:}
    \end{minipage}
    \caption{CPT Induced Action Changes Repeated MP (Last $5,000$ steps)}
    \label{fig:changeRMP}
\end{figure}

Finally, we take a look at the action flip rates in Figure~\ref{fig:changeRMP}, which increased over the action change rates in the one off games. 

With change rates approaching $10\%$, CPT is asserting a significant influence on the games. It is interesting that, across the board, the highest action change rates for the LH were against the AI. Generally speaking, variance within each change rate was low, once again suggesting the robustness of the results. 

\subsubsection{Matching Pennies Concluding Remarks}
Matching Pennies introduced significant anomalous behavior, with departures of up to $40\%$ from established equilibrium. Even so, in both one off and repeated games the deviations in policy were not reflected in the rewards. This result was surprising, but is unlikely to suggest any gaming by the learning agents but rather an evening out of rewards with respect to the equilibrium behavior from the AH --- when the LH plays $90\%$ action 2 against AH, for example, the gains and losses even out against the stability from the AH. It's possible that the AH's stability allowed opposing policies to reach such extremes. 

With respect to the AI/LH interactions specifically, there was no clear advantage in terms of rewards for either agent type. The AI/LH variance was relatively high in the one off games, but relatively stable in the repeated games, possibly because the increase state space allowed anomalous valuations to emerge that were kept at equilibrium by the AH stability. It is worth noting, however, that \textit{the highest CPT induced action change rates for LH came against the AI}. We leave interpretation of that result as an open question. 

\subsection{Stag Hunt}
 Stag Hunt is a coordination game that forces players to weigh social coordination and trust (stag) with safety and sure gain (hare). Classic equilibria are at $(0, 0)$, $(1, 1)$, and, for this paper's payoff matrix, $(\frac{1}{2}, \frac{1}{2})$. Stag Hunt is especially interesting for PT analysis as an inquiry into the impact of loss aversion into the PT player decision making. 

\subsubsection{One Off Game (State History = $0$)}
\begin{figure}[H]
    \centering
    \begin{minipage}[t]{0.32\linewidth}
        \centering
        \includegraphics[width=\linewidth]{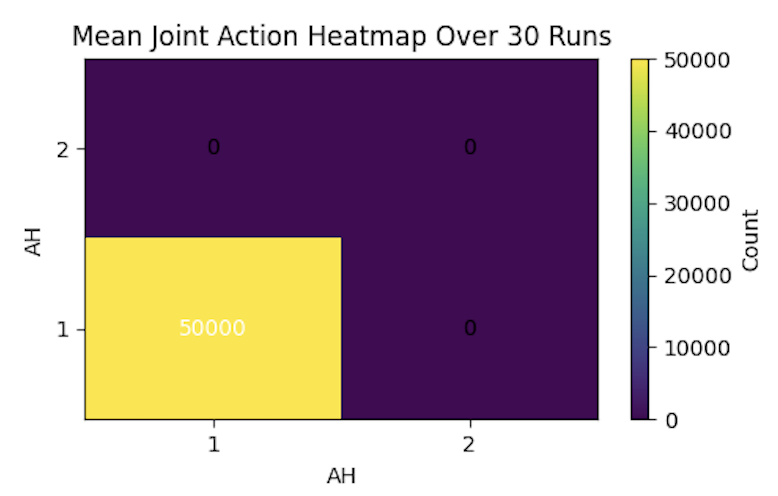}
        \caption{EMA}
        \label{fig:}
    \end{minipage}
    \begin{minipage}[t]{0.32\linewidth}
        \centering
        \includegraphics[width=\linewidth]{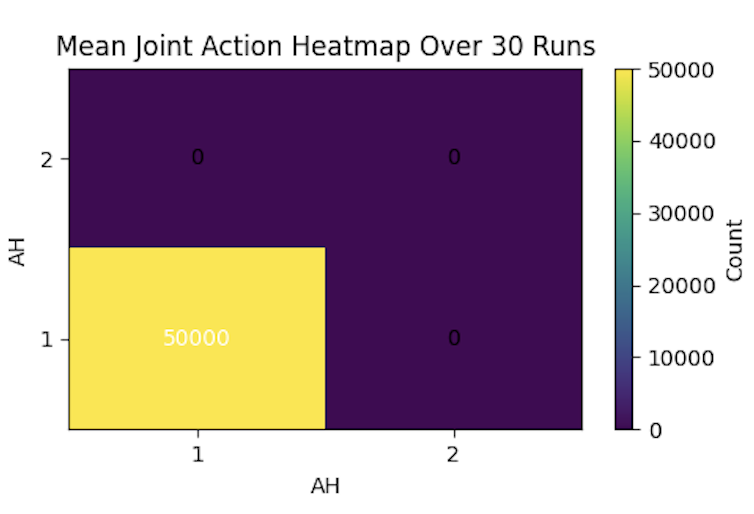}
        \caption{EMAOR}
        \label{fig:}
    \end{minipage}
    \begin{minipage}[t]{0.32\linewidth}
        \centering
        \includegraphics[width=\linewidth]{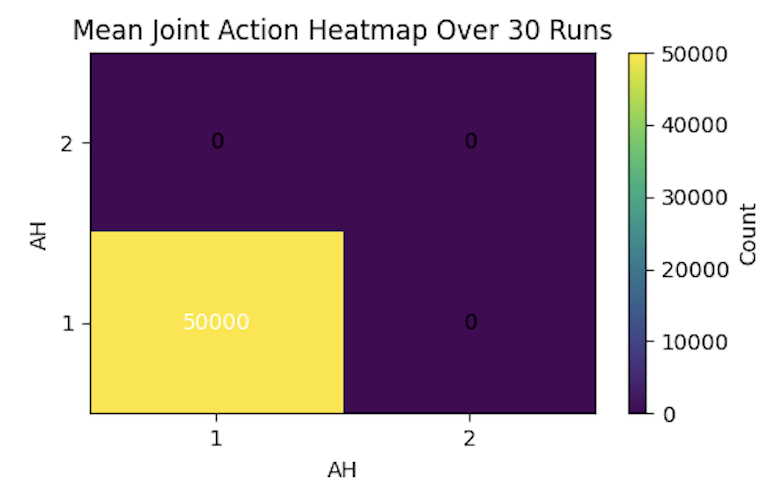}
        \caption{V-based}
        \label{fig:}
    \end{minipage}
    \caption{AH vs AH Baseline Stag Hunt}
    \label{fig:SHAHAH}
\end{figure}

We start with the AH vs AH baseline, visualized in Figure~\ref{fig:SHAHAH}, which all converged to the Stag $(1, 1)$ equilibrium, the highest reward equilibrium that we interpret as social coordination and trust. 

\begin{figure}[H]
    \centering
    \begin{minipage}[t]{0.32\linewidth}
        \centering
        \includegraphics[width=\linewidth]{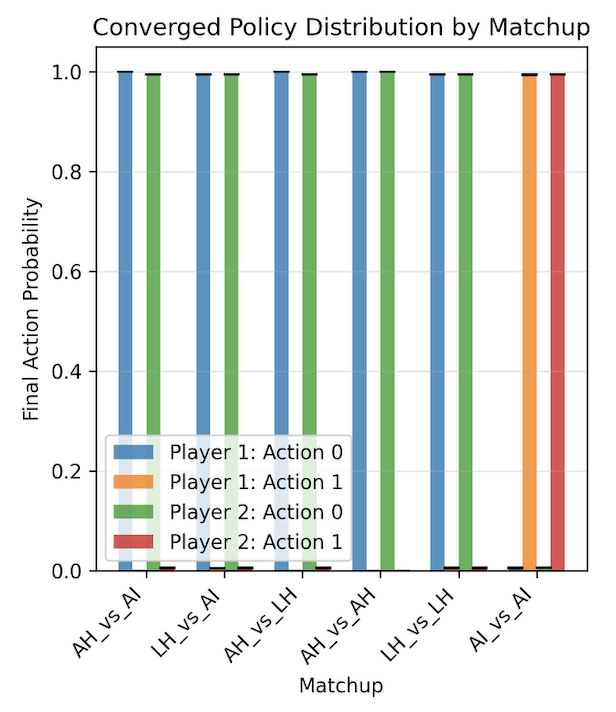}
        \caption{EMA}
        \label{fig:}
    \end{minipage}
    \begin{minipage}[t]{0.32\linewidth}
        \centering
        \includegraphics[width=\linewidth]{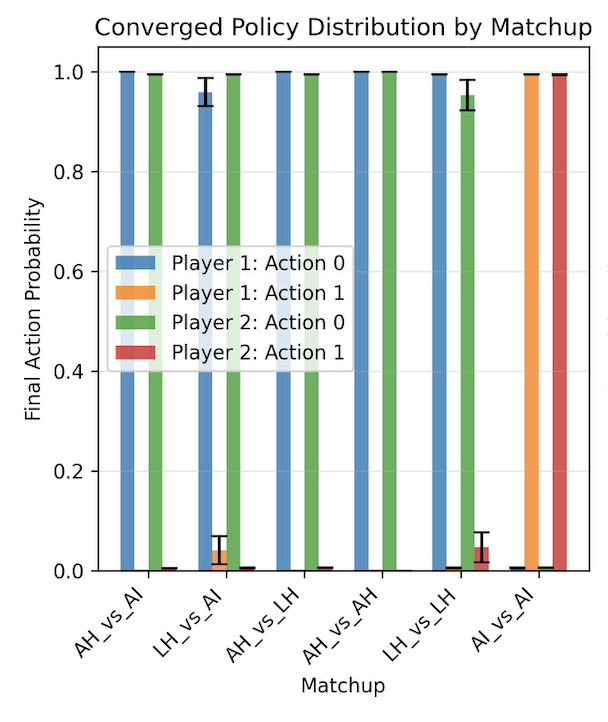}
        \caption{EMAOR}
        \label{fig:}
    \end{minipage}
    \begin{minipage}[t]{0.32\linewidth}
        \centering
        \includegraphics[width=\linewidth]{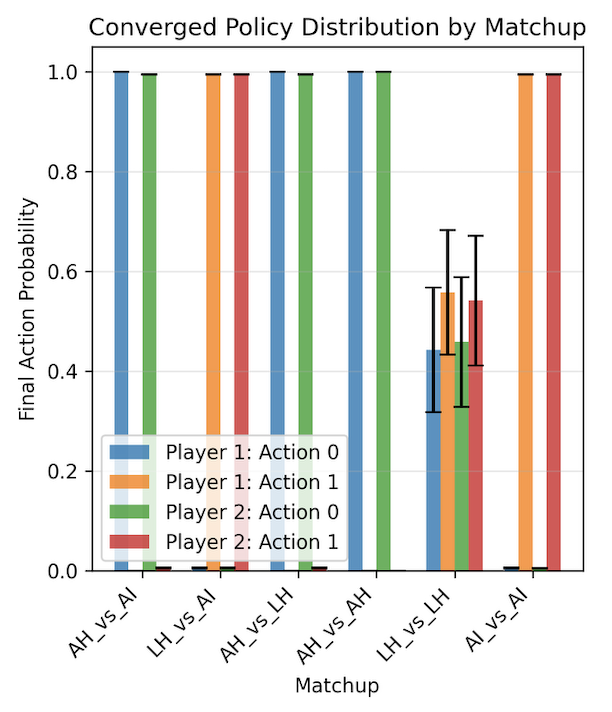}
        \caption{V-based}
        \label{fig:}
    \end{minipage}
    \caption{One Off Stag Hunt Matchup Policies (Last $5,000$ steps)}
    \label{fig:piSH1}
\end{figure}

Next we analyze the policies each matchup converged to, visualized in Figure~\ref{fig:piSH1}. Most matchups found equilibrium at Stag $(1, 1)$, with the exception of AI vs. AI at every reference setting, and LH vs. Lh and LH vs. AI at V-based reference setting.  

AI vs. AI and LH vs. AI (V-based) settled into the safe, less optimal equilibrium at Hare $(0, 0)$. Interestingly, LH vs. LH (V-based) found the mixed equilibrium $(\frac{1}{2}, \frac{1}{2})$, which corresponded with the highest CPT action change rates of any Stag Hunt matchup with mean change rates at $\approx 0.004$. Change rates that low are not statistically meaningful, however the correlation between the LH vs. LH quasi-anomalous mixed equilibrium and the heightened change rates suggest the possibility that CPT, or factors that influence CPT action changes, played a part. 

\subsubsection{Repeated Game (State History = $2$)}
\begin{figure}[H]
    \centering
    \begin{minipage}[t]{0.32\linewidth}
        \centering
        \includegraphics[width=\linewidth]{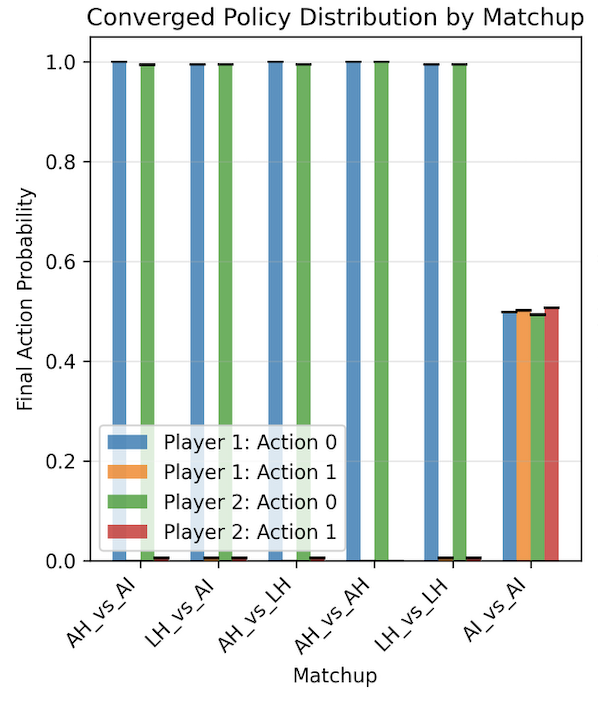}
        \caption{EMA}
        \label{fig:}
    \end{minipage}
    \begin{minipage}[t]{0.32\linewidth}
        \centering
        \includegraphics[width=\linewidth]{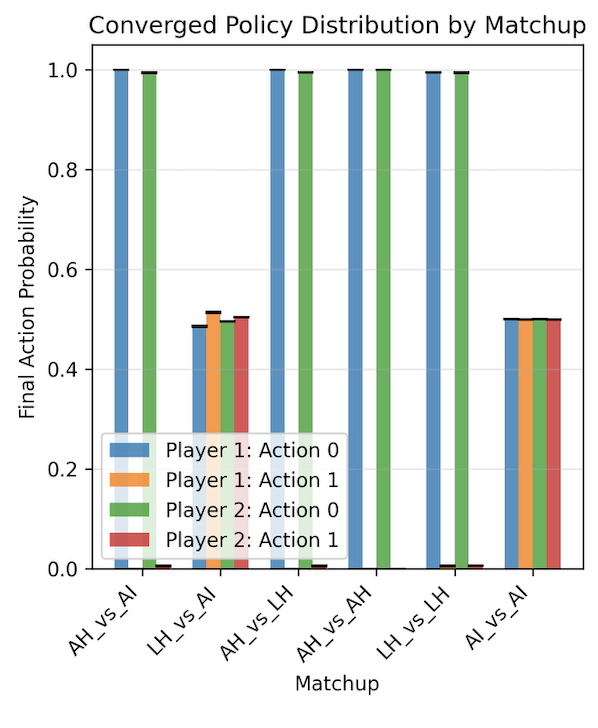}
        \caption{EMAOR}
        \label{fig:}
    \end{minipage}
    \begin{minipage}[t]{0.32\linewidth}
        \centering
        \includegraphics[width=\linewidth]{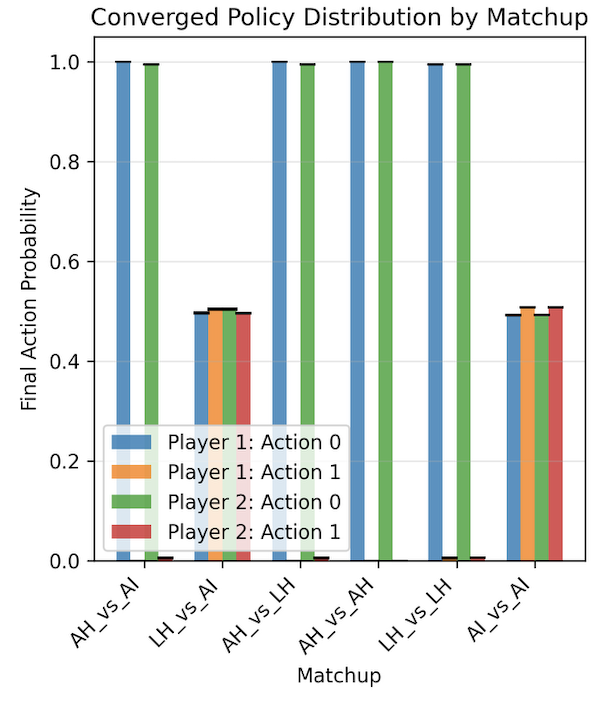}
        \caption{V-based}
        \label{fig:}
    \end{minipage}
    \caption{Repeated Stag Hunt Matchup Policies (Last $5,000$ steps)}
    \label{fig:piSHR}
\end{figure}

Repeated games offer an additional avenue for agents to coordinate towards an equilibrium. However, results generally mirror the one off game, with most matchups finding the Stag $(1, 1)$ equilibrium. However, Figure~\ref{fig:piSHR} shows that AI gravitated towards the $(\frac{1}{2}, \frac{1}{2})$ equilibrium in repeated games. 

Interestingly, the LH vs. AI matchup in EMAOR found the $(\frac{1}{2}, \frac{1}{2})$ equilibrium, indicating that the reference point setting does not determine the conditions for LH to find the $(\frac{1}{2}, \frac{1}{2})$ equilibrium. 

CPT action change rates were negligible in the repeated games, hovering around $0.0001$ when present. The pattern continued from previous games where the max CPT/EU L2 distance came from LH when facing AH. Closer inspection of AI Q values revealed equal valuation of each action, reflecting the policy that formed. 

\subsubsection{Stag Hunt Concluding Remarks}
Stag Hunt matchups generally gravitated towards Stag $(1, 1)$, but Hare $(0, 0)$ and the mixed equilibrium $(\frac{1}{2}, \frac{1}{2})$ both showed up in the AI and LH matchups. Unlike Prisoner's Dilemma and Matching Pennies, Stag Hunt did not exhibit anomalous behavior, but instead matched the theory very closely. 

\subsection{Battle of the Sexes}
Battle of the Sexes (BoS) has two pure equilibria at Opera $(0, 0)$ and Football $(1, 1)$, and a mixed equilibrium at $(3/5, 2/5)$. Notably, the mixed equilibrium is suboptimal for both players. BoS is a coordination game where both agents benefit from cooperating, but the terms of their cooperation determine the winner. Here, we will be looking for which agent is able to induce equilibrium to their advantage, and whether cooperation ever fails.

\subsubsection{One Off Game (State History = $0$)}
\begin{figure}[H]
    \centering
    \begin{minipage}[t]{0.32\linewidth}
        \centering
        \includegraphics[width=\linewidth]{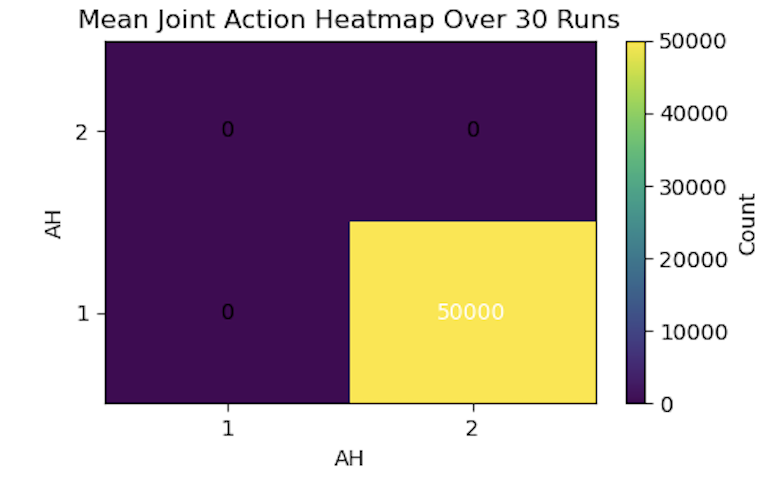}
        \caption{EMA}
        \label{fig:}
    \end{minipage}
    \begin{minipage}[t]{0.32\linewidth}
        \centering
        \includegraphics[width=\linewidth]{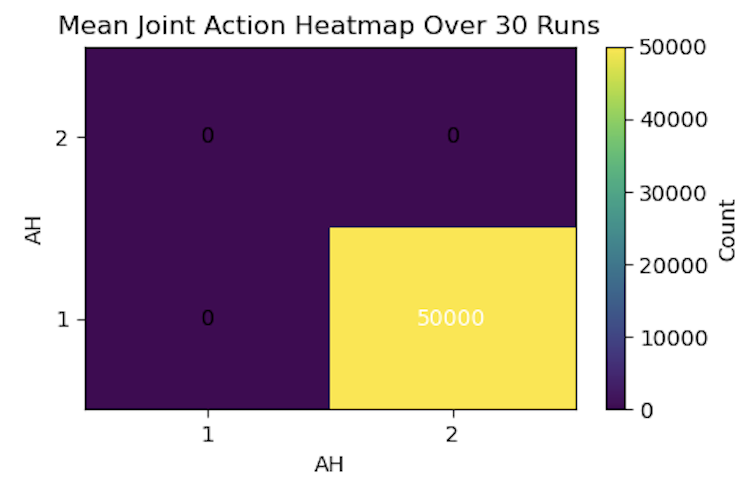}
        \caption{EMAOR}
        \label{fig:}
    \end{minipage}
    \begin{minipage}[t]{0.32\linewidth}
        \centering
        \includegraphics[width=\linewidth]{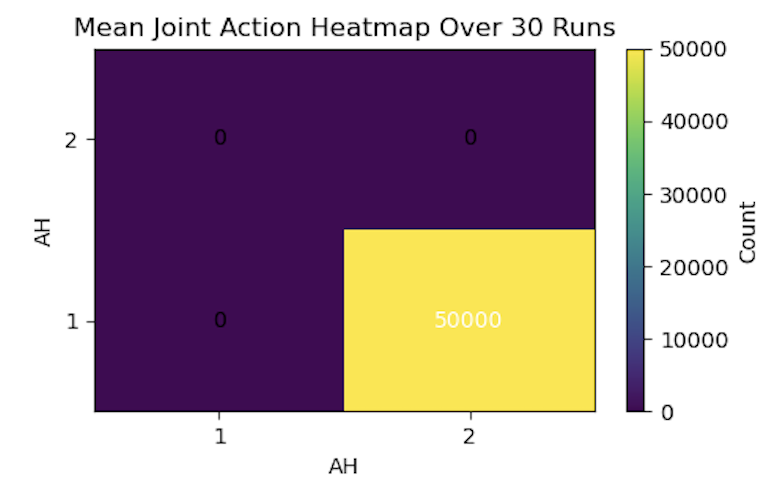}
        \caption{V-based}
        \label{fig:}
    \end{minipage}
    \caption{One Off Battle of the Sexes AH vs AH Baseline}
    \label{fig:piBoS1}
\end{figure}

The AH vs. AH baseline matchup converged to $(0, 1)$, where both players stuck to their preferred equilibrium and failed to coordinate, yielding 0 rewards for both AH players across all reference point types. Figure~\ref{fig:piBoS1} visualizes the joint action graphs. 

\begin{figure}[H]
    \centering
    \begin{minipage}[t]{0.32\linewidth}
        \centering
        \includegraphics[width=\linewidth]{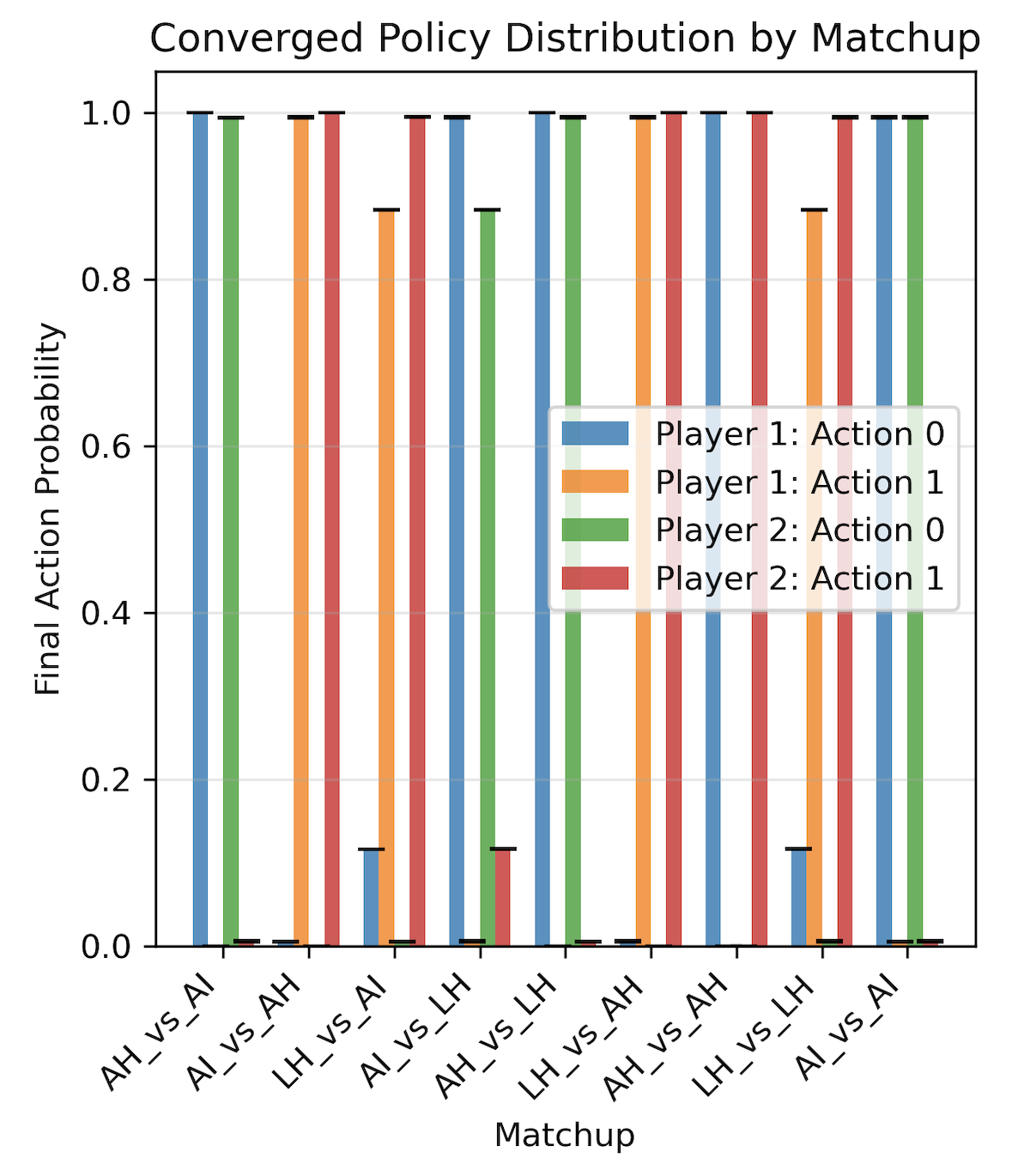}
        \caption{EMA}
        \label{fig:}
    \end{minipage}
    \begin{minipage}[t]{0.32\linewidth}
        \centering
        \includegraphics[width=\linewidth]{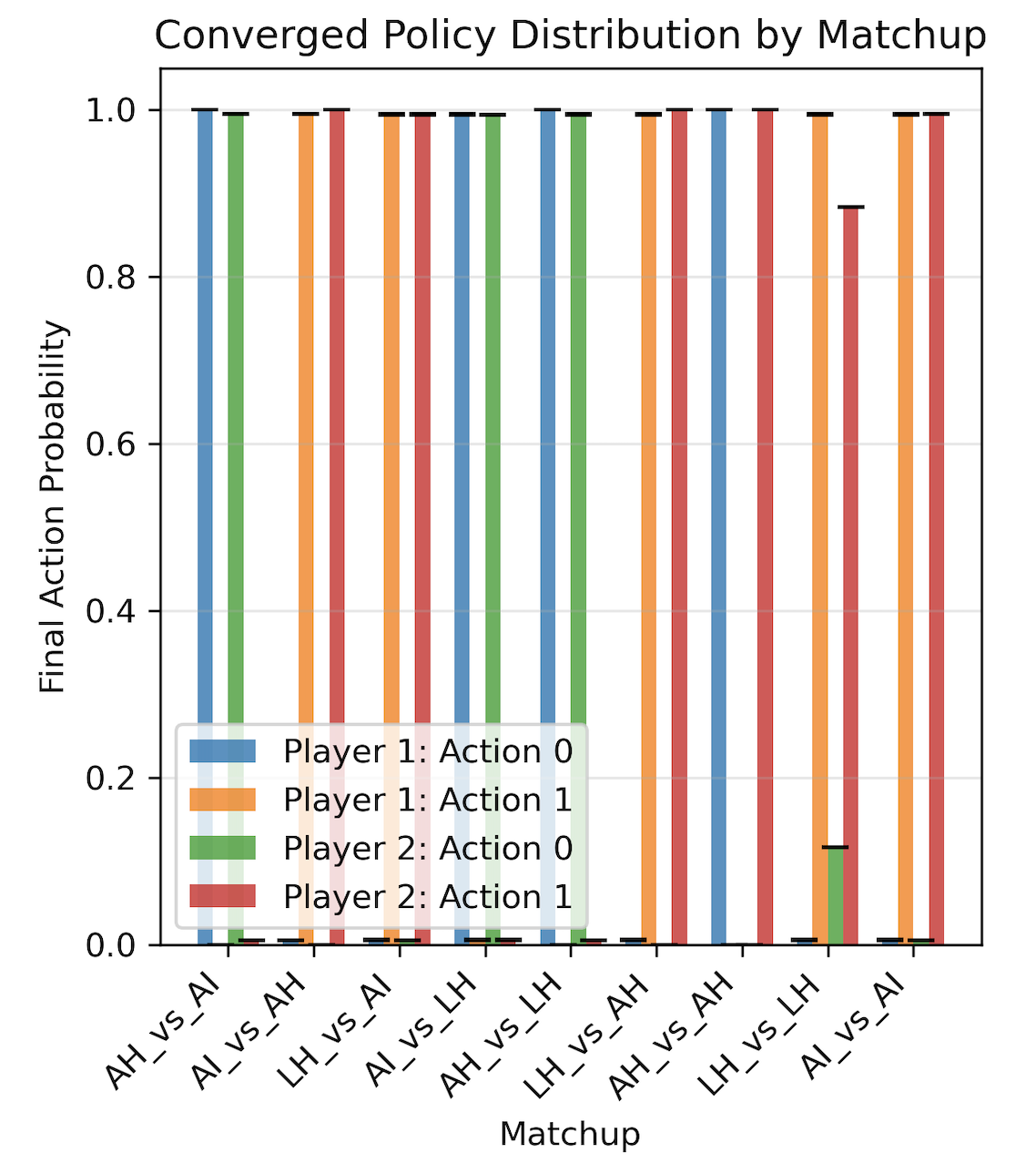}
        \caption{EMAOR}
        \label{fig:}
    \end{minipage}
    \begin{minipage}[t]{0.32\linewidth}
        \centering
        \includegraphics[width=\linewidth]{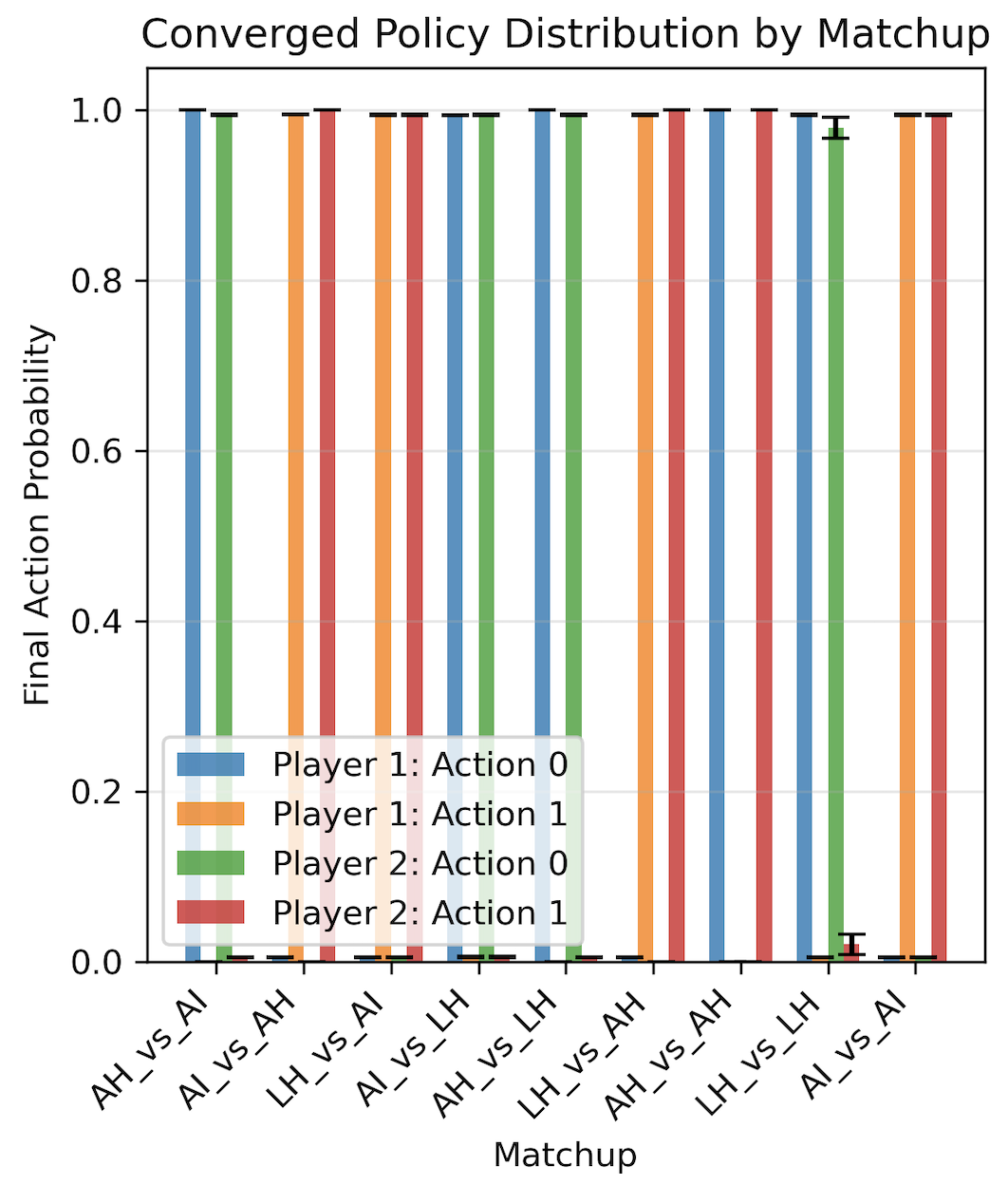}
        \caption{V-based}
        \label{fig:}
    \end{minipage}
    \caption{One Off Battle of the Sexes Matchup Policies (Last $5,000$ steps)}
    \label{fig:BoS1pi}
\end{figure}

Because the AH agents play a best response policy, they had no reason to change their behavior when receiving no reward. This proved to be a strength against learning agents, who were forced to adapt to their behavior. All matchup policies are visualized in Figure~\ref{fig:BoS1pi}.

Close inspection reveals AH winning the optimal equilibrium vs every learning agent, and standard equilibrium behavior from the AI. However, LH exhibits an unexpected and interesting behavior: against learning agents in EMA or EMAOR reference types, it finds a $9:1$ suboptimal, mixed policy that concedes most of the reward, but not all of it. 

The policy is odd because the $10\%$ of times it plays its optimal action against an opponent playing at $100\%$ their optimal action, which is all but guaranteed to be 0 reward. Other agents find suboptimal pure equilibria because they concede (knowingly or unknowingly) that they cannot receive the optimal reward. However, the LH converged to a behavior where $10\%$ of the time it plays an action that it \textit{knows} is worse. 

\begin{figure}[H]
    \centering
    \begin{minipage}[t]{0.32\linewidth}
        \centering
        \includegraphics[width=\linewidth]{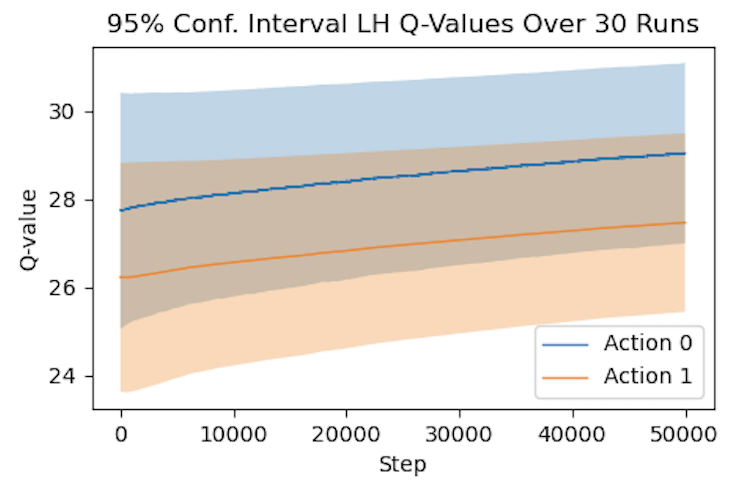}
        \caption{EMA}
        \label{fig:}
    \end{minipage}
    \begin{minipage}[t]{0.32\linewidth}
        \centering
        \includegraphics[width=\linewidth]{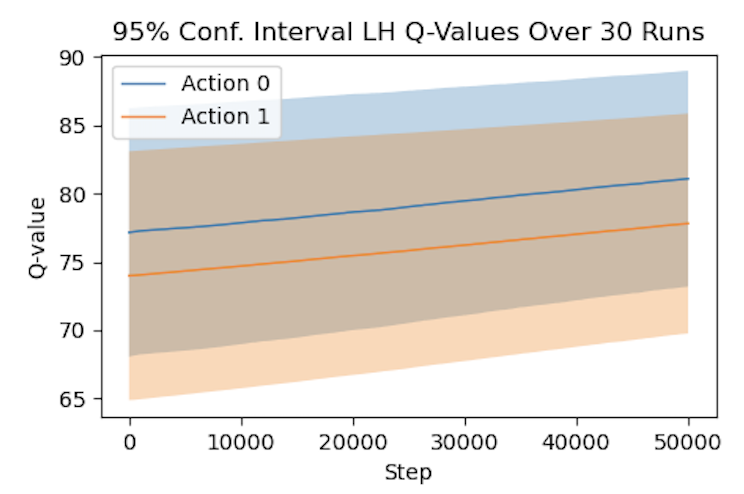}
        \caption{EMAOR}
        \label{fig:}
    \end{minipage}
    \begin{minipage}[t]{0.32\linewidth}
        \centering
        \includegraphics[width=\linewidth]{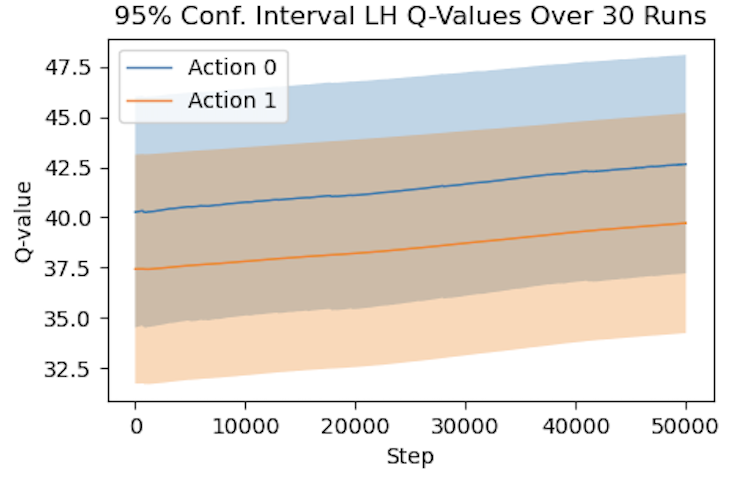}
        \caption{V-based}
        \label{fig:}
    \end{minipage}
    \caption{One Off Battle of the Sexes LH Q Values (vs. AI)}
    \label{fig:BoS1LHQ}
\end{figure}

To support the knowledge claim, consider the LH Q values visualized in Figure~\ref{fig:BoS1LHQ}. 

In each reference type, action 0 is the higher value action. While the exploration parameter $\epsilon$ does get clamped at $0.01$, that is not enough to determine long term convergence behavior. In EMA settings, where the reference point is in $[2, 3]$ for BoS, each Q value is so much higher than the reference point that the CPT transformation diminishes sensitivity to the value difference. Meanwhile, with V-based reference points where the reference point is determined by the Q values, the reference point is close enough to the Q values that CPT does not diminish sensitivity to the value difference. 
\vspace{-.5em}
\subsubsection{Repeated Game (State History = $2$)}
\begin{figure}[H]
    \centering
    \begin{minipage}[t]{0.32\linewidth}
        \centering
        \includegraphics[width=\linewidth]{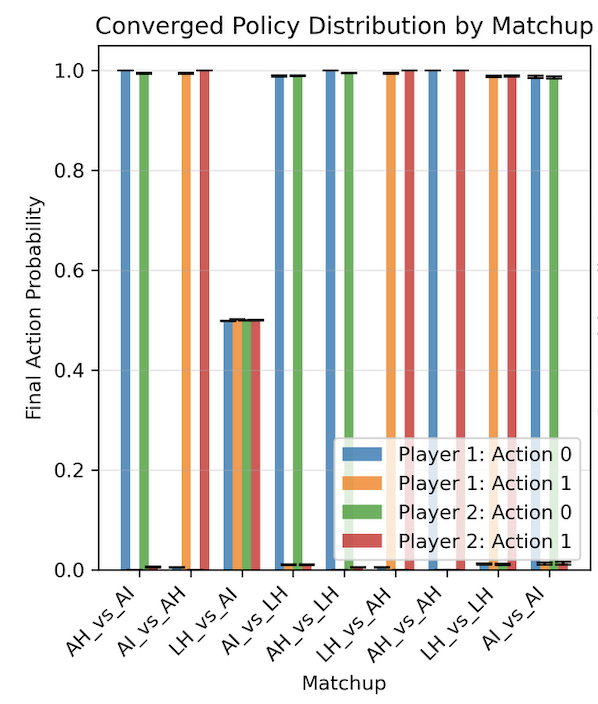}
        \caption{EMA}
        \label{fig:}
    \end{minipage}
    \begin{minipage}[t]{0.32\linewidth}
        \centering
        \includegraphics[width=\linewidth]{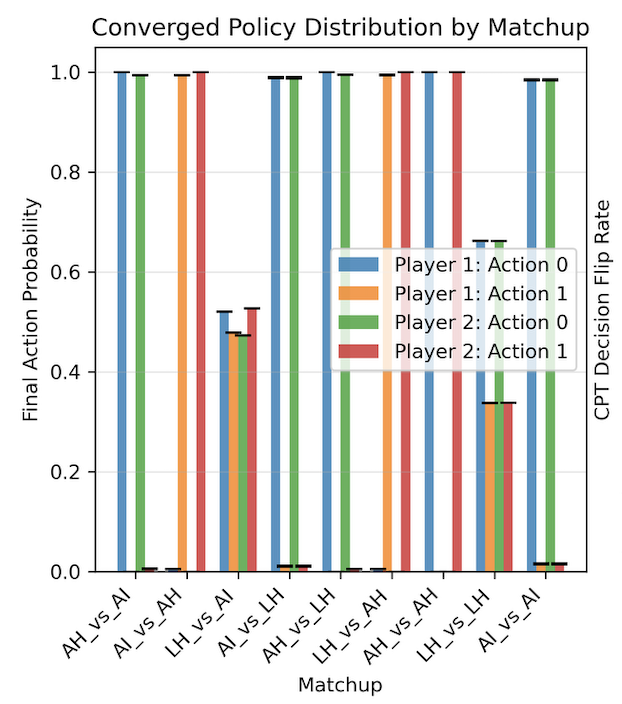}
        \caption{EMAOR}
        \label{fig:}
    \end{minipage}
    \begin{minipage}[t]{0.32\linewidth}
        \centering
        \includegraphics[width=\linewidth]{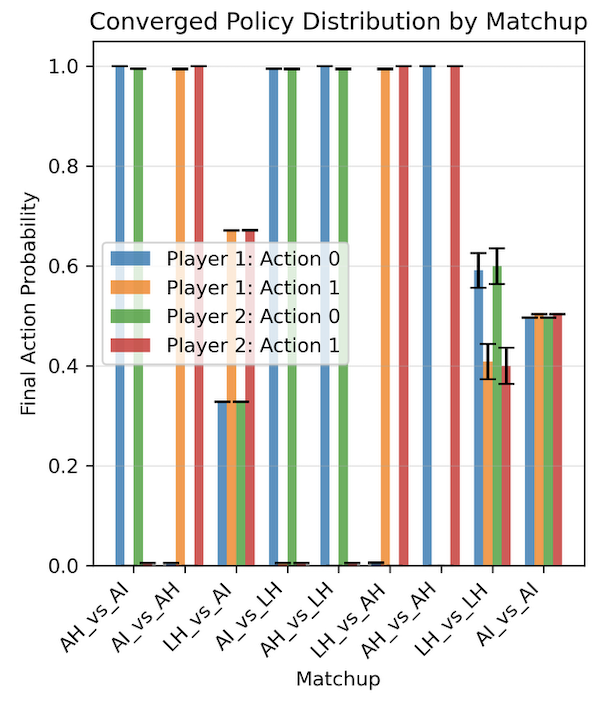}
        \caption{V-based}
        \label{fig:vbasedpiBoS}
    \end{minipage}
    \caption{Repeated Battle of the Sexes Policies (Last $5,000$ Steps)}
    \label{fig:BoSRPI}
\end{figure}

In the repeated game, we open up the convergence space from the one off equilibria and expect to find new attractors. Specifically, we find a convergence point at $(\frac{1}{2}, \frac{1}{2})$, as visualized in Figure~\ref{fig:BoSRPI}. 
\vspace{-1em}

With LH as the first player, in both EMA settings LH vs. AI converges to $(\frac{1}{2}, \frac{1}{2})$. However, when AI is the first player, the equilibrium collapses back to Football. In the V-based reference point setting, LH and AI find a strange convergence point at $\approx (0.35, 0.35)$ that yields slightly better rewards for the AI. Despite the odd behavior with LH, who also finds an $\approx (0.6, 0.6)$ convergence point with itself, the CPT action change and CPT/EU L2 distance are both negligible. 

\begin{figure}[h]
    \centering
    \includegraphics[width=0.75\linewidth]{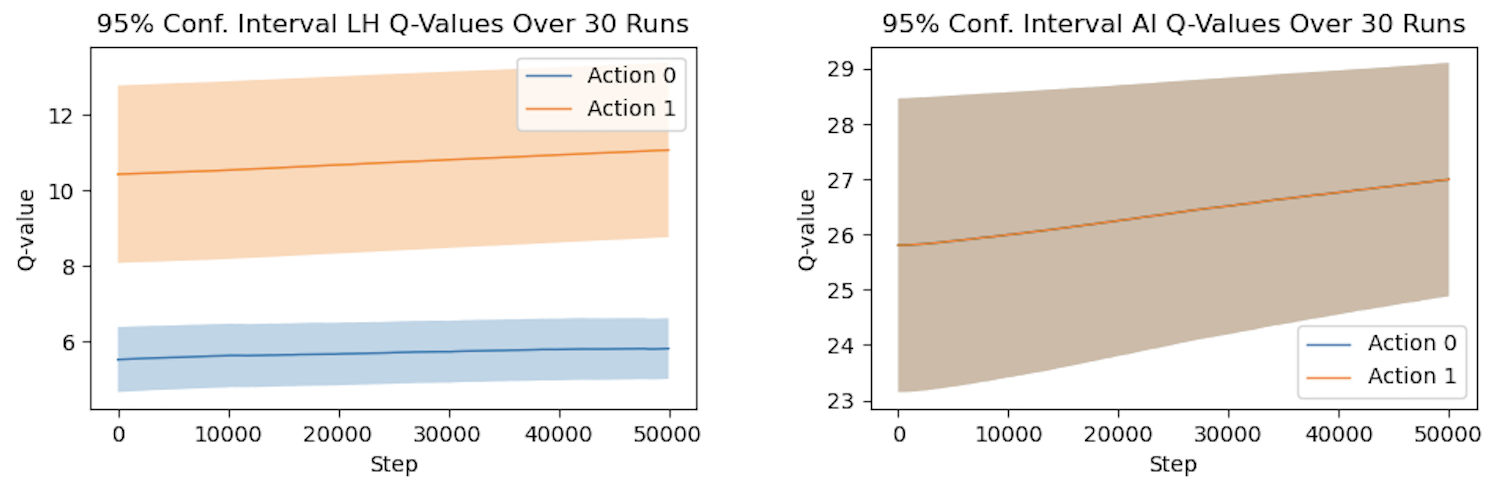}
    \caption{LH vs. AI Q values: Clear preference vs. Indifference (V-based)}
    \label{fig:LHAIQBoSR}
\end{figure}

Interestingly, in both of the anomalous cases (LH vs. AI $(0.35, 0.35)$ and LH vs. LH $(0.6, 0.6)$) the policies diverge from their Q values. For example, in LH vs AI, the same policy is generated from two very different Q values (Figure~\ref{fig:LHAIQBoSR}). Importantly, the reference point setting was V-based, and split the Q values for the LH in half, so action 1 should be a gain and action 0 should be a loss. And yet, the policy mixed. With this information and the lack of variance between runs in Figure~\ref{fig:vbasedpiBoS}, we conjecture that this could be an equilibrium, applying to the similarly structured (i.e. Q value and policy disparity) LH vs. LH in EMAOR and V-based reference point settings. 

\subsubsection{Battle of the Sexes Concluding Remarks}
The Battle of the Sexes drew out some interesting convergence behavior idiosyncratic to CPT (LH vs. AI in one off games) and suggestive of equilibrium (LH vs AI in repeated V-based games). Most matchups gravitated towards pure equilibria, but in one off games we saw signs of diminished sensitivity changing the structure of equilibrium for LH agents. Furthermore, we saw convergence to $(\frac{1}{2}, \frac{1}{2})$ among several matchups in repeated games. 

The LH did not perform well in the Battle of the Sexes, as it never chose a pure optimal equilibrium and for that reason never outperformed its opponent (excluding self play). As our primary object of inquiry, this was the first game result that suggested that the LH has a competitive disadvantage. Whether that is actually the case is an open question.

\subsection{Chicken}
Chicken is a brinksmanship game specifically designed to model catastrophic mutual actions. It has two pure equilibria at $(0, 1)$ and $(1, 0)$ and a mixed equilibrium specific to our payoff matrix at $(\frac{9}{10}, \frac{9}{10})$. In our analysis, we are specifically looking for signs that the LH is more susceptible to suboptimal pure or mixed equilibria due to loss aversion. 

\subsubsection{One Off Game (State History = $0$}
First we start with the AH vs. AH baseline, where all three reference types yielded a mutually catastrophic convergence point where both AH play $(0, 0)$, resulting in a perpetual negative reward for both agents. 

\begin{figure}[H]
    \centering
    \begin{minipage}[t]{0.32\linewidth}
        \centering
        \includegraphics[width=\linewidth]{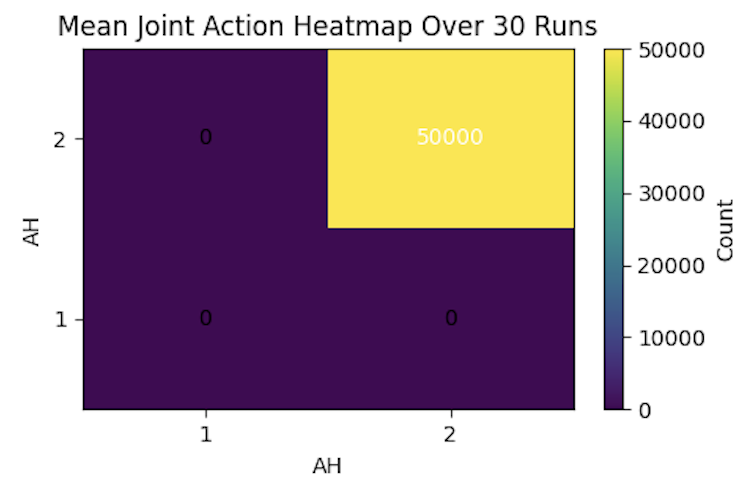}
        \caption{EMA}
        \label{fig:}
    \end{minipage}
    \begin{minipage}[t]{0.32\linewidth}
        \centering
        \includegraphics[width=\linewidth]{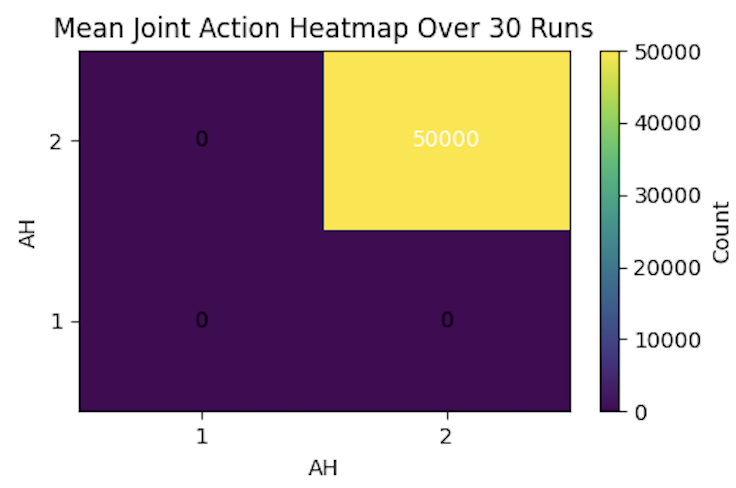}
        \caption{EMAOR}
        \label{fig:}
    \end{minipage}
    \begin{minipage}[t]{0.32\linewidth}
        \centering
        \includegraphics[width=\linewidth]{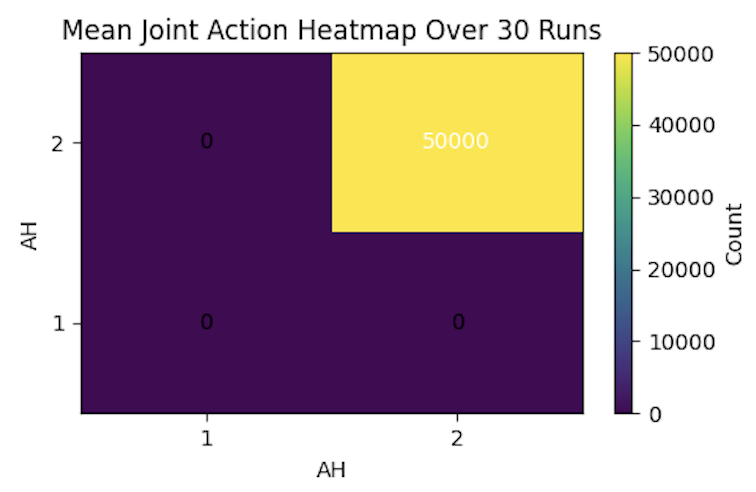}
        \caption{V-based}
        \label{fig:}
    \end{minipage}
    \caption{AH vs. AH baseline Chicken}
    \label{fig:CAHAH}
\end{figure}

Next we observe the matchup policies across the three reference types, where we observe a general adherence to pure strategies with a few unstable matchups playing a policy $\approx (0.95, 0.5)$.

\begin{figure}[H]
    \centering
    \begin{minipage}[t]{0.32\linewidth}
        \centering
        \includegraphics[width=\linewidth]{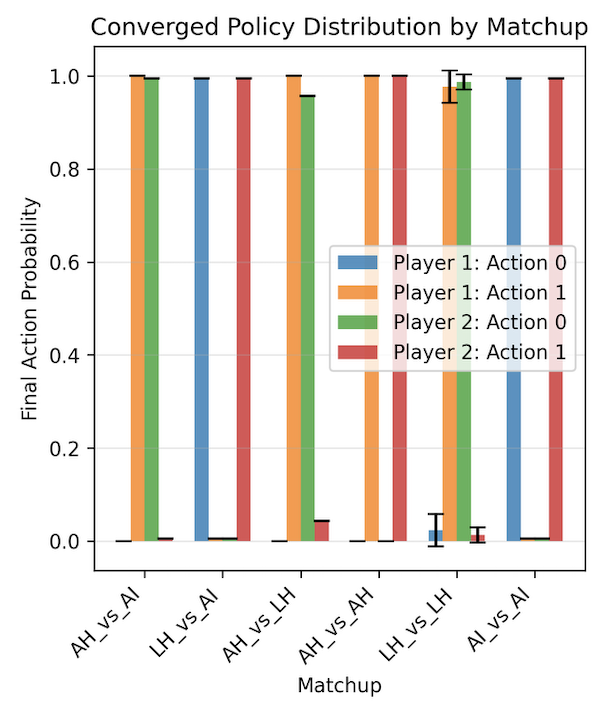}
        \caption{EMA}
        \label{fig:}
    \end{minipage}
    \begin{minipage}[t]{0.32\linewidth}
        \centering
        \includegraphics[width=\linewidth]{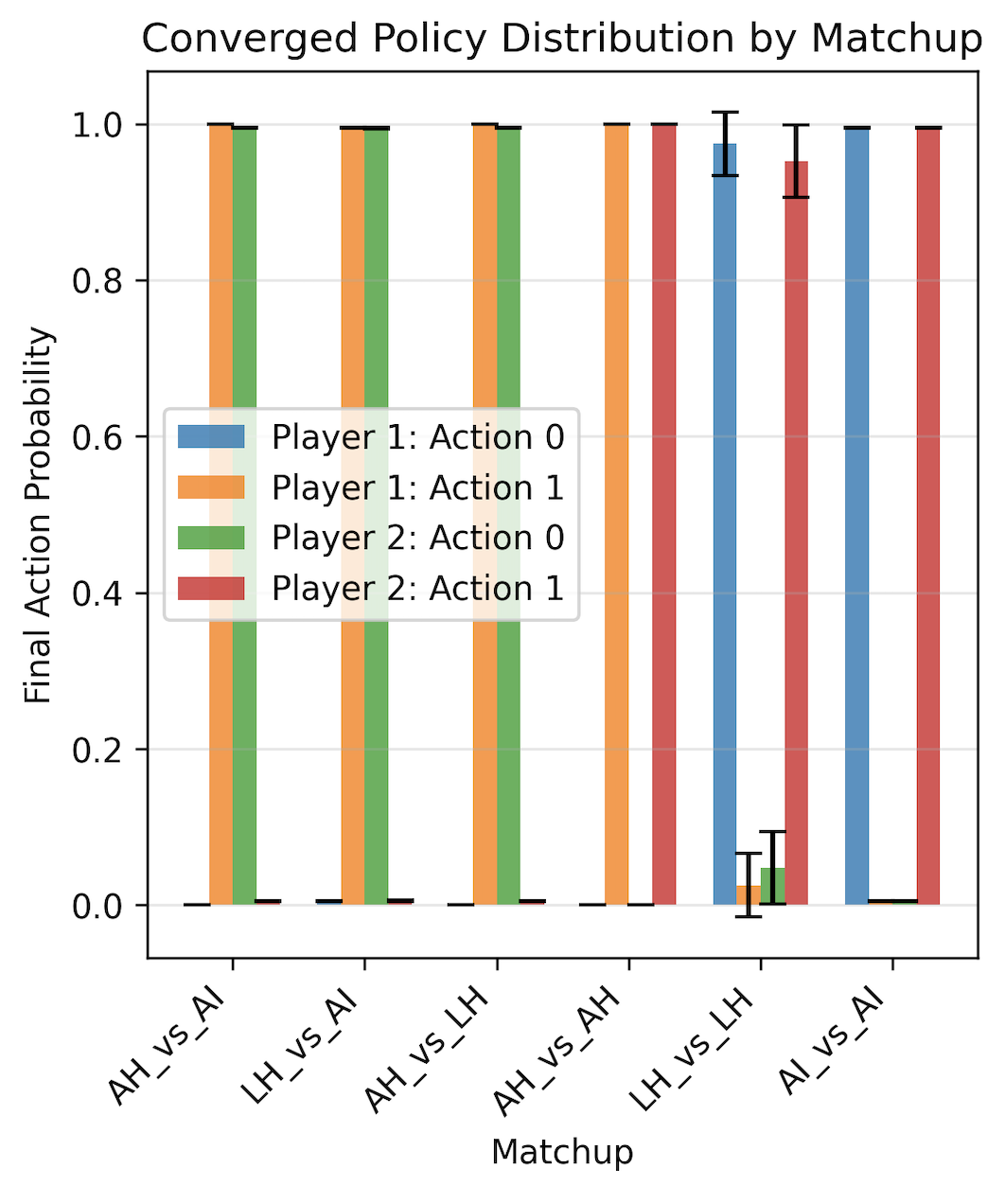}
        \caption{EMAOR}
        \label{fig:}
    \end{minipage}
    \begin{minipage}[t]{0.32\linewidth}
        \centering
        \includegraphics[width=\linewidth]{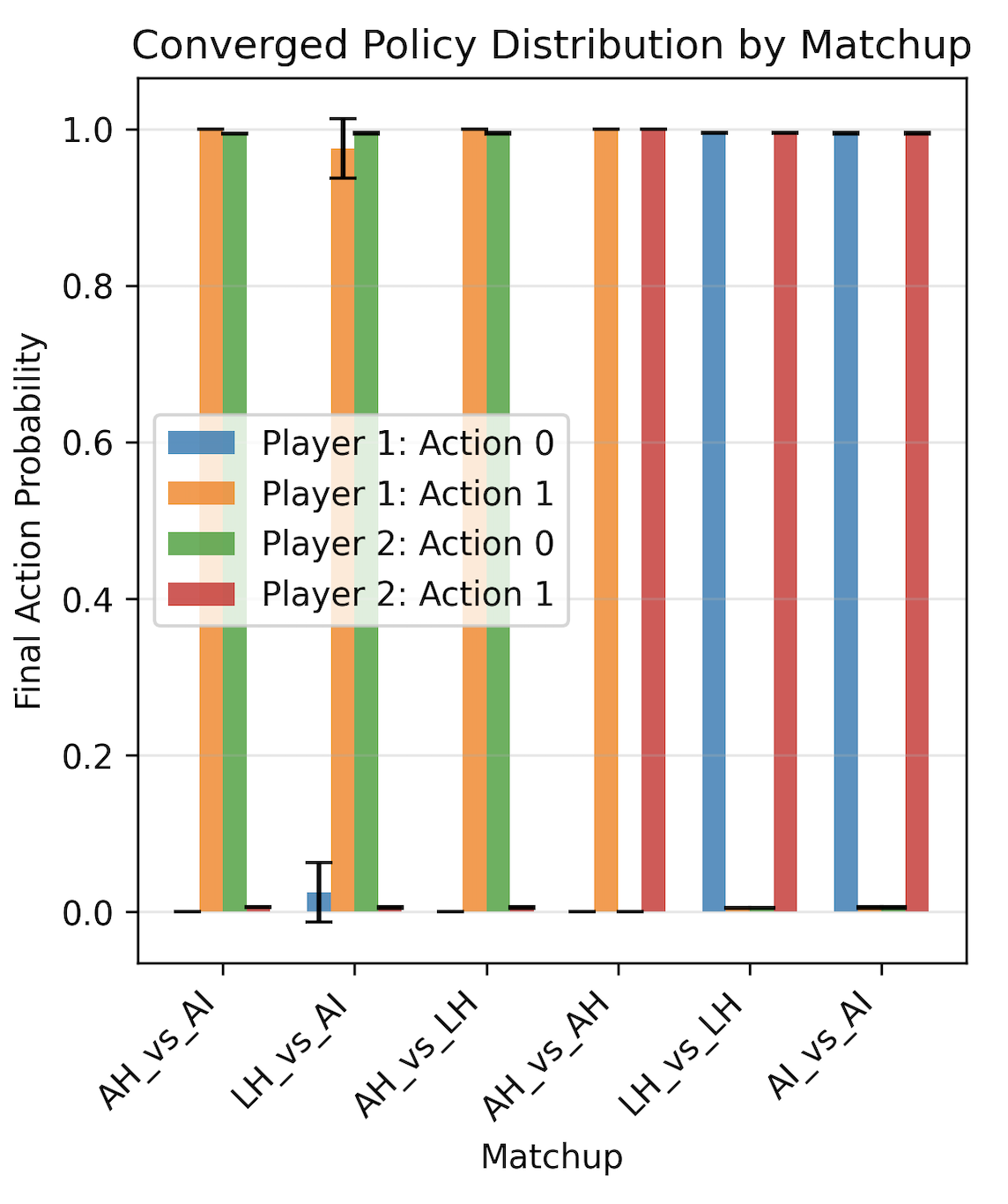}
        \caption{V-based}
        \label{fig:}
    \end{minipage}
    \caption{One Off Chicken Policies (Last $5,000$ Steps)}
    \label{fig:C1Pi}
\end{figure}

Most of the results were standard and unsurprising. The AH had a significant advantage against the learners who were forced to adapt to its brute force strategy. LH outperformed AI in the EMAOR and V-based reference point settings. LH had a bit of uncertainty in its strategies at times, potentially caused by its relatively high CPT change action rates that got as high as $0.01$ in the EMA setting. 

\subsubsection{Repeated Game (State History = $2$)}
\begin{figure}[H]
    \centering
    \begin{minipage}[t]{0.32\linewidth}
        \centering
        \includegraphics[width=\linewidth]{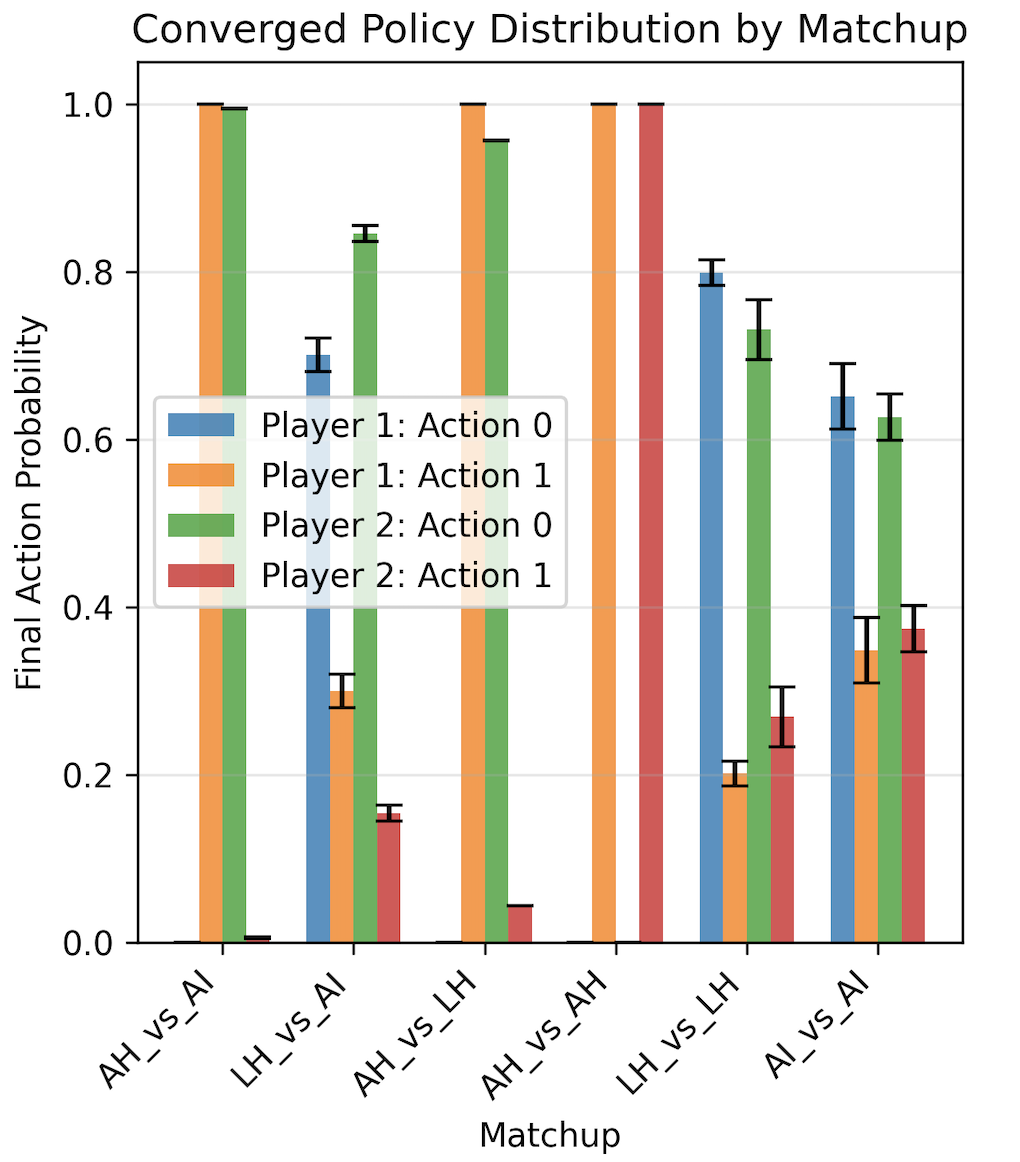}
        \caption{EMA}
        \label{fig:}
    \end{minipage}
    \begin{minipage}[t]{0.32\linewidth}
        \centering
        \includegraphics[width=\linewidth]{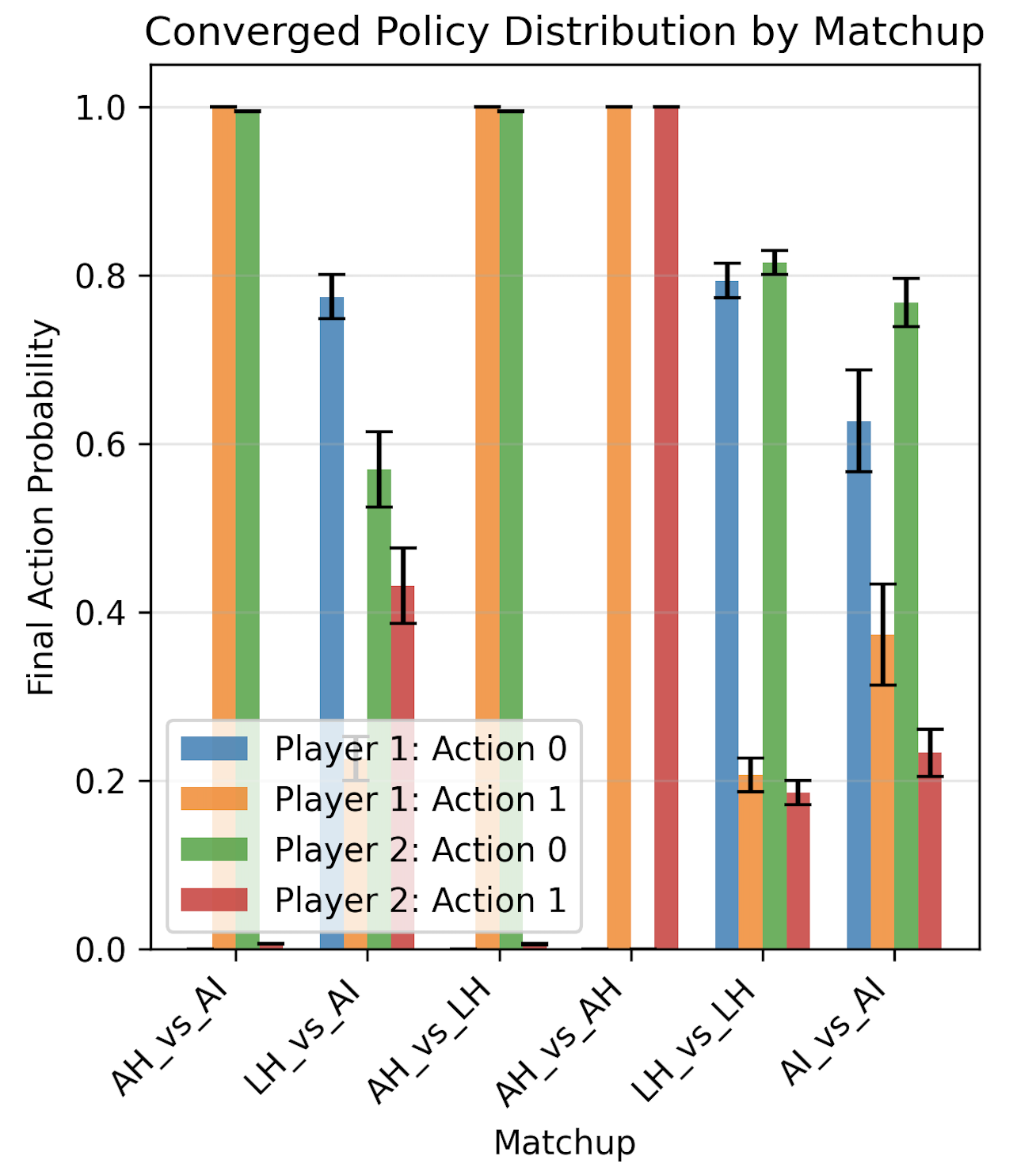}
        \caption{EMAOR}
        \label{fig:}
    \end{minipage}
    \begin{minipage}[t]{0.32\linewidth}
        \centering
        \includegraphics[width=\linewidth]{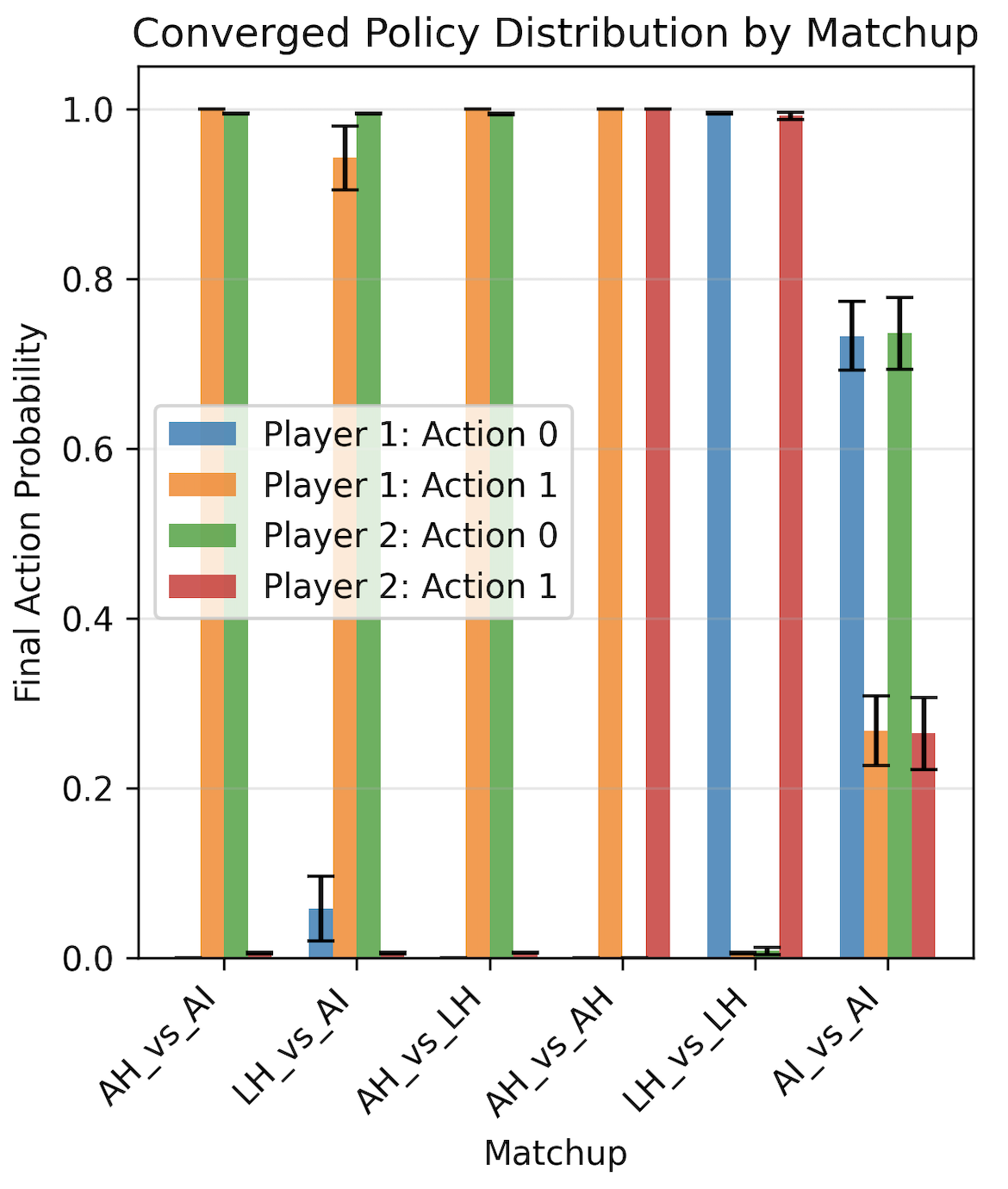}
        \caption{V-based}
        \label{fig:}
    \end{minipage}
    \caption{Repeated Chicken Policies (Last $5,000$ Steps)}
    \label{fig:CRpi}
\end{figure}

The repeated game tells a very different story from the one off games. Where the one off games were standard and stuck to the expectation, the repeated games mixed significantly, had high CPT action change rates, and converged to unanticipated regions. Figure~\ref{fig:CRpi} visualizes the coverged policies for each reference point. 

Policies in the repeated Chicken game group into pure strategies, mixed strategies $\approx (0.8, 0.8)$, and mixed strategies $\approx (0.6, 0.6)$. Variance exists between runs, but not to the extent that the basins of attraction are in question.  
\vspace{-1em}
\begin{figure}[H]
    \centering
    \begin{minipage}[t]{0.32\linewidth}
        \centering
        \includegraphics[width=\linewidth]{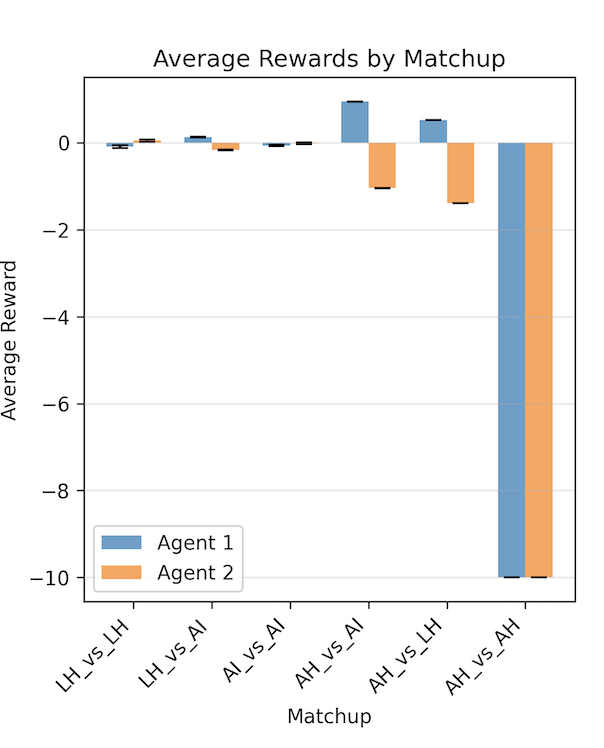}
        \caption{EMA}
        \label{fig:}
    \end{minipage}
    \begin{minipage}[t]{0.32\linewidth}
        \centering
        \includegraphics[width=\linewidth]{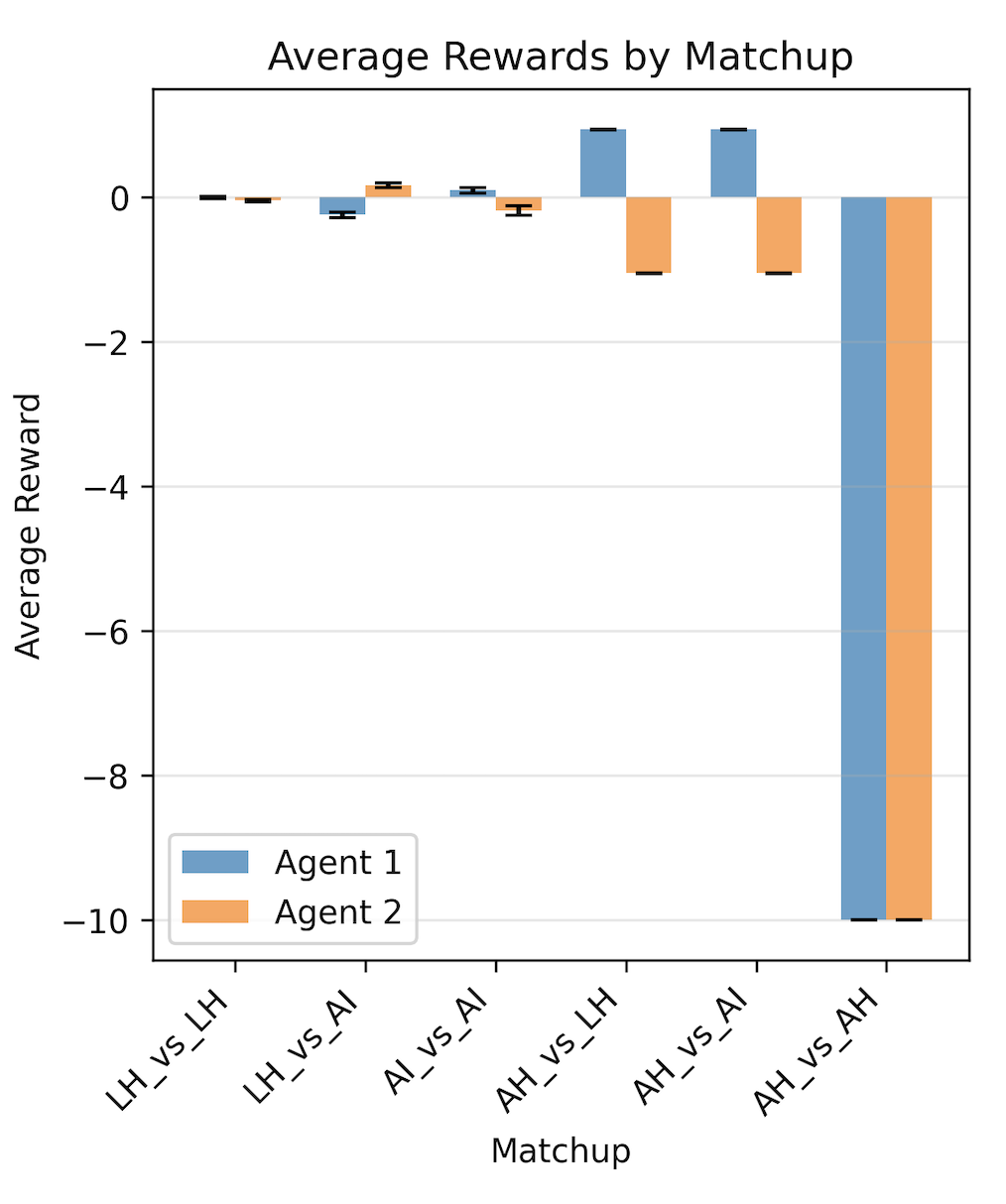}
        \caption{EMAOR}
        \label{fig:}
    \end{minipage}
    \begin{minipage}[t]{0.32\linewidth}
        \centering
        \includegraphics[width=\linewidth]{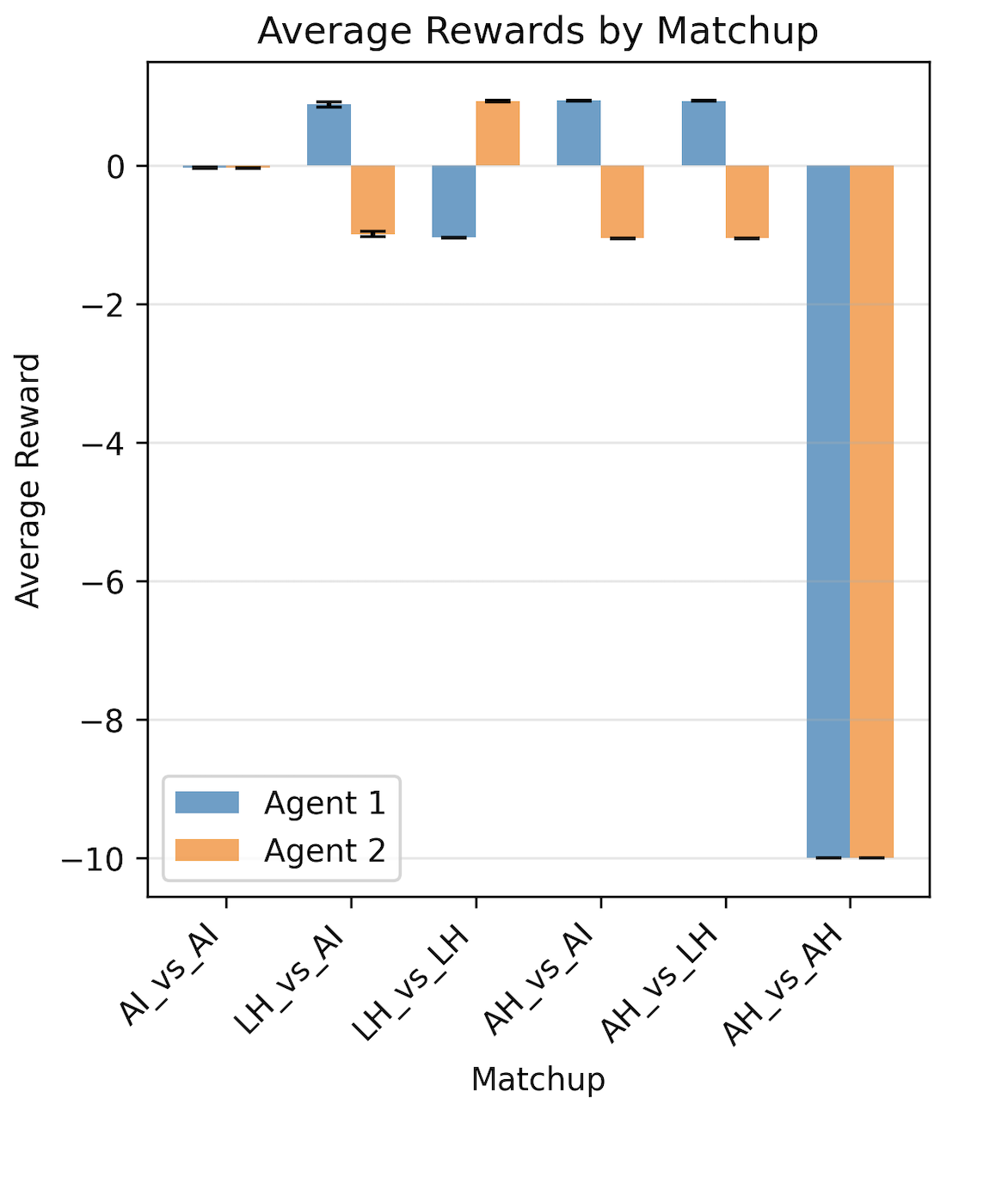}
        \caption{V-based}
        \label{fig:}
    \end{minipage}
    \caption{Repeated Chicken Avg. Rewards (Last $5,000$ Steps)}
    \label{fig:CRrew}
\end{figure}
Within the EMA and EMAOR runs, rewards for learning agent matchups (no AH) converge to 0 on both sides as agents minimize losses. However, in matchups with the AH agent they are forced into pure strategies at risk of significant losses. Figure~\ref{fig:} visualizes the rewards.

Finally, the CPT change of action rate exceeds $7\%$ for LH in LH vs. AI and exceeds $5\%$ for LH vs. LH in the EMA reference point setting, and exceeds $7\%$ for LH vs. LH in the EMAOR setting. However, there is no correlation between runs with high CPT action change rate and unexpected mixing. 

\subsubsection{Chicken Concluding Remarks}
One off Chicken mirrored the EU expectations, with a heavy attraction towards pure equilibria that gave an edge to the best responders. LH struggled in the one off games, never adhering to its optimal action and instead mixing frequently. 

Repeated Chicken, on the other hand, found multiple mixed strategy basins of attraction, including points $\approx (0.8, 0.8)$ and $(0.6, 0.6)$. While CPT action change rates spiked, the CPT/EU L2 norm declined aside from the LH agent when playing AH --- its L2 magnitude shot up into the hundreds. 

Finally, with respect to the competition between the LH and the AI, there was no clear edge for the AI over the LH. In fact, the human found frequently found optimal or at least neutral equilibria. 

\subsection{Crawford's Counterexample}
Recall that Crawford's Counterexample (CC) is explicitly designed to expose equilibrium non existence pathologies for players with non-EU preferences. The experiment we ran with CC can then be framed as a numerical verification and analysis of the results from CPT players playing CC. Recall once more that the Nash Equilibrium for CC is $(\frac{1}{2}, \frac{1}{2})$. 

\subsubsection{Pathology Existence}
\begin{figure}[H]
    \centering
    \begin{minipage}[t]{0.32\linewidth}
        \centering
        \includegraphics[width=\linewidth]{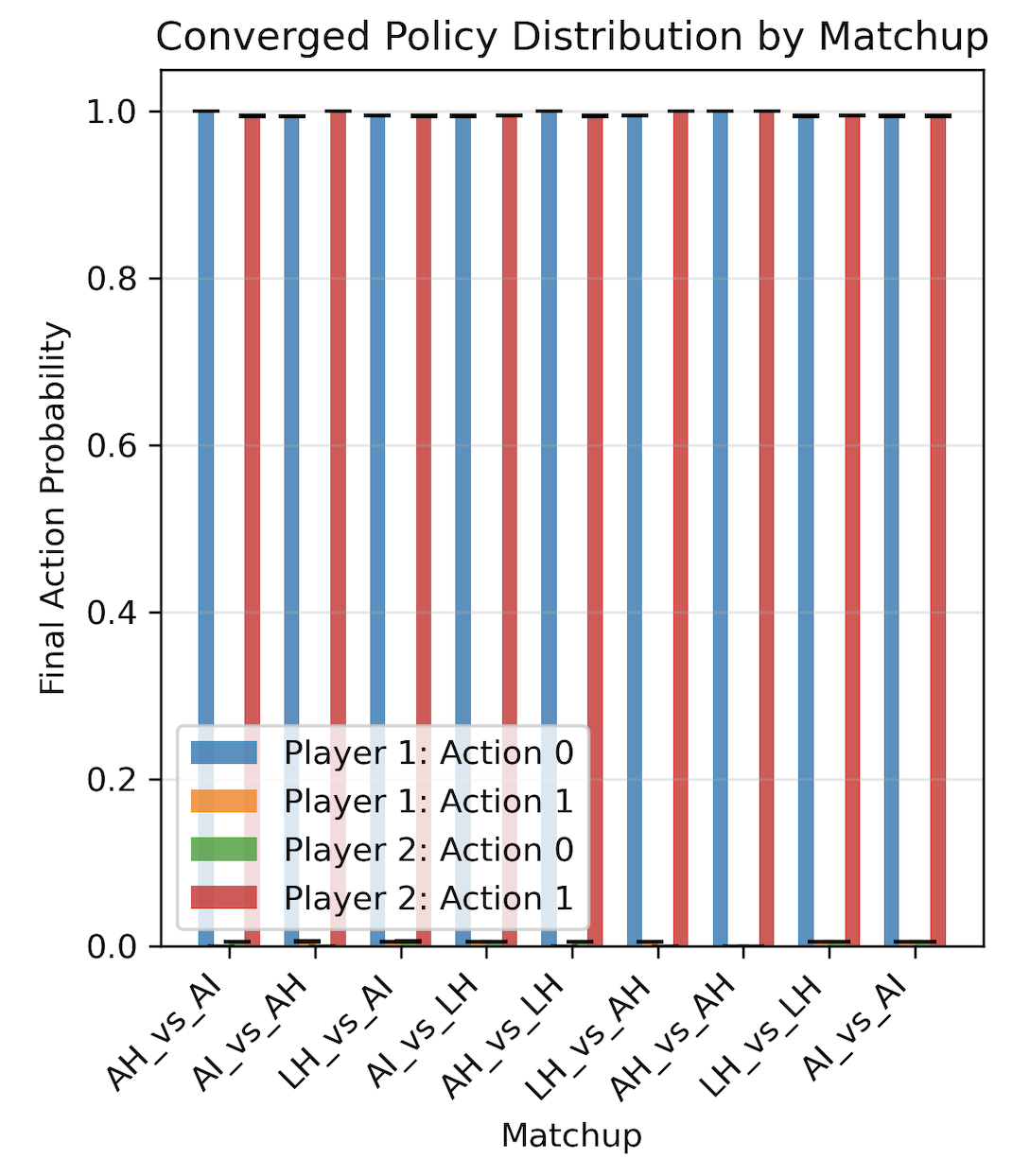}
        \caption{EMA}
        \label{fig:}
    \end{minipage}
    \begin{minipage}[t]{0.32\linewidth}
        \centering
        \includegraphics[width=\linewidth]{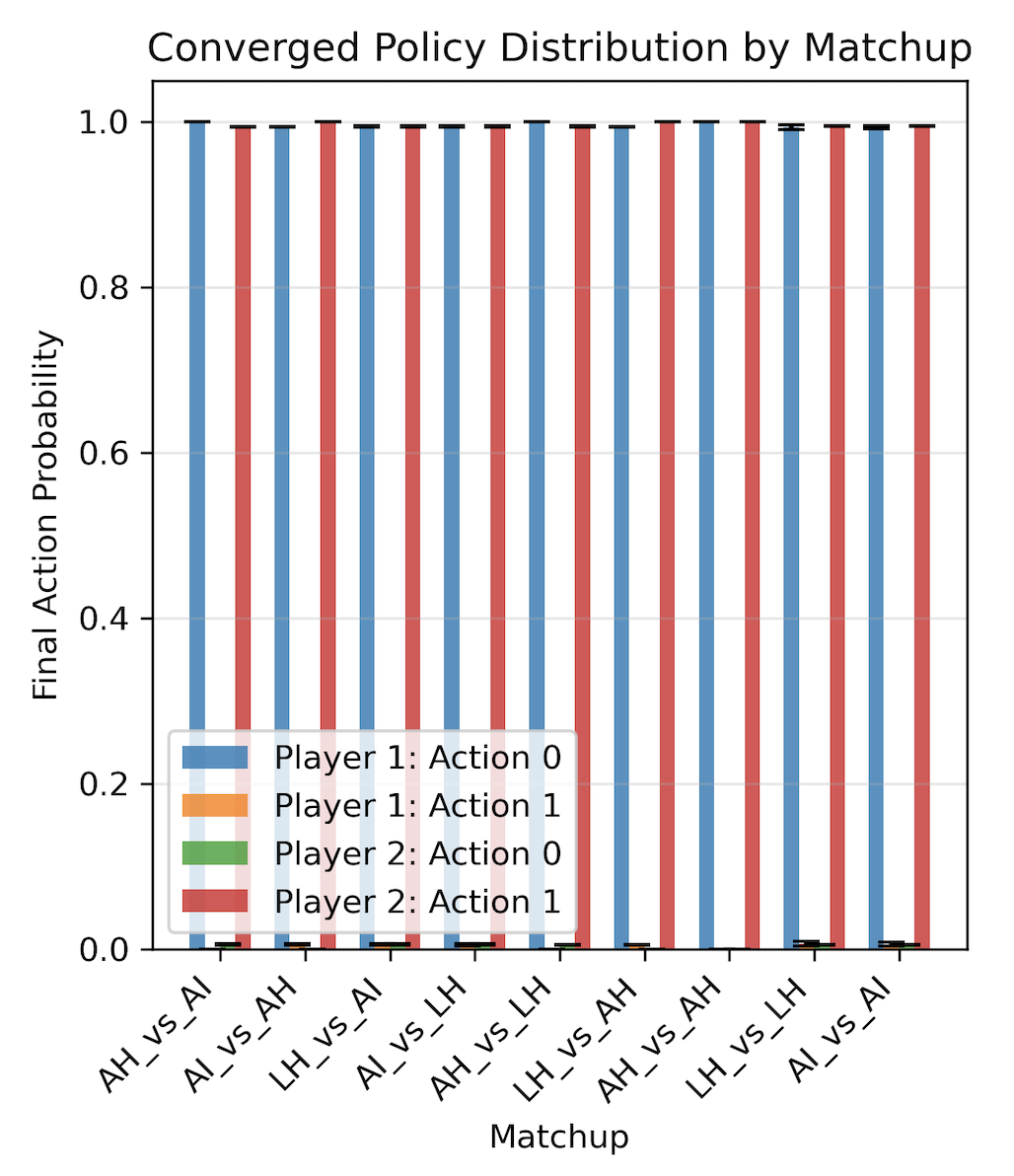}
        \caption{EMAOR}
        \label{fig:}
    \end{minipage}
    \begin{minipage}[t]{0.32\linewidth}
        \centering
        \includegraphics[width=\linewidth]{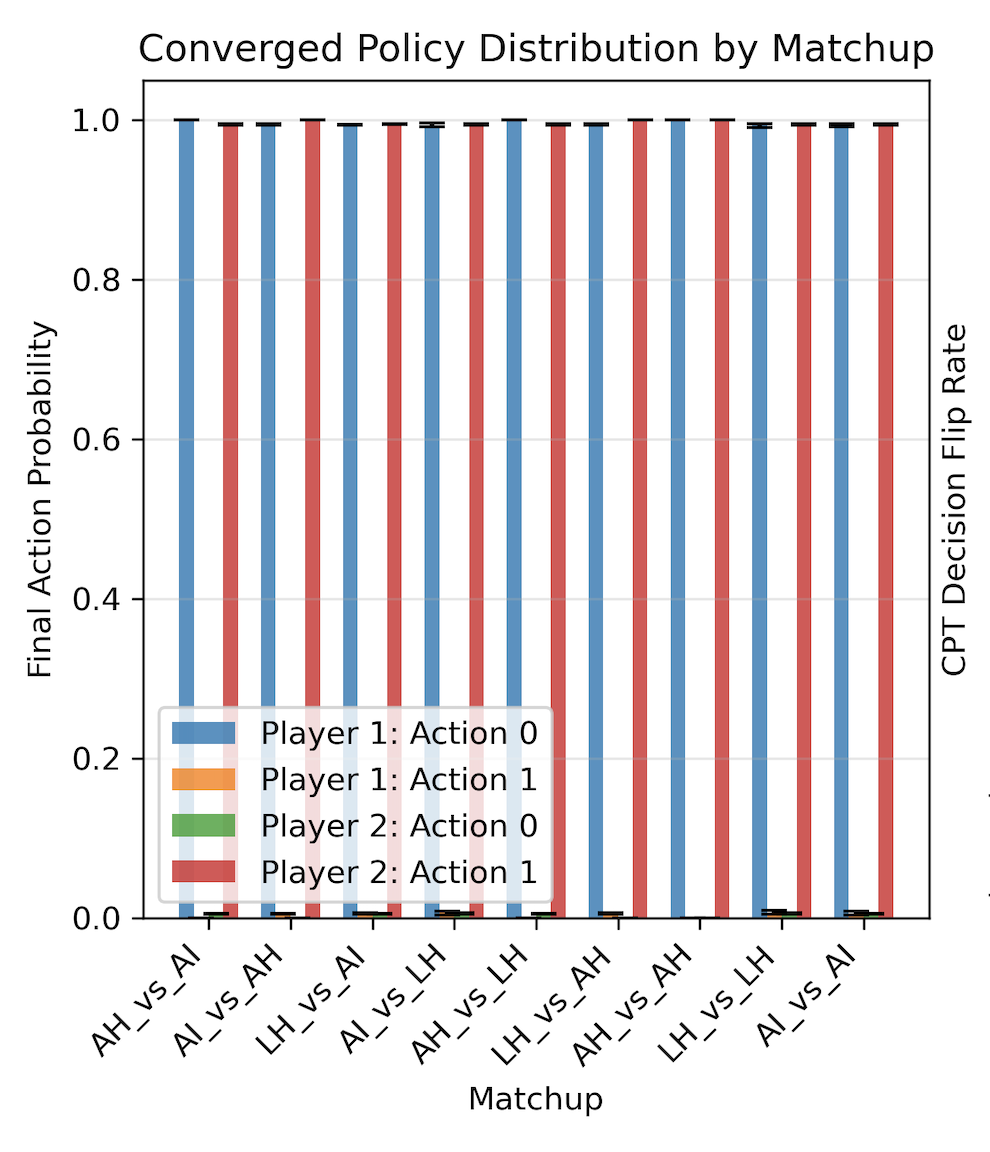}
        \caption{V-based}
        \label{fig:}
    \end{minipage}
    \caption{One Off Crawford's Counterexample Policies by Matchup (Last $5,000$ Steps)}
    \label{fig:CC1PI}
\end{figure}

In lieu of presenting the AH vs. AH baseline and state history ablations, we present the entire CC analysis in this section as a confirmation of non EU players' preference for pure strategies. First, we present the one off game policies in Figure~\ref{fig:CC1PI}.

Each matchup consists of pure strategies, failing to find the true equilibrium. Notably, AI vs. AI also converges to a pure strategy, potentially caused by noise in the learning process pushing the policy towards one action or another. 

\begin{figure}[H]
    \centering
    \begin{minipage}[t]{0.32\linewidth}
        \centering
        \includegraphics[width=\linewidth]{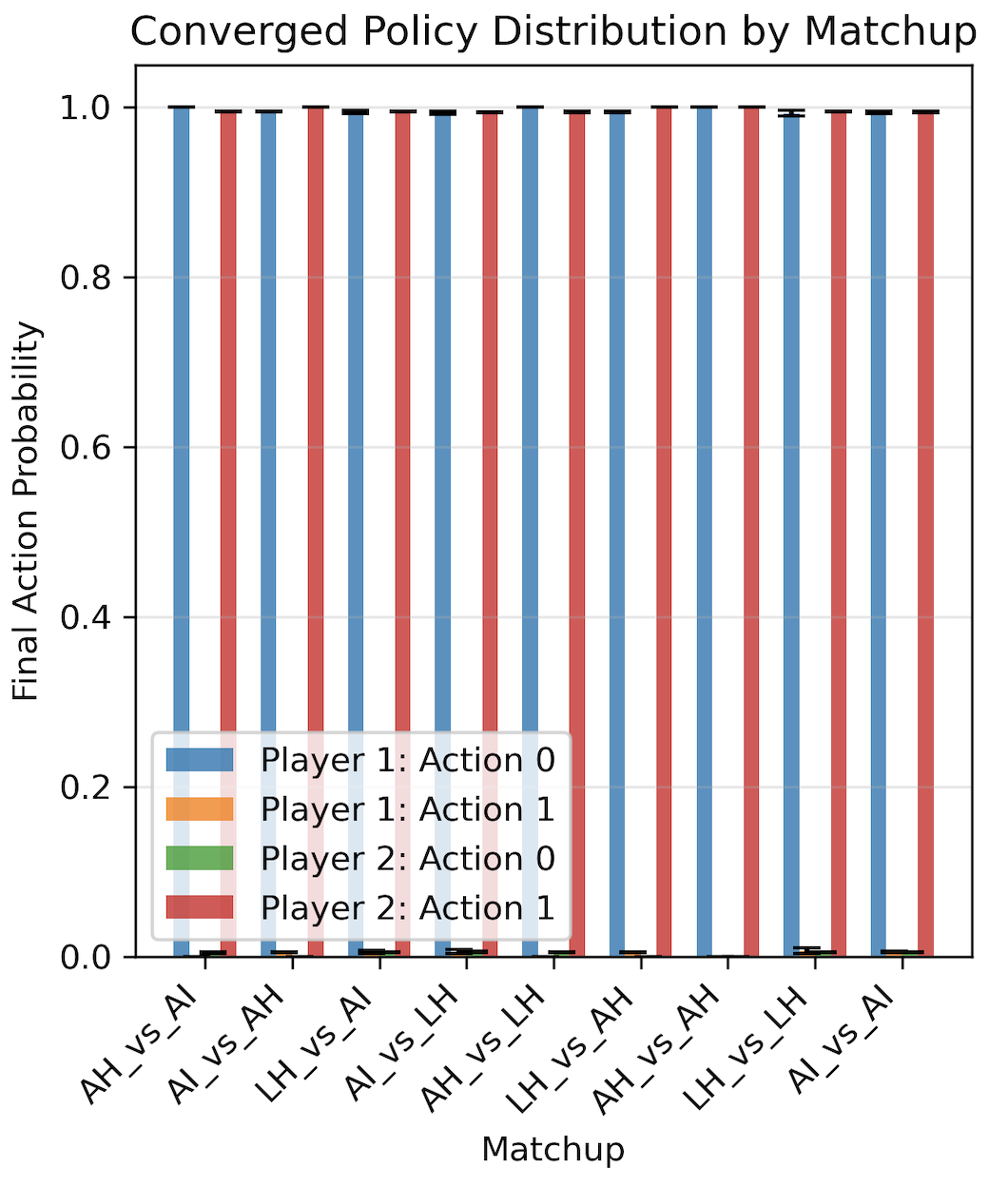}
        \caption{EMA}
        \label{fig:}
    \end{minipage}
    \begin{minipage}[t]{0.32\linewidth}
        \centering
        \includegraphics[width=\linewidth]{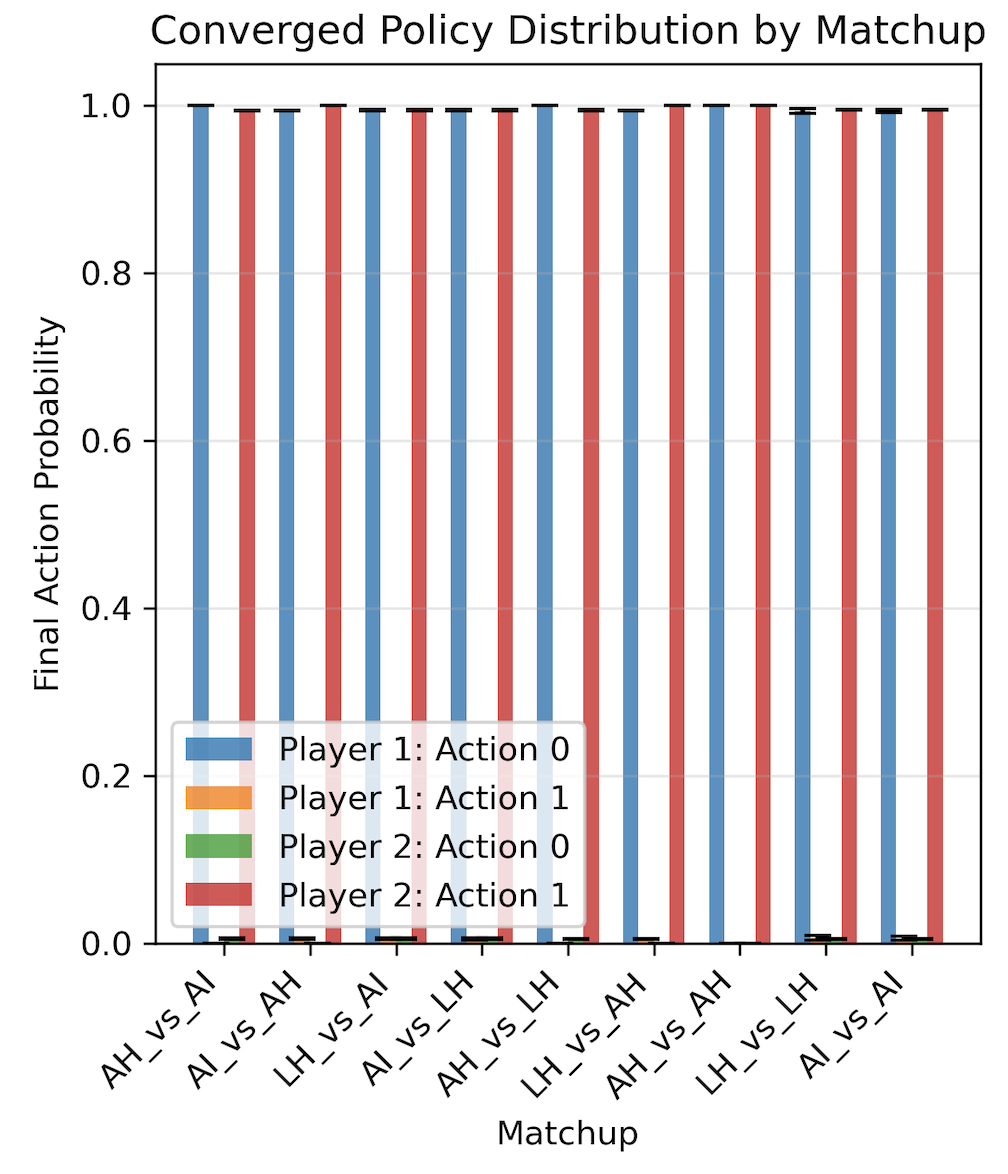}
        \caption{EMAOR}
        \label{fig:}
    \end{minipage}
    \begin{minipage}[t]{0.32\linewidth}
        \centering
        \includegraphics[width=\linewidth]{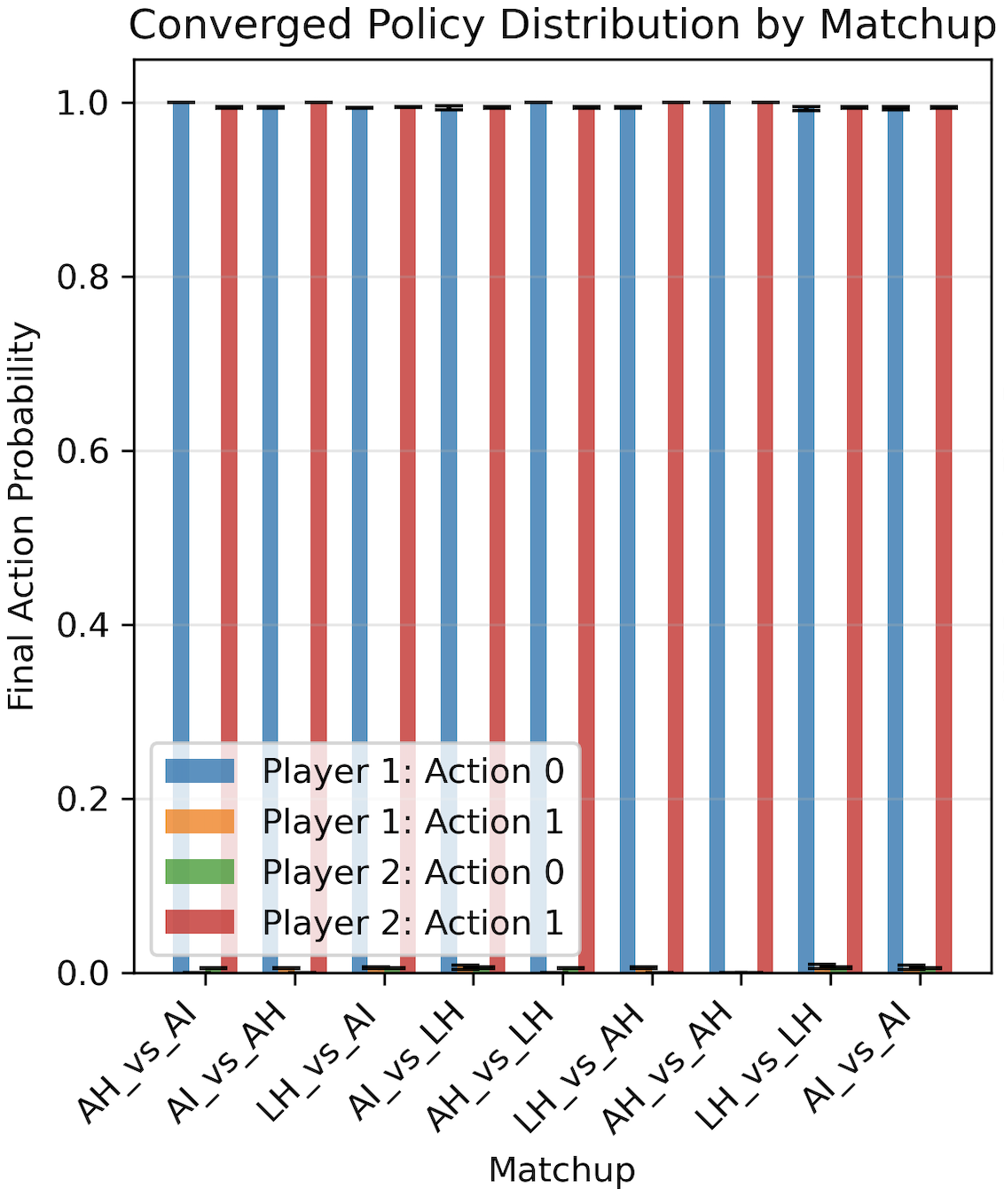}
        \caption{V-based}
        \label{fig:}
    \end{minipage}
    \caption{Repeated Crawford's Counterexample Policies by Matchup (Last $5,000$ Steps)}
    \label{fig:CCRPI}
\end{figure}

Next, we present the repeated CC policies by matchup in Figure~\ref{fig:CCRPI}. 

Once again, every matchup converges to pure strategies, demonstrating the non existence pathology that CC sets out to show. 

\subsubsection{Crawford's Counterexample Concluding Remarks}
Crawford's Counterexample sets out to show that players that violate the Von Neumann-Morgenstern independence axiom, as CPT players do, will be unable to find the mixed equilibrium in CC. Our numerical findings support his claim and verify it. We acknowledge that the AI vs. AI matchup failed to find equilibrium, and point to learning noise as a potential cause of the collapse.

\subsection{Ochs' Game}
Ochs' game is designed to expose equilibrium pathologies in games with PT player. Leclerc shows that no equilibrium exists in PT-NE, but that an equilibrium exists at $\approx (0.5, 0.05)$ for PT-EB. Our analysis will focus on identifying the basins of attraction that agents converge to and the upstream learning and preference dynamics that may have caused them. 

\subsubsection{One Off Game (State History = $0$}
\begin{figure}
    \centering
    \includegraphics[width=0.8\linewidth]{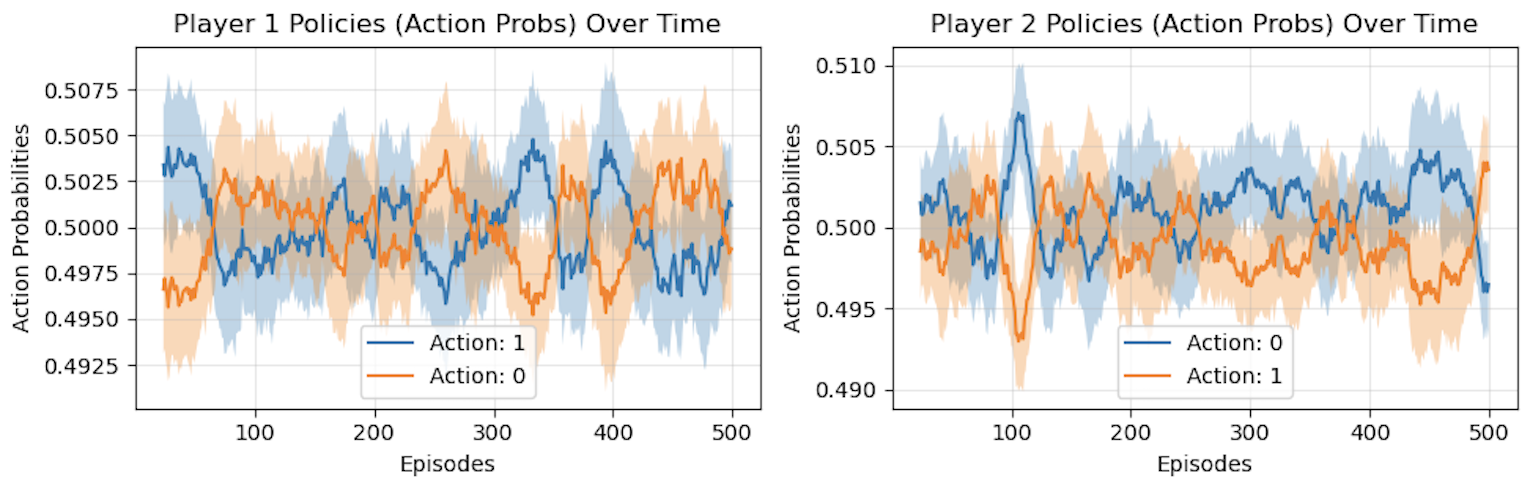}
    \caption{AH vs. AH Baseline Policy (EMA)}
    \label{fig:AHAHPIOchs}
\end{figure}

The AH vs. AH baselines all converged to $(0.5, 0.5)$ across all three reference points. Figure~\ref{fig:AHAHPIOchs} visualizes the AH vs. AH policies in the EMA setting as representative of all three trials. 

\begin{figure}[H]
    \centering
    \begin{minipage}[t]{0.32\linewidth}
        \centering
        \includegraphics[width=\linewidth]{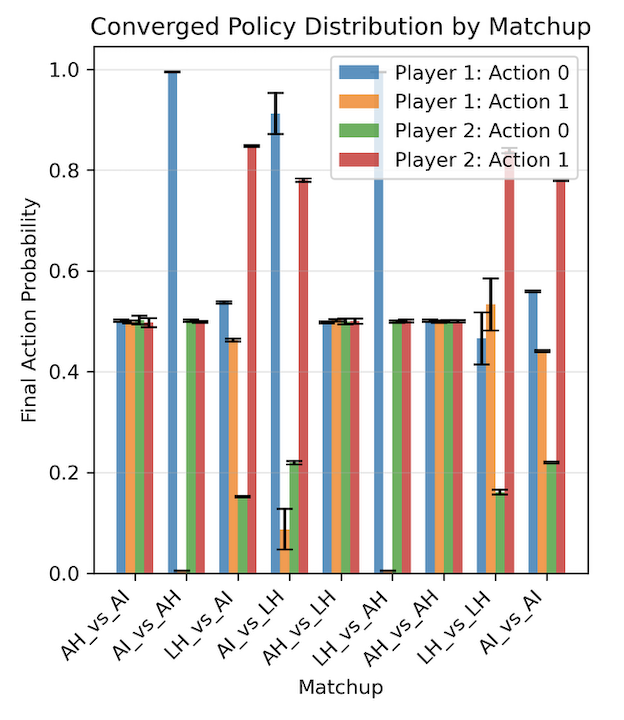}
        \caption{EMA}
        \label{fig:}
    \end{minipage}
    \begin{minipage}[t]{0.32\linewidth}
        \centering
        \includegraphics[width=\linewidth]{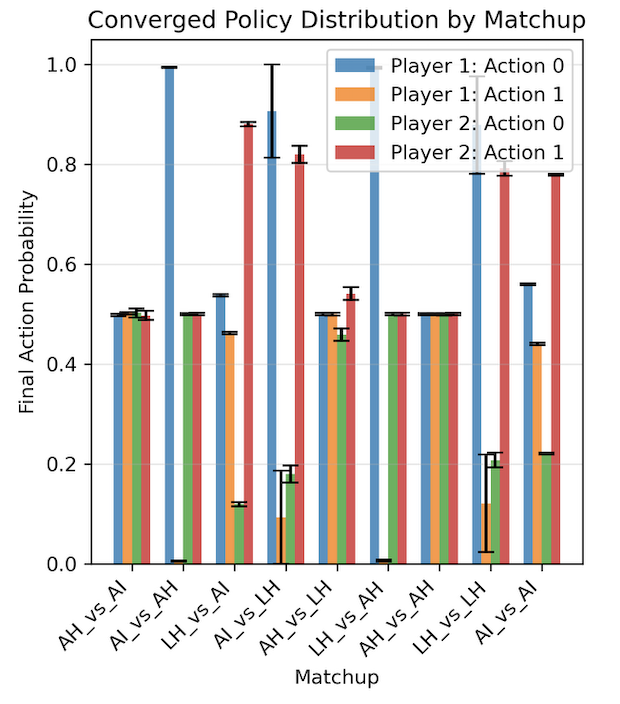}
        \caption{EMAOR}
        \label{fig:}
    \end{minipage}
    \begin{minipage}[t]{0.32\linewidth}
        \centering
        \includegraphics[width=\linewidth]{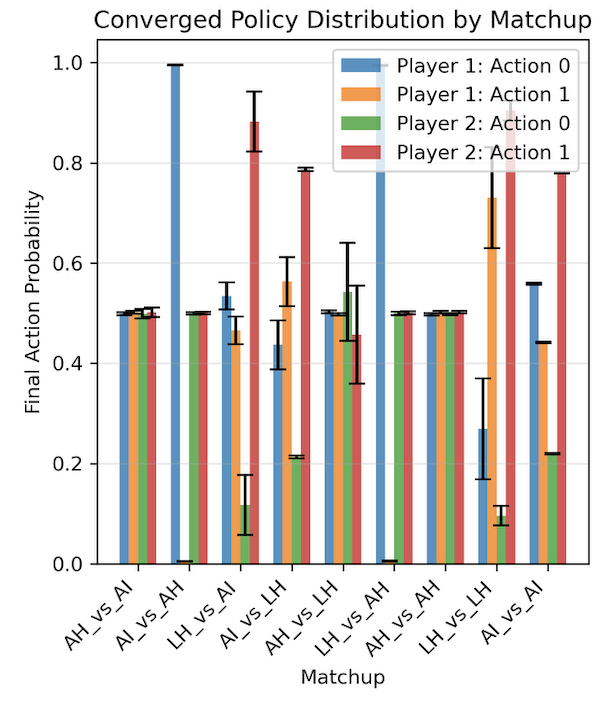}
        \caption{V-based}
        \label{fig:}
    \end{minipage}
    \caption{One Off Ochs' Game Policies by Matchup (Last $5,000$ Steps)}
    \label{fig:OG1Pi}
\end{figure}

The one off policies split by reference point type are visualized in Figure~\ref{fig:OG1Pi}. Matchups split convergence between $(0.5, 0.5)$ basins and $\approx (0.5, 0.1)$ basins, with some variance and transitivity between the values of $p$ and $q$ for $(0.5, 0.1)$. 
\vspace{-1.2em}

The only notable difference between the EMA and EMAOR runs was the switch from player 1 from $p = 0.5$ to $p \approx 0.8$. The V-based run told a similar story w.r.t. basins of attraction, but exhibited more variance between runs. 

\begin{figure}[H]
    \centering
    \begin{minipage}[t]{0.32\linewidth}
        \centering
        \includegraphics[width=\linewidth]{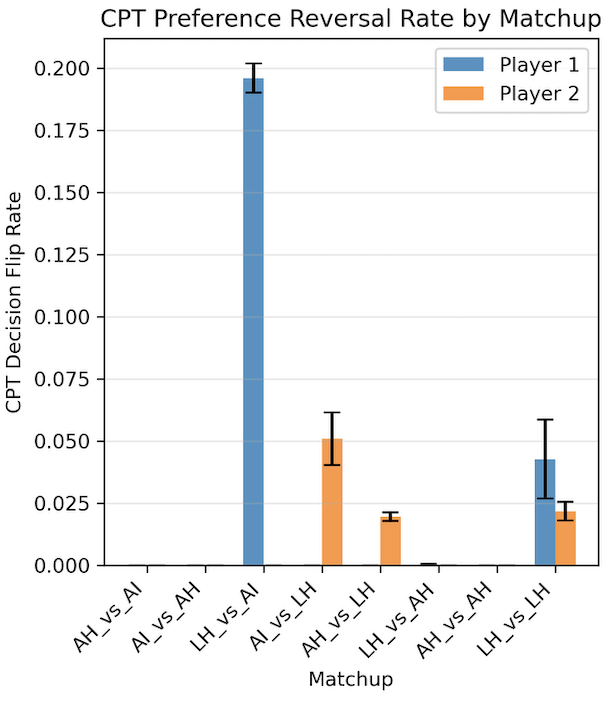}
        \caption{EMA}
        \label{fig:}
    \end{minipage}
    \begin{minipage}[t]{0.32\linewidth}
        \centering
        \includegraphics[width=\linewidth]{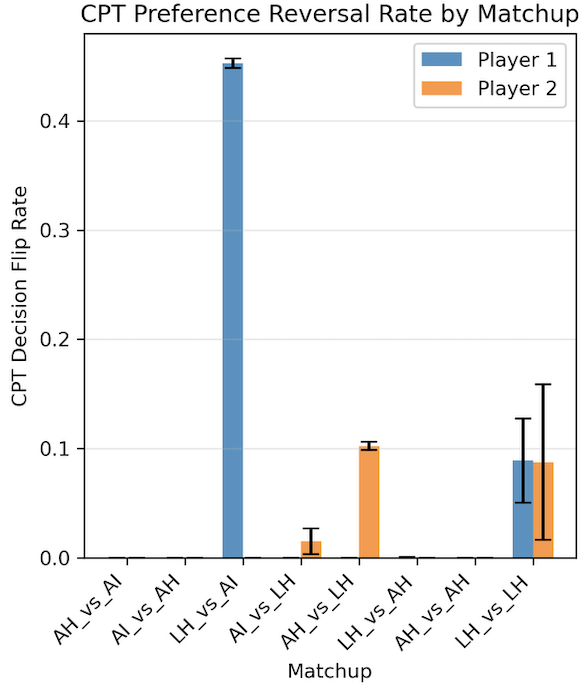}
        \caption{EMAOR}
        \label{fig:}
    \end{minipage}
    \begin{minipage}[t]{0.32\linewidth}
        \centering
        \includegraphics[width=\linewidth]{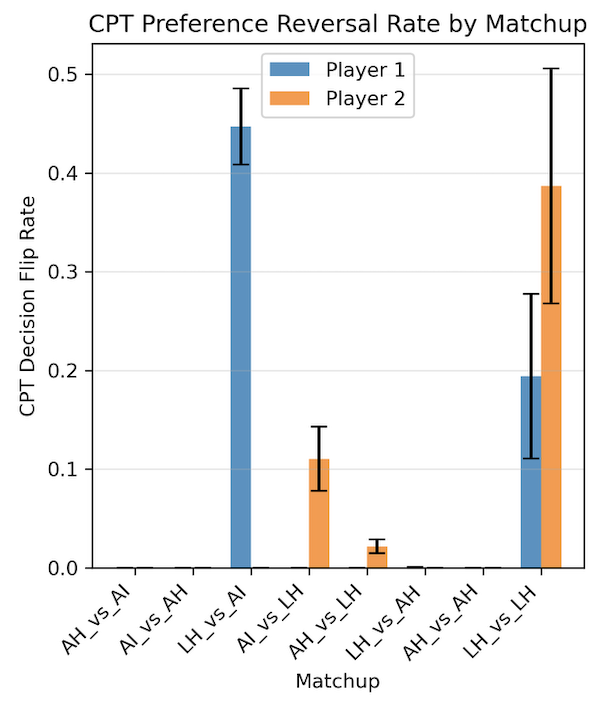}
        \caption{V-based}
        \label{fig:}
    \end{minipage}
    \caption{One Off Ochs' Game CPT Action Change Rates by Matchup (Last $5,000$ Steps)}
    \label{fig:OG1CPTA}
\end{figure}

Critically, the most interesting finding from Och's game was the influence of CPT on behavior, visualized in Figure~\ref{fig:OG1CPTA}.

CPT transformations accounted for nearly $50\%$ of the action decisions for EMAOR and V-Based reference types, two orders of magnitude higher than any other game in the study. 

\subsubsection{Repeated Game (State History =$2$)}
Repeated games told a similar story to the one off games, with basins of attraction emerging around $(0.5, 0.5)$ and $\approx (0.5, 0.2)$ and $(0.8, 0.8)$. 
\vspace{-1em}
\begin{figure}[H]
    \centering
    \begin{minipage}[t]{0.32\linewidth}
        \centering
        \includegraphics[width=\linewidth]{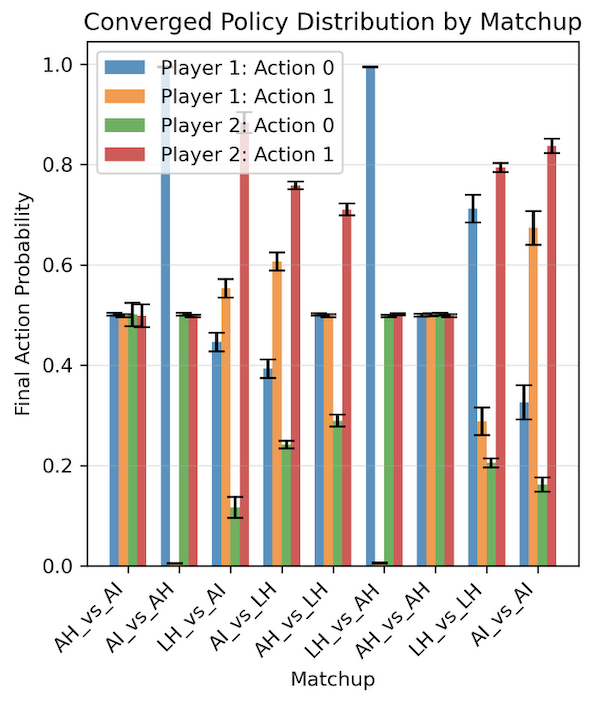}
        \caption{EMA}
        \label{fig:}
    \end{minipage}
    \begin{minipage}[t]{0.32\linewidth}
        \centering
        \includegraphics[width=\linewidth]{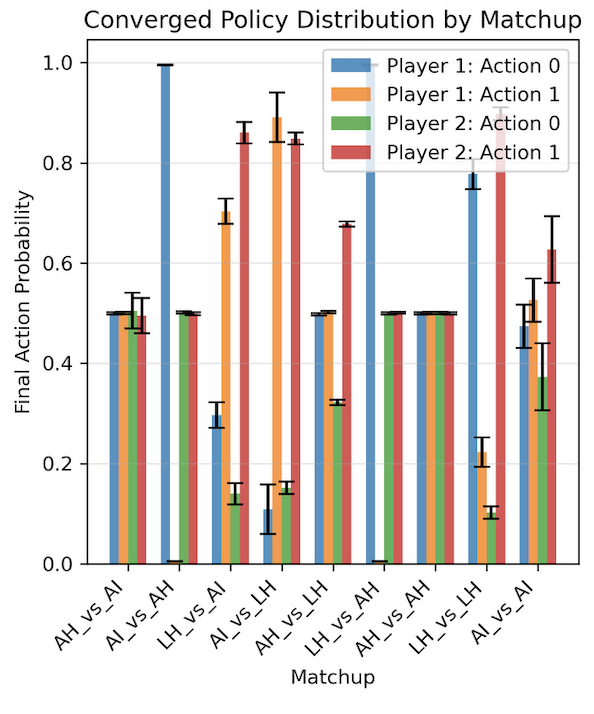}
        \caption{EMAOR}
        \label{fig:}
    \end{minipage}
    \begin{minipage}[t]{0.32\linewidth}
        \centering
        \includegraphics[width=\linewidth]{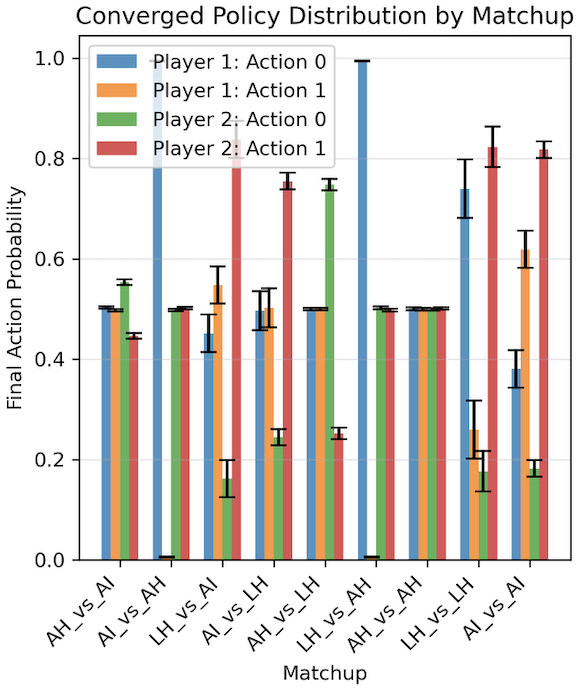}
        \caption{V-based}
        \label{fig:}
    \end{minipage}
    \caption{Repeated Ochs' Game Policies by Matchup (Last $5,000$ Steps)}
    \label{fig:OGRPi}
\end{figure}
The matchups did not all coincide to any specific basin of attraction, and the variance between mixed strategies suggests that pathologies were present. 

\subsubsection{Ochs' Game Concluding Remarks}
Ochs' Game was designed to draw out equilibrium pathologies in games with PT players, and our analysis revealed significant variance between strategies that do not correlate to the expected equilibrium at $(0.5, 0.05)$. Some matchups approached this basin, but none explictly landed in it, suggesting that equilibrium pathologies were present and verified in our analysis. 
\section{Conclusion}\label{sec:conc}
Our model updates classical game theoretic decision making with empirical human preferences to explore what convergence points may exist between human and AI competitive games, and finds that human agents typically find suboptimal, mixed basins when deviating from classical theory. 
The anomalous behaviors from the human agents mixed their strategies in suboptimal, but PT motivated (e.g. diminishing sensitivity) ways. Specifically, basins of attraction emerged in the Battle of the Sexes and Chicken that were not predicted by classical theory, but suggested the influence of PT preference warping. Importantly, the behavioral deviations did not necessarily correlate with changes in reward received, so broader conclusions about PT preferences effect on competitive outcomes is not clear.  

\paragraph{LH CPT/EU L2 Norm Spike During AH vs LH matchups}
An open question this paper leaves is regarding the behavior of the CPT transformation in AH vs. LH games with respect to the LH agent. Nearly every run, the LH agent, when matched against AH, dominated the L2 magnitude. Despite that exploding metric, its behavior in these matchups was unremarkable. Understanding the spike will be key to understanding how CPT is interacting on top of EU learning, and explaining why massive CPT transformations don't necessarily correlate with downstream behavioral changes. 

\subsection{Future Work}
Future work will focus on refining the model towards increasingly accurate representations of human behavior and preferences. For example, future extensions could consider beliefs as a part of the state representation for RL agents, and ablations where the reference point is not. Modeling the cognitive biases from PT in an RL setting requires significant subjective decision making, and more extensive ablations across these decisions can surface the tradeoffs and influence of each PT component with gameplay. 

Future work will also include evolutionary game theoretic analysis of PT and EU players with the intention of further inquiry into the competitive influence of PT preferences for humans competing with AIs. The point is that we are not only interested in classical game analysis, but in forecasting and planning for anticipated increased competition between humans and AIs. This line of research is largely in line with AI safety concerns and research. To turn the descriptive analysis into prescriptive, functional research we may impose mechanism design to study the effect of different policy choices on the proliferation and competitiveness of AI agents. 

Finally, future work may impose PT preferences onto public goods games and in trading/autonomous driving simulations to better account for the influence of human cognitive biases on the success and behavior of AI agents interacting with humans in the world. Embedding cognitive biases into AI decision making will help them accommodate human imperfections adn integrate into society more easily (e.g. by driving differently when around humans vs other AIs).

\newpage
\bibliographystyle{plainnat}
\bibliography{refs}

@Article{kahneman2013prospect,
journal={Econometrica},
author={Kahneman, Daniel and Tversky, Amos},
title={Prospect Theory: An Analysis of Decision under Risk},
year={1979},
month={March},
pages={263-291},
volume={47},
number={2},
doi={None},
url={https://ideas.repec.org/a/ecm/emetrp/v47y1979i2p263-91.html},
}

@book{page2022optimally,
  title={Optimally irrational: The good reasons we behave the way we do},
  author={Page, Lionel},
  year={2022},
  publisher={Cambridge University Press}
}

@book{leclerc2014prospect,
  title={Prospect theory preferences in noncooperative game theory},
  author={Leclerc, Phillip},
  year={2014},
  publisher={Virginia Commonwealth University}
}

@article{phade2019geometry,
  title={On the geometry of Nash and correlated equilibria with cumulative prospect theoretic preferences},
  author={Phade, Soham R and Anantharam, Venkat},
  journal={Decision Analysis},
  volume={16},
  number={2},
  pages={142--156},
  year={2019},
  publisher={INFORMS}
}

@article{cadre2025irrationality,
  title={How Irrationality Shapes Nash Equilibria: A Prospect-Theoretic Perspective},
  author={Cadre, H{\'e}l{\`e}ne Le and Bu{\v{s}}i{\'c}, Ana and others},
  journal={arXiv preprint arXiv:2504.16556},
  year={2025}
}

@article{wang2022cooperative,
  title={Cooperative and competitive multi-agent systems: From optimization to games},
  author={Wang, Jianrui and Hong, Yitian and Wang, Jiali and Xu, Jiapeng and Tang, Yang and Han, Qing-Long and Kurths, J{\"u}rgen},
  journal={IEEE/CAA Journal of Automatica Sinica},
  volume={9},
  number={5},
  pages={763--783},
  year={2022},
  publisher={IEEE}
}

@article{keskin2016equilibrium,
  title={Equilibrium notions for agents with cumulative prospect theory preferences},
  author={Keskin, Kerim},
  journal={Decision Analysis},
  volume={13},
  number={3},
  pages={192--208},
  year={2016},
  publisher={INFORMS}
}

@inproceedings{lalmohammed2025modeling,
  title={Modeling Human Behavior Without Humans: Prospect Theoretic Multi-Agent Reinforcement Learning},
  author={Lalmohammed, Sheyan and Gupta, Khush and Shah, Alok and Ramji, Keshav},
  booktitle={Proceedings of the ICML Workshop on Multi-Agent Security},
  year={2025},
  url={https://openreview.net/pdf?id=XDZdrTMWbM}
}

@misc{a2016cumulativeprospecttheorymeets,
      title={Cumulative Prospect Theory Meets Reinforcement Learning: Prediction and Control}, 
      author={Prashanth L. A. and Cheng Jie and Michael Fu and Steve Marcus and Csaba Szepesvári},
      year={2016},
      eprint={1506.02632},
      archivePrefix={arXiv},
      primaryClass={cs.LG},
      url={https://arxiv.org/abs/1506.02632}, 
}

@misc{lepel2025prospecttheoreticpolicygradientframework,
      title={A Prospect-Theoretic Policy Gradient Framework for Behaviorally Nuanced Reinforcement Learning}, 
      author={Olivier Lepel and Anas Barakat},
      year={2025},
      eprint={2410.02605},
      archivePrefix={arXiv},
      primaryClass={cs.LG},
      url={https://arxiv.org/abs/2410.02605},
}

@misc{ramasubramanian2021reinforcementlearningexpectation,
      title={Reinforcement Learning Beyond Expectation}, 
      author={Bhaskar Ramasubramanian and Luyao Niu and Andrew Clark and Radha Poovendran},
      year={2021},
      eprint={2104.00540},
      archivePrefix={arXiv},
      primaryClass={cs.LG},
      url={https://arxiv.org/abs/2104.00540}, 
}

@inproceedings{ghaemi2024namg, author = {Ghaemi, Hafez and Kebriaei, Hamed and Ramezani Moghaddam, Alireza and Nili Ahmadabadi, Majid}, title = {Risk-Sensitive Multi-Agent Reinforcement Learning in Network Aggregative Markov Games}, year = {2024}, isbn = {9798400704864}, publisher = {International Foundation for Autonomous Agents and Multiagent Systems}, address = {Richland, SC},  booktitle = {Proceedings of the 23rd International Conference on Autonomous Agents and Multiagent Systems}, pages = {2282–2284}, numpages = {3}, location = {Auckland, New Zealand}, series = {AAMAS '24} }

@misc{phade2020blackboxstrategiesequilibriumgames,
      title={Black-Box Strategies and Equilibrium for Games with Cumulative Prospect Theoretic Players}, 
      author={Soham R. Phade and Venkat Anantharam},
      year={2020},
      eprint={2004.09592},
      archivePrefix={arXiv},
      primaryClass={econ.TH},
      url={https://arxiv.org/abs/2004.09592}, 
}

@article{HOTA2016135,
title = {Fragility of the commons under prospect-theoretic risk attitudes},
journal = {Games and Economic Behavior},
volume = {98},
pages = {135-164},
year = {2016},
issn = {0899-8256},
doi = {https://doi.org/10.1016/j.geb.2016.06.003},
url = {https://www.sciencedirect.com/science/article/pii/S0899825616300458},
author = {Ashish R. Hota and Siddharth Garg and Shreyas Sundaram},
}

@article{METZGER2019396,
title = {Non-cooperative games with prospect theory players and dominated strategies},
journal = {Games and Economic Behavior},
volume = {115},
pages = {396-409},
year = {2019},
issn = {0899-8256},
doi = {https://doi.org/10.1016/j.geb.2019.04.001},
url = {https://www.sciencedirect.com/science/article/pii/S089982561930051X},
author = {Lars Peter Metzger and Marc Oliver Rieger},
}

@article{RIEGER20141,
title = {Evolutionary stability of prospect theory preferences},
journal = {Journal of Mathematical Economics},
volume = {50},
pages = {1-11},
year = {2014},
issn = {0304-4068},
doi = {https://doi.org/10.1016/j.jmateco.2013.11.002},
url = {https://www.sciencedirect.com/science/article/pii/S0304406813001055},
author = {Marc Oliver Rieger},
}

@article{CRAWFORD1990127,
title = {Equilibrium without independence},
journal = {Journal of Economic Theory},
volume = {50},
number = {1},
pages = {127-154},
year = {1990},
issn = {0022-0531},
doi = {https://doi.org/10.1016/0022-0531(90)90088-2},
url = {https://www.sciencedirect.com/science/article/pii/0022053190900882},
author = {Vincent P Crawford}
}

@article{Borkar_2021,
   title={Prospect-theoretic Q-learning},
   volume={156},
   ISSN={0167-6911},
   url={http://dx.doi.org/10.1016/j.sysconle.2021.105009},
   DOI={10.1016/j.sysconle.2021.105009},
   journal={Systems \&amp; Control Letters},
   publisher={Elsevier BV},
   author={Borkar, Vivek S. and Chandak, Siddharth},
   year={2021},
   month=oct, pages={105009} }

@misc{danisMARL2023,
author = {Dominic Danis and Parker Parmacek and David Dunajsky and Bhaskar Ramasubramanian},
title = {Multi-Agent Reinforcement Learning with Prospect Theory},
booktitle = {2023 Proceedings of the Conference on Control and its Applications (CT)},
chapter = {},
pages = {9-16},
doi = {10.1137/1.9781611977745.2},
URL ={https://epubs.siam.org/doi/abs/10.1137/1.9781611977745.2},
eprint = {https://epubs.siam.org/doi/pdf/10.1137/1.9781611977745.2},
}

@misc{phade2020learninggamescumulativeprospect,
      title={Learning in Games with Cumulative Prospect Theoretic Preferences}, 
      author={Soham R. Phade and Venkat Anantharam},
      year={2020},
      eprint={1804.08005},
      archivePrefix={arXiv},
      primaryClass={cs.GT},
      url={https://arxiv.org/abs/1804.08005}, 
}

@article{Tversky1992,
  author    = {Tversky, Amos and Kahneman, Daniel},
  title     = {Advances in Prospect Theory: Cumulative Representation of Uncertainty},
  journal   = {Journal of Risk and Uncertainty},
  year      = {1992},
  volume    = {5},
  number    = {4},
  pages     = {297--323},
  doi       = {10.1007/BF00122574},
  url       = {https://doi.org/10.1007/BF00122574},
  issn      = {1573-0476}
}

@article{OCHS1995202,
title = {Games with Unique, Mixed Strategy Equilibria: An Experimental Study},
journal = {Games and Economic Behavior},
volume = {10},
number = {1},
pages = {202-217},
year = {1995},
issn = {0899-8256},
doi = {https://doi.org/10.1006/game.1995.1030},
url = {https://www.sciencedirect.com/science/article/pii/S0899825685710305},
author = {Jack Ochs}
}

@book{Sutton1998,
  author = {Sutton, Richard S. and Barto, Andrew G.},
  edition = {Second},
  publisher = {The MIT Press},
  title = {Reinforcement Learning: An Introduction},
  url = {http://incompleteideas.net/book/the-book-2nd.html},
  year = {2018 }
}

@book{Bellman:DynamicProgramming,
  author = {Bellman, Richard},
  isbn = {9780486428093},
  publisher = {Dover Publications},
  title = {{Dynamic Programming}},
  year = 1957
}

@article{Shapley,
author = {L. S. Shapley },
title = {Stochastic Games*},
journal = {Proceedings of the National Academy of Sciences},
volume = {39},
number = {10},
pages = {1095-1100},
year = {1953},
doi = {10.1073/pnas.39.10.1095},
URL = {https://www.pnas.org/doi/abs/10.1073/pnas.39.10.1095},
eprint = {https://www.pnas.org/doi/pdf/10.1073/pnas.39.10.1095}}

@book{Owen1982,
  author    = {Guillermo Owen},
  title     = {Game Theory},
  edition   = {2},
  year      = {1982},
  publisher = {Academic Press},
  address   = {Orlando, Florida}
}

@inproceedings{Littman1994MarkovGA,
  title={Markov Games as a Framework for Multi-Agent Reinforcement Learning},
  author={Michael L. Littman},
  booktitle={International Conference on Machine Learning},
  year={1994},
  url={https://api.semanticscholar.org/CorpusID:8108362}
}

@Article{disposition,
journal={Journal of Finance},
author={ Shefrin, Hersh and Statman, Meir},
title={The Disposition to Sell Winners Too Early and Ride Losers Too Long: Theory and Evidence},
year={1985},
month={July},
pages={777-790},
volume={40},
number={3},
doi={None},
url={https://ideas.repec.org/a/bla/jfinan/v40y1985i3p777-90.html},
}

@article{SadighAutonomous,
author = {Sadigh, Dorsa and Landolfi, Nick and Sastry, Shankar S. and Seshia, Sanjit A. and Dragan, Anca D.},
title = {Planning for cars that coordinate with people: leveraging effects on human actions for planning and active information gathering over human internal state},
year = {2018},
issue_date = {Oct 2018},
publisher = {Kluwer Academic Publishers},
address = {USA},
volume = {42},
number = {7},
issn = {0929-5593},
url = {https://doi.org/10.1007/s10514-018-9746-1},
doi = {10.1007/s10514-018-9746-1},
journal = {Auton. Robots},
month = oct,
pages = {1405–1426},
numpages = {22}
}

@ARTICLE{fridmanmitcars,
  author={Fridman, Lex and Brown, Daniel E. and Glazer, Michael and Angell, William and Dodd, Spencer and Jenik, Benedikt and Terwilliger, Jack and Patsekin, Aleksandr and Kindelsberger, Julia and Ding, Li and Seaman, Sean and Mehler, Alea and Sipperley, Andrew and Pettinato, Anthony and Seppelt, Bobbie D. and Angell, Linda and Mehler, Bruce and Reimer, Bryan},
  journal={IEEE Access}, 
  title={MIT Advanced Vehicle Technology Study: Large-Scale Naturalistic Driving Study of Driver Behavior and Interaction With Automation}, 
  year={2019},
  volume={7},
  number={},
  pages={102021-102038},
  doi={10.1109/ACCESS.2019.2926040}}

@article{investorreluctantodean,
author = {Odean, Terrance},
title = {Are Investors Reluctant to Realize Their Losses?},
journal = {The Journal of Finance},
volume = {53},
number = {5},
pages = {1775-1798},
doi = {https://doi.org/10.1111/0022-1082.00072},
url = {https://onlinelibrary.wiley.com/doi/abs/10.1111/0022-1082.00072},
eprint = {https://onlinelibrary.wiley.com/doi/pdf/10.1111/0022-1082.00072},
year = {1998}
}

@inproceedings{secureGrosslags,
author = {Grossklags, Jens and Christin, Nicolas and Chuang, John},
title = {Secure or insure? a game-theoretic analysis of information security games},
year = {2008},
isbn = {9781605580852},
publisher = {Association for Computing Machinery},
address = {New York, NY, USA},
url = {https://doi.org/10.1145/1367497.1367526},
doi = {10.1145/1367497.1367526},
booktitle = {Proceedings of the 17th International Conference on World Wide Web},
pages = {209–218},
numpages = {10},
location = {Beijing, China},
series = {WWW '08}
}

@article{Shogren1990,
  author    = {Shogren, J. F.},
  title     = {On increased risk and the voluntary provision of public goods},
  journal   = {Social Choice and Welfare},
  year      = {1990},
  volume    = {7},
  number    = {3},
  pages     = {221--229},
  doi       = {10.1007/BF01395723}
}

@article{aggRiskAverse,
author = {Palvi Aggarwal and Omkar Thakoor and Aditya Mate and Milind Tambe and Edward A. Cranford and Christian Lebiere and Cleotilde Gonzalez},
title ={An Exploratory Study of a Masking Strategy of Cyberdeception Using CyberVAN},

journal = {Proceedings of the Human Factors and Ergonomics Society Annual Meeting},
volume = {64},
number = {1},
pages = {446-450},
year = {2020},
doi = {10.1177/1071181320641100},

URL = { 
    
    
        https://journals.sagepub.com/doi/abs/10.1177/1071181320641100
    

},
eprint = { 
    
    
        https://journals.sagepub.com/doi/pdf/10.1177/1071181320641100
    

}
,

}

@article{AGGARWAL2022102671,
title = {Designing effective masking strategies for cyberdefense through human experimentation and cognitive models},
journal = {Computers \& Security},
volume = {117},
pages = {102671},
year = {2022},
issn = {0167-4048},
doi = {https://doi.org/10.1016/j.cose.2022.102671},
url = {https://www.sciencedirect.com/science/article/pii/S0167404822000700},
author = {Palvi Aggarwal and Omkar Thakoor and Shahin Jabbari and Edward A. Cranford and Christian Lebiere and Milind Tambe and Cleotilde Gonzalez},

}

@article{FehrFair,
    author = {Fehr, Ernst and Schmidt, Klaus M.},
    title = {A Theory of Fairness, Competition, and Cooperation*},
    journal = {The Quarterly Journal of Economics},
    volume = {114},
    number = {3},
    pages = {817-868},
    year = {1999},
    month = {08},
    doi = {10.1162/003355399556151},
    url = {https://doi.org/10.1162/003355399556151},
    eprint = {https://academic.oup.com/qje/article-pdf/114/3/817/5228571/114-3-817.pdf},
}

@article{KimKankanhalli2009,
  author    = {Kim, Hee-Woong and Kankanhalli, Atreyi},
  title     = {Investigating User Resistance to Information Systems Implementation: A Status Quo Bias Perspective},
  journal   = {MIS Quarterly},
  year      = {2009},
  volume    = {33},
  number    = {3},
  pages     = {567--582}
}
\end{document}